\documentclass[aps,prd,twocolumn,showpacs,notitlepage,eqsecnum,
superscriptaddress]{revtex4-1}

\usepackage[centertags]{amsmath}
\usepackage{amssymb}
\usepackage{latexsym}
\usepackage{enumerate}
\usepackage{enumitem}
\usepackage{graphicx}
\usepackage{mathrsfs}
\usepackage{hyperref}
\usepackage{stmaryrd}
\usepackage{import}
\usepackage{tensor}
\usepackage{color}
\usepackage{bm}
\usepackage{multirow}
\usepackage{breakurl}
\usepackage{float}
\usepackage{placeins}
\usepackage{autobreak}
 
\newcommand{\bi}{\begin{itemize}}
\newcommand{\ei}{\end{itemize}}
\newcommand{\be}{\begin{equation}}
\newcommand{\ee}{\end{equation}}

\renewcommand{\l}{\left(}
\renewcommand{\r}{\right)}
\renewcommand{\a}{\alpha}

\renewcommand{\O}{\Omega}
\renewcommand{\o}{\omega}

\newcommand{\q}{\quad}
\newcommand{\qq}{\qquad}

\newcommand{\vp}{\varphi}

\newcommand{\pa}{\partial}

\allowdisplaybreaks

\begin{document}

\title{Determination of new coefficients in the angular momentum and energy fluxes at infinity to 9PN
for eccentric Schwarzschild extreme-mass-ratio inspirals using mode-by-mode fitting}

\author{Christopher Munna}
\author{Charles R. Evans}
\affiliation{Department of Physics and Astronomy, University of North 
Carolina, Chapel Hill, North Carolina 27599, USA}
\author{Seth Hopper}
\affiliation{Department of Physics and Astronomy, Earlham College, Richmond, 
IN 47374, USA}
\author{Erik Forseth}
\affiliation{Graham Capital Management, Rowayton, CT 06853, USA}

\begin{abstract}
We present an extension of work in an earlier paper showing high precision comparisons between black hole perturbation 
theory and post-Newtonian (PN) theory in their region of overlapping validity for bound, eccentric-orbit, Schwarzschild 
extreme-mass-ratio inspirals.  As before we apply a numerical fitting scheme to extract eccentricity coefficients in 
the PN expansion of the gravitational wave fluxes, which are then converted to exact analytic form using an 
integer-relation algorithm.  In this work, however, we fit to individual $lmn$ modes to exploit simplifying 
factorizations that lie therein.  Since the previous paper focused solely on the energy flux, here we concentrate 
initially on analyzing the angular momentum flux to infinity.  A first step involves finding convenient forms for 
hereditary contributions to the flux at low-PN order, analogous to similar terms worked out previously 
for the energy flux.  We then apply the upgraded techniques to find new PN terms through 9PN order and (at many 
PN orders) to $e^{30}$ in the power series in eccentricity.  With the new approach applied to angular momentum 
fluxes, we return to the energy fluxes at infinity to extend those previous results.  Like before, the underlying 
method uses a \textsc{Mathematica} code based on use of the Mano-Suzuki-Takasugi (MST) function expansion formalism to 
represent gravitational perturbations and spectral source integration (SSI) to find numerical results at arbitrarily 
high precision.  
\end{abstract}


\maketitle

\section{Introduction}
\label{sec:intro}

At present the two-body problem is the subject of ongoing investigation in gravitational physics \cite{BaraETC18}.  
As two massive objects orbit one another, gravitational waves are emitted that carry off energy and angular momentum 
and drive an inspiral and eventual merger of the binary.  Of particular interest is the class of binaries known as 
extreme-mass-ratio inspirals (EMRIs), in which the primary black hole of mass $M$ is much heavier than the secondary 
compact object of mass $\mu$ (i.e., $\mu/M\ll 1$) \cite{AmarETC07}.  Such systems will be sources of gravitational 
waves observable by the LISA detector set to launch in 2034 \cite{eLISA,LISA}.  EMRIs can be understood using black 
hole perturbation theory (BHPT), i.e., expansion in powers of the small mass ratio, and subsequent calculation of the 
gravitational self-force (GSF) \cite{PoisPounVega11}. 

An orthogonal approach for approximating the motion of binary orbits is post-Newtonian (PN) theory \cite{Blan14}, 
which is most suited for widely separated orbits or, equivalently, slowly-orbiting systems.  In PN theory corrections 
are computed in powers of the small velocity $v/c\ll 1$.  Additionally, a fruitful region of overlap exists between 
BHPT and PN regimes, where simultaneously $\mu/M\ll 1$ and $v/c\ll 1$ and where both formalisms may be applied and 
compared \cite{Detw08,SagoBaraDetw08,BaraSago09,BlanETC09,BlanETC10,Fuji12a,Fuji12b,ShahFrieWhit14,Shah14,DolaETC14b,
JohnMcDaShahWhit15,BiniDamoGera15,AkcaETC15,SagoFuji15,ForsEvanHopp16,HoppKavaOtte16,BiniDamoGera16a,BiniDamoGera16b,
BiniDamoGera16c,MunnEvan19a}.

Some of these BHPT studies have involved application of the MST (for Mano, Suzuki, and Takasugi) formalism 
\cite{ManoSuzuTaka96a,ManoSuzuTaka96b} for obtaining homogeneous solutions to the radial Teukolsky equation.
In \cite{GanzETC07}, the authors used the MST formalism, along with a PN expansion of the geodesic equation, to 
develop PN expansions for the rates of change of the constants of motion for arbitrarily inclined, eccentric
orbits on Kerr backgrounds.  That work was recently extended in \cite{SagoFuji15}, producing results accurate to 
4PN order relative to the Newtonian result and accurate to $\mathcal{O}(e^6)$ in a power series expansion in the 
orbital eccentricity.  Others \cite{Fuji12a,ShahFrieWhit14,Shah14,Fuji15} applied MST to arrive at extremely 
high-order PN expansions for the rates of change of orbital constants as well as self-force quantities in the case of 
circular orbits on both Schwarzschild and Kerr backgrounds.

In an earlier paper \cite{ForsEvanHopp16} (hereafter referred to as Paper I), several of us extended the methods of
\cite{ShahFrieWhit14,Shah14,JohnMcDaShahWhit15} to compute the energy flux radiated to infinity by eccentric EMRIs on
Schwarzschild backgrounds, using the method of spectral source integration \cite{HoppETC15} to treat the eccentric 
source.  By using a \textsc{Mathematica} code and computing the flux to high precision (up to 200 significant digits) 
for a variety of orbital parameters, it proved possible to fit out multiple terms in the high-order PN expansion (to 7PN 
relative order) to varying depths in the power series expansion in eccentricity $e$.  Furthermore, by applying 
experimental mathematics techniques \cite{JohnMcDaShahWhit15} such as the PSLQ algorithm \cite{FergBailArno99} 
(an integer-relation algorithm), it was possible to determine exact, analytic forms for a number of those expansion 
coefficients. 

In the present paper, we take those methods a step further by performing a separate fit on each individual $lmn$ 
mode.  Past work \cite{JohnMcDa14} revealed that for circular orbits the $lm$ modes of the energy flux have 
patterns in the PN expansion that allow factorization and simplification, which in turn helps improve the numerical 
fit.  Motivated by those discoveries in circular orbits, we found similar, generalized simplifications and 
factorizations in the $lmn$ modes for eccentric orbits, which also allows sharply improved fitting.  With these 
techniques, we are able to push the analytical understanding of the gravitational wave fluxes to 9PN order and to as 
far as $\mathcal{O}(e^{30})$ in eccentricity for most PN terms.  At some PN orders it has been possible 
\cite{MunnEvan19a,MunnEvan19b} to develop an underlying physical explanation for the high PN order flux contributions 
in terms of lower order multipole moments of the source motion.

The first application we make of this new technique is to gravitational wave angular momentum emission to 
infinity.  Most of the prior understanding of angular momentum emission in eccentric EMRIs, out to 3PN relative 
order, is found in \cite{ArunETC09a}, which extended comparable analysis on energy fluxes found in 
\cite{ArunETC08a, ArunETC08b}.  Beyond that, as noted above, Sago and Fujita \cite{SagoFuji15} showed 
expansions for the angular momentum flux to 4PN relative order through $\mathcal{O}(e^6)$ in the eccentricity 
expansion.  As part of our analysis, we verify all of these previous results.  We then apply the new approach to take 
a renewed look at the energy fluxes, and present an improved determination of that expansion that augments the 
results found in Paper I.

The outline of this paper is as follows.  We start in Sec.~\ref{sec:BHPT} by briefly reviewing the MST and SSI 
methods that are used to calculate the first order gravitational field perturbation and thence the 
first-order-in-mass-ratio angular momentum and energy fluxes.  In Sec.~\ref{sec:LcurrPN} we review the current 
state of knowledge on the fluxes as determined by PN theory.  We continue in that section with further investigation 
of the angular momentum flux finding original arbitrary-order expansions of the low-PN order enhancement functions 
via decomposition into Fourier modes.  Like the corresponding section in Paper I (Sec.~IV), we focus on the hereditary 
or ``tail'' terms, briefly deriving their asymptotic forms to verify the eccentricity singular behavior as 
$e \rightarrow 1$.  We then in Sec.~\ref{sec:lmnfitting} detail the improved method of fitting by $lmn$ mode, which 
involves the use of a novel eccentric-orbit analog of the well-known ``eulerlog'' function 
\cite{DamoIyerNaga09,JohnMcDa14}.  Finally, in Secs.~\ref{sec:LResults} and \ref{sec:Eresults} we present the new 
coefficients we have found using this method, first giving angular momentum flux terms through 9PN order (some of 
these angular momentum flux terms were found earlier by the methods of Paper I and reported in \cite{Fors16}) and then 
updating the energy flux results of Paper I to the same level.  The now-known length of many of these expressions 
precludes giving them fully in this paper.  Though where we have truncated expressions for brevity in 
Secs.~\ref{sec:LResults} and \ref{sec:Eresults}, the full results can be found in a \textsc{Mathematica} notebook 
posted on the archival Black Hole Perturbation Toolkit website \cite{BHPTK18}.

Throughout this paper we use units in which $c = G = 1$ and metric signature $(-+++)$.  Our notation for the RWZ 
formalism follows that found in Paper I, which in part derives from notational changes for tensor spherical harmonics 
and perturbation amplitudes made by Martel and Poisson \cite{MartPois05}.  For the MST formalism, we largely 
make use of the discussion and notation found in the review by Sasaki and Tagoshi \cite{SasaTago03}.

\section{BHPT background and formalism}
\label{sec:BHPT}

The angular momentum and energy radiated to infinity by a small body in eccentric orbit about a Schwarzschild black 
hole can be described using the methods of first-order BHPT.  At this order the small body can be treated as a 
point mass.  In this section we briefly summarize the notation used for describing bound eccentric orbits and review
the formalism behind our flux calculations. Full details can be found in Paper I. 

\subsection{Bound orbits on a Schwarzschild background}
\label{sec:orbits}

We consider generic bound motion of a point mass $\mu$ around a Schwarzschild black hole of mass $M$ in 
the equatorial plane, with $\mu/M \ll 1$.  At lowest order in the mass ratio the primary can be regarded as a stationary 
black hole.  Schwarzschild coordinates $x^{\mu} = (t,r,\theta, \varphi )$ are used, with the line element given by
\be
ds^2 = -f dt^2 + f^{-1} dr^2
+ r^2 \left( d\theta^2 + \sin^2\theta \, d\varphi^2 \right) ,
\ee
where $f(r) = 1 - 2M/r$.  Likewise at lowest order, the motion of the small body will approximate that of a geodesic in 
this background.  Several first integrals can be exploited allowing the four-velocity to be written as
\be
\label{eqn:four_velocity}
u^\a(\tau) = \frac{dx_p^{\alpha}(\tau)}{d\tau} = \l \frac{{\mathcal{E}}}{f_{p}}, u^r, 0, \frac{{\mathcal{L}}}{r_p^2} \r ,
\ee
where $\mathcal{E}$ is the specific energy, $\mathcal{L}$ is the angular momentum, and the subscript $p$
indicates evaluation at the location of the particle.  The radial motion is given by
\be
\label{eqn:rpDots}
\dot r_p^2(t) = f_{p}^2 \left[ 1 - \frac{f_p}{{\mathcal{E}}^2} 
\l 1 + \frac{{\mathcal{L}}^2}{r_p^2} \r \right] ,
\ee
where dot refers to a derivative with respect to coordinate time $t$. 

As usual, we reparameterize these equations using Darwin's geometric quantities $p$ (the semi-latus rectum)
and $e$ (the eccentricity) \cite{Darw61,CutlKennPois94,BaraSago10}, which are more conducive to PN 
expansion.  These parameters are related to $\mathcal{E}$ and $\mathcal{L}$ by
\be
\label{eqn:defeandp}
{\mathcal{E}}^2 = \frac{(p-2)^2-4e^2}{p(p-3-e^2)},
\q
{\mathcal{L}}^2 = \frac{p^2 M^2}{p-3-e^2}.
\ee
We transform similarly the curve parameter of the orbital motion from $\tau$ to Darwin's relativistic anomaly $\chi$,
casting the radial position into the Keplerian-like form
\be
r_p \l \chi \r = \frac{pM}{1+ e \cos \chi} ,
\ee
where one radial libration corresponds to a $2 \pi$ advance in $\chi$ \cite{Darw61}.  This equation can be combined 
with previous ones to generate simple, singularity-free ordinary differential equations (ODEs) for each remaining 
Schwarzschild coordinate location \cite{CutlKennPois94}.  In this way, $\vp_p(\chi)$ can be integrated analytically 
in terms of elliptic functions and $t_p(\chi)$ can be determined numerically.   

In addition, this representation provides simple means to compute the two fundamental frequencies, $\O_r$ and $\O_\vp$.  
Explicitly, the radial libration period is found to be
\begin{gather}
\label{eqn:O_r}
T_r = \int_{0}^{2 \pi}  \frac{r_p \l \chi \r^2}{M (p - 2 - 2 e \cos \chi)}
 \left[\frac{(p-2)^2 -4 e^2}{p -6 -2 e \cos \chi} \right]^{1/2}    d \chi ,
 \notag
\end{gather}
with $\Omega_r = 2 \pi/T_r$.  The frequency of (mean) azimuthal advance is 
\be
\label{eqn:O_phi}
\O_\varphi = \frac{4}{T_r} \left(\frac{p}{p - 6 - 2 e}\right)^{1/2} \, 
K\left(-\frac{4 e}{p - 6 - 2 e}  \right) ,
\ee
where $K(m)$ is the complete elliptic integral of the first kind \cite{GradETC07}.  Finally, the compactness parameter 
$y$, which is the PN expansion parameter we use, is given by $y = (M \O_\vp)^{2/3}$. 

As the expressions above show, $y$ is a function (through $\O_\varphi$ and $T_r$) of $p$ and $e$.  For a given $p$ 
and $e$, it proved useful to compute $y$ to 600 decimal places for the numerical fitting work in this paper.  Such 
precision is straightforward to obtain because the integrand in the integral for $T_r$ is periodic and smooth in 
$\chi$, leading to exponential convergence in a Riemann sum (as summarized in \cite{HoppETC15}).

\subsection{The RWZ master equation}
\label{sec:TDmasterEq}

The geodesic motion of the small body provides the source of a perturbation to the background Schwarzschild metric.  
Finding this perturbation is convenient in the Regge-Wheeler-Zerilli (RWZ) formalism, and for purposes of computing 
fluxes we need not go beyond calculating the master functions.  In a spherical harmonic decomposition, for each 
$l,m$ the master function satisfies a single inhomogeneous time domain (TD) equation of the form
\begin{align}
\begin{split}
\label{eqn:masterEqTD}
&\left[-\frac{\pa^2}{\pa t^2} + \frac{\pa^2}{\pa r_*^2} - V_l (r) \right]
\Psi_{lm}(t,r)  = S_{lm} (t,r) .
\end{split}
\end{align}
Here $r_{*} = r + 2 M \ln | r/2 M - 1 |$ is the tortoise coordinate and the source is a distribution of the form
$ S_{lm} (t,r) \equiv G_{lm}(t) \, \delta[r - r_p(t)] + F_{lm}(t) \, \delta'[r - r_p(t)]$.  Both the source term and 
the potential $V_{l}(r)$ are ($l+m$) parity-dependent.

While this equation can be solved directly in the TD (e.g. \cite{Mart04}), our method works in the frequency domain 
(FD) and utilizes the MST formalism to produce solutions at extremely high accuracy.  Transforming the field and 
source to the FD involves Fourier series
\begin{align}
\label{eqn:psiSeries}
\Psi_{lm}(t,r) &= \sum_{n=-\infty}^\infty X_{lmn}(r) \, e^{-i \o t} , \\
S_{lm}(t,r) &= \sum_{n=-\infty}^\infty Z_{lmn}(r) \, e^{-i \o t} , 
\end{align}
since discrete frequencies $\o \equiv \omega_{mn} = m\Omega_\vp + n\Omega_r$ arise as part of the bi-periodicity of 
the bound motion.  

In this way, the TD master equation is reduced in the FD to a set of inhomogeneous ODEs now tagged by harmonic $n$ 
(from the eccentric motion) as well as spherical harmonic indices $l,m$,
\be
\label{eqn:masterInhomogFD}
\left[\frac{d^2}{dr_*^2} +\omega^2 -V_l (r) \right]
X_{lmn}(r) = Z_{lmn} (r) .
\ee
The homogeneous version to this equation yields two independent solutions: $X_{lmn}^{-}$, with downgoing causal behavior 
at the horizon, and $X_{lmn}^{+},$ with outgoing causal behavior at infinity.  

It is possible to determine the homogeneous solution for the odd-parity master function using the MST formalism first, 
and then to recover the even-parity counterpart using the Detweiler-Chandrasekar transformation 
\cite{Chan75,ChanDetw75,Chan83,Bern07}.  Though a direct MST formalism for the (odd-parity) RWZ functions 
$X_{lmn}^{-}$ and $X_{lmn}^{+}$ exists \cite{ManoSuzuTaka96a}, we instead computed the MST solutions to the related 
Bardeen-Press-Teukolsky equation \cite{BardPres73,CutlKennPois94}.  Then $X_{lmn}^{\pm}$ are recovered using another 
version of the Detweiler-Chandrasekar transformation.  We refer the reader to Paper I for details.

Once the RWZ functions are computed, the resulting $X^{\pm}_{lmn}$ will not be unit normalized at infinity nor at the 
horizon, since the MST solution involves iterating a recurrence relation starting with an arbitrary value $a_0$.  
We separately and precisely determine the resulting amplitudes at infinity and at the horizon and divide these off to 
produce unit-normalized functions
\be
\hat{X}^{\pm}_{lmn} \sim e^{\pm i \o r_*} , \qquad r_* \rightarrow \pm\infty .
\ee
These unit-normalized homogeneous solutions are then used \cite{HoppEvan10} in an integration over the source to 
determine the proper normalization amplitudes $C^{\pm}_{lmn}$, as we summarize next.

\subsection{The TD solution $\Psi_{lm} (t,r)$ and the angular momentum flux}

With the unit-normalized solutions computed, the TD function $\Psi_{lm} (t,r)$ can be directly constructed 
using the method of extended homogeneous solutions (EHS) \cite{BaraOriSago08,HoppEvan10}.  This process involves a 
pair of homogeneous solutions of the TD equation \eqref{eqn:masterEqTD}
\be
\label{eqn:TD_EHS}
\Psi^\pm_{lm} (t,r) 
\equiv \sum_n C^{\pm}_{lmn} \hat X_{lmn}^\pm (r) \, e^{-i \o t}, \q \q r > 2M ,
\ee
where $C^{\pm}_{lmn}$ are the two key sets of normalization constants (determined below).  The full (particular) 
solution to the RWZ equation \eqref{eqn:masterEqTD} is then formed by abutting the two TD EHS $\Psi^\pm (t,r)$ at the 
particle's location
\begin{align}
\label{eqn:TD_PS}
\begin{split}
\Psi_{lm} (t,r) &= \Psi^{+}_{lm}(t,r) \theta \left[ r - r_p(t) \right] \\
& \hspace{10ex}
+ 
\Psi^{-}_{lm}(t,r) \theta \left[ r_p(t) - r \right] .
\end{split}
\end{align}
Unlike the standard procedure of constructing a solution to \eqref{eqn:masterEqTD} by summing the inhomogeneous FD 
solutions $X_{lmn}(r)$ of \eqref{eqn:masterInhomogFD} (found by variation of parameters or equivalently use of the Green 
function), the EHS method experiences no Gibbs behavior near $r_p(t)$ nor within the radial libration region.  Instead 
the sums in \eqref{eqn:TD_EHS} converge exponentially for all $r$.
 
The only remaining issue is finding the particular values of $C^{\pm}_{lmn}$, equivalent to incorporating the internal 
boundary condition at the discontinuity at $r_p(t)$.  As shown in \cite{HoppEvan10}, these coefficients are given by
\begin{align}
\label{eqn:EHSC}	
C_{lmn}^\pm  
&=   \frac{1}{W_{lmn} T_r} \int_0^{T_r}
\Bigg[ 
 \frac{1}{f_{p}} \hat X^\mp_{lmn}
 G_{lm} \hspace{5ex} \\
&\hspace{5ex} 
+ \l \frac{2M}{r_{p}^2 f_{p}^{2}} \hat X^\mp_{lmn}
 - \frac{1}{f_{p}} 
 \frac{d \hat X^\mp_{lmn}}{dr} \r F_{lm}
 \Bigg]  e^{i \o t}  \, dt, \notag
\end{align}
where $W_{lmn}$ is the Wronskian
\be
W_{lmn} = f \hat{X}^-_{lmn} \frac{d \hat{X}^+_{lmn}}{dr}
- f \hat{X}^+_{lmn} \frac{d \hat{X}^-_{lmn}}{dr} .
\ee
The integral \eqref{eqn:EHSC} is computed using spectral source integration (SSI) \cite{HoppETC15}, in which the 
integral is replaced by a sum over equally-spaced samples.  Because the integrand is periodic in $t$ and smooth, this 
produces exponential convergence of the result (see \cite{HoppETC15} and Paper I for more details).  This rapid 
convergence has permitted the MST calculation of the $lmn$ modes of our eccentric-orbit fluxes described in this paper 
to as many as 450 decimal places of precision.

Once the $C^{\pm}_{lmn}$ coefficients have been determined, the angular momentum flux at infinity is calculated as
\begin{align}
\label{eqn:fluxNumeric}
\left\langle \frac{dL}{dt} \right\rangle^{\infty} = 
\sum_{lmn}\frac{m \o}{64\pi}\frac{(l+2)!}{(l-2)!} |C^{+}_{lmn}|^2 ,
\end{align} 
with the analogous, also standard expression for energy flux at infinity given in Paper I.

\begin{widetext} 

\section{Angular momentum radiated to infinity: current complete PN theory and added analysis of hereditary terms}
\label{sec:LcurrPN}

Here we review the state of complete PN theory for angular momentum fluxes (known up to 3PN relative order 
\cite{ArunETC09a,Blan14}), from which we build new results at higher PN order later in this paper.  Additionally, in 
keeping with a corresponding section in Paper I, we analyze the hereditary (tail) terms of the angular momentum 
flux expansion, and determine arbitrary-order expansions in eccentricity for those terms.  Furthermore, we utilize 
asymptotic analysis to identify and confirm the singular behavior of enhancement functions as the eccentricity nears 
unity.  Brief summaries of the methods and the quasi-Keplerian representation of the orbital motion are also given 
in Paper I \cite{ForsEvanHopp16}. 

\subsection{Instantaneous angular momentum flux terms}
\label{sec:AngMomInf1}

Just as in the original articles and Paper I, we use the quasi-Keplerian time eccentricity, $e_t$, which differs 
from the Darwin eccentricity, $e$, described earlier in the paper.  Focusing only on the terms that are lowest order 
in the mass ratio $\nu$ (in keeping with present BHPT), the instantaneous contributions to the orbit-averaged angular 
momentum flux through 3PN can be written as
\be
\left\langle\frac{dL}{dt}\right\rangle_\infty^\text{inst} = \frac{32}{5}\frac{\mu^2}{M}y^{7/2}
\left(\mathcal{N}_0+y\mathcal{N}_1+y^2\mathcal{N}_2+y^3\mathcal{N}_3\right) ,
\ee
where again $y=(M\Omega_{\varphi})^{2/3}$ is the compactness parameter, and where
\begin{align}
\label{eqn:instFlux3PN}
\mathcal{N}_0 &= \frac{1}{(1-e_t^2)^2}\left(1+\frac{7}{8} e_t^2\right),\\
\mathcal{N}_1 &= \frac{1}{(1-e_t^2)^3}\left(-\frac{1247}{336}+\frac{3019}{336}e_t^2
+ \frac{8399}{2688}e_t^4\right),\\
\mathcal{N}_2 &= \frac{1}{(1-e_t^2)^4}\biggl[-\frac{135431}{9072} -
\frac{598435}{6048}e_t^2 + \frac{30271}{3456}e_t^4 + \frac{30505}{16128}e_t^6
+\sqrt{1-e_t^2}\left(10+\frac{335}{8} e_t^2+\frac{35}{8} e_t^4\right)
\biggr],\\
\mathcal{N}_3 &= \frac{1}{(1-e_t^2)^5}\Biggl[
\frac{2017023341}{9979200} + \frac{270214177}{623700}e_t^2
-\frac{6350078491}{13305600}e_t^4 - \frac{272636461}{4435200}e_t^6
-\frac{10305073}{5677056}e_t^8\notag\\&\qquad+
\sqrt{1-e_t^2}\left(-\frac{379223}{5040} + \frac{309083}{2520}e_t^2
+ \frac{13147661}{40320}e_t^4+\frac{35}{4} e_t^6\right)\Biggr]
+\frac{1712}{105}\tilde{F}(e_t)
\log\left[\frac{y}{y_0}\frac{1+\sqrt{1-e_t^2}}{2(1-e_t^2)}\right],
\end{align}
are the functions of time eccentricity derived in prior work \cite{ArunETC09a}.  Note that $y_0$ depends on $r_0$ 
(a parameter defined in the original paper), which is an arbitrary length scale and which cancels in the total flux.  
The expressions above are similar to (4.11) of \cite{ArunETC09a}, but with a different overall scaling and are 
expressed in modified harmonic gauge.  The function $\tilde{F}(e_t)$ is given by
\be
\label{eqn:FTilde}
\tilde{F}(e_t) = 
\frac{1}{(1-e_t^2)^5}
\left(
1+\frac{229}{32}e_t^2+\frac{327}{64}e_t^4+\frac{69}{256}e_t^6
\right) .
\ee
Any appearance of a tilde over a function name, as in $\tilde{F}(e_t)$, refers to angular momentum, while a 
corresponding function (in this case $F(e_t)$) without a tilde appears in the energy flux expansion.

\subsection{Hereditary angular momentum flux terms through 3PN order}

As with the energy flux (c.f. Paper I), the hereditary portion of the angular momentum 
flux can be defined in terms of a set of enhancement functions \cite{ArunETC08a,ArunETC08b,ArunETC09a, 
BlanScha93, RietScha97},
\begin{align}
\label{eqn:heredFlux3PN}
\left\langle\frac{dL}{dt}\right\rangle_\infty^\text{hered} &=
\frac{32}{5}\frac{\mu^2}{M}y^{7/2}
\biggl\{
4\pi y^{3/2}\tilde{\phi}(e_t)
-\frac{8191}{672}\pi y^{5/2}\tilde{\psi}(e_t)
\notag\\&\qquad+y^3
\left(
-\frac{1712}{105}\tilde{\chi}(e_t)
+\left[-\frac{116761}{3675}+
\frac{16}{3}\pi^2-\frac{1712}{105}\gamma_\text{E}
-\frac{1712}{105}\log\left(\frac{4 y^{3/2}}{y_0}\right)
\right]
\tilde{F}(e_t)
\right)
\biggr\} .
\end{align}
Here, $\tilde{\varphi}(e_t)$ is the 1.5PN tail term; $\tilde{\psi}(e_t)$ is the 1PN correction to the tail; and
the 3PN portion involves $\tilde{\chi}(e_t)$ and $\tilde{F}(e_t)$ and emerges upon combining the angular momentum 
tail-of-tails and tail$^2$ contributions.  Unlike the instantaneous terms $\mathcal{N}_i$, the functions 
$\tilde{\varphi}(e_t), \tilde{\psi}(e_t),$ and $\tilde{\chi}(e_t)$ admit no simple closed forms.  Arun \textit{et al.} 
originally calculated these contributions numerically but also produced a low-order expansion through $e_t^4$ for 
each enhancement function \cite{ArunETC09a}.

These expressions are written as functions of the time eccentricity $e_t$.  However, as we will see, the 1.5PN tail 
$\tilde{\vp}$ and the functions $\tilde{F}$ and $\tilde{\chi}$ depend only on Newtonian order quantities.  Hence, for 
these functions (as well as $\mathcal{N}_0$) there is no distinction between using $e_t$ and the usual Keplerian 
eccentricity.  Nevertheless, we will keep the notation consistent by expressing everything here in terms of $e_t$.  
Finally, each of these functions is defined to equal 1 in the case of a circular orbit (in keeping with the meaning 
of an enhancement function) except for $\tilde{\chi}$, which limits to 0 as $e_t \rightarrow 0$. 
\end{widetext}

\subsection{Arbitrary-order expansions for hereditary terms}
\label{sec:heredExps}

Previous expressions for the expansions of the tail terms were too limited for our purposes (with the exception of 
$\tilde{F}(e_t)$ which is exact).  We sought arbitrary-order expansions of these terms and applied the methods used 
in Paper I.  The majority of these functions are best handled in the FD.  There the Fourier decomposition of the 
(dimensionless) Newtonian trace-free quadrupole moment \cite{ArunETC08a,ArunETC08b,ArunETC09a} can be used with the 
leading-order angular momentum flux to calculate $\mathcal{N}_0(e_t)$
\begin{align}
\label{eqn:pmsum}
\mathcal{N}_0(e_t)& = \frac{-i}{16} \epsilon_{ijk} \hat{L}_i  \left\langle \ddot{\hat{I}}_{\!ja}^{\,\,(\mathrm{N})} 
 \dddot{\hat{I}}\strut_{\!ka}\strut^{\!\!\!\!\!(\mathrm{N})} \right\rangle \notag \\&
 = \frac{-i}{8} \epsilon_{ijk} \hat{L}_i \sum_{n=1}^\infty n^5 
 \! \mathop{\hat{I}}_{(n)}{}_{\!ja}^{\!\!(\mathrm{N})} \mathop{\hat{I}}_{(n)}{}_{\!ka}^{\!\!\!* (\mathrm{N})}  = 
\sum_{n=1}^\infty \tilde{g}(n,e_t) \notag \\
&= \tilde{f}(e_t) =
\frac{1}{(1-e_t^2)^{2}}
{\left(1+\frac{7}{8}~e_t^2 \right)} .
\end{align}
Here, $\tilde{f}(e_t)$ is just alternate notation for $\mathcal{N}_0(e_t)$ and is the angular momentum analogue of the 
traditional Peters-Mathews function $f(e_t)$ (first derived by Peters in \cite{Pete64}) for the quadrupole energy 
flux.  In this expression, ${}_{(n)} \hat{I}_{ij}^{(\mathrm{N})}$ is the $n$th Fourier harmonic of the 
dimensionless (Newtonian) quadrupole moment (see Sections III through V of \cite{ArunETC08a}).  The product of terms 
yielding the angular momentum flux radiated into each harmonic is compactly expressed as the function 
$\tilde{g}(n,e_t)$.  This power spectrum for angular momentum flux is given by
\begin{flalign}
\begin{autobreak}
\label{eqn:gfunc}
\tilde{g}(n,e_t) \equiv
\sqrt{1-e_t^2} \bigg\{\left(-\frac{2}{e_t^2}+2\right) n^2 J_n'(n e_t)^2 
+\left[-\frac{2}{e_t^4}+\frac{3}{e_t^2}-1\right] n^2 J_n(ne_t)^2 
 +\left[2 e_t n^2+\frac{2}{e_t^3}\left(1+n^2\right)-\frac{1}{e_t}\left(1+4 n^2\right)\right] n
J_n(ne_t) J_n'(ne_t)\bigg\} .
\end{autobreak}
\end{flalign}

As Arun \textit{et al.} \cite{ArunETC09a} make clear, three of the desired hereditary functions -- 
$\tilde{F}(e_t)$, $\tilde{\vp}(e_t)$, and $\tilde{\chi}(e_t)$ -- follow immediately from knowledge of just this 
quadrupole spectrum
\begin{align}
\label{eqn:capFeSum}
\tilde{F}(e_t) &= \frac{1}{4} \sum_{n=1}^\infty n^2 \, \tilde{g}(n,e_t),  \\
\label{eqn:phi2}
\tilde{\vp}(e_t) &=\frac{1}{2} \sum_{n=1}^\infty n \, \tilde{g}(n,e_t), \\
\label{eqn:chiSum}
\tilde{\chi}(e_t) &= \frac{1}{4}  \sum_{n=1}^\infty n^2  
\log\left(\frac{n}{2}\right)  \tilde{g}(n,e_t) .
\end{align}
Unfortunately, unlike $\tilde{f}(e_t)$ and $\tilde{F}(e_t)$, the latter two sums likely do not admit closed form
expressions.  This stems from the odd power of $n$ in $\tilde{\vp}(e_t)$ and the logarithm in $\tilde{\chi}(e_t)$,
which give the two sums complicated representations in the time domain and preclude the ability to calculate the 
time integral over a libration.

As shown in Paper I, however, it is still possible to extract arbitrary-order expansions for sums of these types.  
Using the Bessel function representation of $\tilde{g}(n,e_t)$ and expanding \eqref{eqn:gfunc} in a Maclaurin series 
in $e_t$, we find
\begin{align}
\label{eqn:gexp}
\begin{autobreak}
\MoveEqLeft
\tilde{g}(n,e_t)=\left(\frac{n}{2}\right)^{2 n-1} e_t^{2 n-4}  \bigg(\frac{1}{\Gamma (n-1)^2}
-\frac{n^3+3 n^2-6 n+2}{2 \Gamma (n)^2}e_t^2
+\frac{2 n^4+15 n^3+6 n^2-16 n+2}{16  \Gamma (n)^2}e_t^4+\cdots \bigg) .
\end{autobreak}
\end{align}
In a sum over $n$, successive harmonics each contribute a series that starts at a progressively higher power of 
$e_t^2$.  Thus, summations like \eqref{eqn:capFeSum}, \eqref{eqn:phi2}, or \eqref{eqn:chiSum} can be determined
to any desired finite order in $e_t^2$ with only a finite sum (of some length) over $n$.  As in the energy case, the 
$e_t^{-2}$ and $e_t^0$ coefficients vanish for $n = 1$, and the $n = 2$ harmonic is the only one that contributes 
at $e_t^0$ (and thus for a circular orbit).

Accordingly, we expand the latter two summations from above and find the leading behavior of these functions to be
\begin{align}
\begin{autobreak}
\label{eqn:vphiExpand}
\MoveEqLeft
\tilde{\varphi}(e_t) = \frac{1}{(1-e_t^2)^{7/2}} \, 
\bigg(1+\frac{97}{32} e_t^2+\frac{49}{128} e_t^4 
-\frac{49}{18432} e_t^6 
-\frac{109}{147456} e_t^8  
-\frac{2567}{58982400} e_t^{10}
+\frac{4649}{707788800} e_t^{12} 
+\frac{418837}{221962567680} e_t^{14} 
+\frac{28447343}{53271016243200} e_t^{16} 
+\frac{5249748289}{19725496300339200} e_t^{18} 
+\cdots \bigg),
\end{autobreak}\\
\begin{autobreak}
\label{eqn:chiTild}
\MoveEqLeft
\tilde{\chi} (e_t) = - \frac{3}{2} \tilde{F}(e_t) \log(1 - e_t^2) +
\frac{1}{(1-e_t^2)^{5}} \, 
\Bigg\{ 
\left[-\frac{3}{2}-\frac{549}{32} \log (2)+\frac{2187}{128} \log (3)\right]e_t^2
+\left[-\frac{735}{64}+\frac{18881}{64}\log(2)-\frac{85293}{512} \log (3)\right]e_t^4
+\biggl[-\frac{433}{32}
-\frac{6159821}{2304}\log (2)
+\frac{5981445}{8192} \log (3)
+\frac{48828125}{73728} \log (5)
\biggr]e_t^6
+
\biggl[-\frac{4193}{512}
+\frac{16811095}{1152}\log (2)
+\frac{56772333}{65536}\log (3)
-\frac{4052734375}{589824} \log (5)
\biggr]e_t^8+\cdots
\Bigg\} .
\end{autobreak}
\end{align}
We give only the first few terms in these power series here.  The much lengthier expressions that we have used in 
our numerical modeling and analytic fitting are given in a \textsc{Mathematica} notebook archived on the Black Hole 
Perturbation Toolkit website \cite{BHPTK18}.  For computational reference, calculation of over 100 terms in these 
series in \textsc{Mathematica} requires under 20 seconds on an average laptop.  The first four terms 
of \eqref{eqn:vphiExpand} are also published in \cite{SagoFuji15}.  

Both $\tilde{\vp}$ and $\tilde{\chi}$ diverge as $e_t \rightarrow 1$; however, as displayed in the above equations, 
both can be written in forms that isolate their divergences.  These singular factors will be justified in Section 
\ref{sec:asymptotic}, where we analyze the asymptotic behavior near $e_t = 1$ using methods developed in Paper I.  Of 
particular note is the fact that the structure of $\tilde{\chi}$ closely mirrors its energy flux counterpart with a 
combination of algebraic and logarithmic divergences.  We find by direct high-order expansion that the two series have 
the following limits near $e_t = 1$
\begin{align}
\label{eqn:tailAsymp}
\tilde{\vp} &\rightarrow  \frac{4.41063}{(1-e_t^2)^{7/2}}, 
\\
\tilde{\chi} &\rightarrow  - \frac{3}{2} \bigg(\frac{3465}{256}\bigg) \frac{\log(1 - e_t^2) }{(1-e_t^2)^{5}}
+ \frac{16.7230}{(1-e_t^2)^{5}}, 
\end{align}
where $3465/256 \simeq (2/3) \, 20.3027$ is simply the value of the polynomial portion of $\tilde{F}(e_t)$ evaluated 
at $e_t = 1$.

As expected from Paper I, the most difficult enhancement function to extract is the 2.5PN term $\tilde{\psi}$.  
Calculating $\tilde{\psi}$ involves not only the Newtonian mass octupole and current quadrupole moments, but also 
the 1PN correction to the mass quadrupole moment.  At 1PN order the orbital motion no longer closes and corrections 
to the quadrupole moment require a biperiodic Fourier expansion.  Arun \textit{et al.} describe a procedure for 
computing $\tilde{\psi}$ in \cite{ArunETC09a}, which they evaluated numerically.  Using a modified form of our 
expansion methods, we were able to obtain $\tilde{\psi}$ in an expansion out to $e_t^{120}$, complete with factoring 
out the relevant eccentricity singular factor.  The procedure for using these 1PN multipole moments will be detailed
in a future paper \cite{MunnEvan19b}, where it is shown that not only $\tilde{\psi}$ but an infinite set of 
1PN-corrected leading logarithm terms can be derived from the next-most important multipole moments beyond the 
mass quadrupole.  Focusing here just on $\tilde{\psi}(e_t)$, the first few terms in the expansion are
\begin{widetext}
\begin{align}
\tilde{\psi}(e_t)  =&  \frac{1}{\left(1-e_t^2\right)^{9/2}}
\biggl(1-\frac{108551}{16382} e_t^2-\frac{5055125 }{524224} e_t^4  
-\frac{4125385}{9436032} e_t^6+\frac{11065099}{603906048} e_t^8
-\frac{68397463 }{30195302400}e_t^{10}  \notag \\&
-\frac{194038163}{1159499612160} e_t^{12}  
+\frac{3310841491}{189384936652800} e_t^{14} 
+\frac{5520081282241}{436342894048051200} e_t^{16}   \notag \\&
+\frac{78911659620611}{14137509767156858880} e_t^{18} 
+\frac{22307748275735593 }{8078577009803919360000} e_t^{20}  +\cdots \biggr) .
\label{eqn:psiExpand}
\end{align}
\end{widetext}
The power series in the parentheses appears to converge to -15.6906.  Again, the more complete expansion is archived 
at \cite{BHPTK18}.

\subsection{Applying asymptotic analysis to determine eccentricity singular 
factors}
\label{sec:asymptotic}

We briefly derive the divergent behavior of the preceding functions as $e_t \rightarrow 1$.  The same asymptotic 
techniques developed in Paper I apply to the angular momentum flux terms.  We refer the reader to Paper I for details 
on the procedure.  We note that a comparable analysis of the energy and angular momentum flux asymptotics was 
undertaken in \cite{LoutYune17}.

Four of the relevant enhancement functions share a dependence on the quadrupole moment spectrum $\tilde{g}(n,e_t)$ 
found in \eqref{eqn:gfunc} and therefore we require the high eccentricity behavior of this function.  To aid our 
efforts near $e_t = 1$, we define $x \equiv 1 - e_t^2$ and rewrite \eqref{eqn:gfunc} as
\begin{align}
\label{eqn:gxfunc}
& \tilde{g}(n,e_t) = 
- n^2 
\frac{x^{3/2}(1+x)}{(1-x)^2} J_n(ne_t)^2 
-n^2\frac{2x^{3/2}}{1-x} J_n'(n e_t)^2 \notag \\
& \hspace{5ex} +n\frac{x^{1/2}(1+x+2n^2x^2)}{(1-x)^{3/2}} J_n(n e_t) J_n'(n e_t) .
\end{align}  
From this point on the procedure of Paper I is followed exactly: $J_n(n e_t)$ and $J_n'(n e_t)$ are expressed in terms
of their uniform asymptotic expansions for large-order (and large-argument) \cite{DLMF}, which have growing importance 
as $x \rightarrow 0$.  This representation involves sums of Airy functions and their derivatives, which must themselves 
be expanded in the reciprocal of the variable
\be
\xi = n \log\left(\frac{1 + \sqrt{x}}{\sqrt{1-x}} \right) - n \sqrt{x} = n \rho(x) .
\ee
The various series are inserted into the enhancement function summations, which are then converted from sums over 
$n$ to integrals over $dn = d\xi/\rho(x)$.  Finally, these integrals can be evaluated to extract not only the divergent 
behavior of the four enhancement functions, but also surprisingly sharp estimates of the numerical limit of the 
coefficients attached to these divergent terms.

We now apply this asymptotic procedure to the four enhancement functions.  The simplest are the two with closed-form 
expressions.  While these terms are already exactly known, they serve as good tests of the asymptotic analysis.  
The functions $\tilde{f}(e_t)$ in \eqref{eqn:pmsum} and $\tilde{F}(e_t)$ in \eqref{eqn:capFeSum} have known singular 
dependences of
\begin{align}
\tilde{f}(e_t) &\sim   \frac{15}{8} \frac{1}{(1-e_t^2)^{2}} 
= \frac{1.875}{(1-e_t^2)^{2}}, \notag \\
\tilde{F}(e_t) &\sim \frac{3465}{256} \frac{1}{(1-e_t^2)^{5}} \simeq \frac{13.5352}{(1-e_t^2)^{5}},
\end{align}
as $e_t \rightarrow 1$.  If we instead make the asymptotic approximations of the sums in \eqref{eqn:pmsum} and 
\eqref{eqn:capFeSum}, we find
\begin{align}
\tilde{f}(& e_t) \sim \frac{1141}{192 \, \pi} \frac{1}{(1-e_t^2)^{2}} \simeq 
\frac{1.8916}{(1-e_t^2)^{2}} , \notag \\
\tilde{F}(e_t &)  \sim  \frac{56429761}{1327104 \,\pi} \frac{1}{(1-e_t^2)^{5}} \simeq 
\frac{13.5348}{(1-e_t^2)^{5}} ,
\end{align}
which extracts the correct eccentricity singular functions and yields close estimates of the exact coefficients. 

Next we move to $\tilde{\varphi}(e_t)$ and $\tilde{\chi}(e_t)$, which are not known analytically.  For 
$\tilde{\varphi}(e_t)$, the sum in \eqref{eqn:phi2} leads to the following asymptotic estimate
\begin{align}
\begin{autobreak}
\MoveEqLeft
\tilde{\varphi}(e_t) \sim 
\frac{191287}{13824 \, \pi} \frac{1}{(1-e_t^2)^{7/2}} -
\frac{386929}{34560 \, \pi} \frac{1}{(1-e_t^2)^{5/2}} + \cdots
\simeq 
\frac{4.40455}{(1-e_t^2)^{7/2}} -
\frac{3.56375}{(1-e_t^2)^{5/2}} + \cdots ,
\end{autobreak}
\end{align}
where in this case we retained the first and second terms in the expansion about $e_t = 1$.  The leading singular 
factor matches that chosen in \eqref{eqn:vphiExpand} and its coefficient approximates the 4.41063 value found in 
\eqref{eqn:tailAsymp}.

The last function of this kind is $\tilde{\chi}(e_t)$, whose definition involves $\log(n/2)$.  Using the same asymptotic 
expansions and integral approximation for the sum, and retaining the first two divergent terms, we find
\begin{align}
\begin{autobreak}
\label{eqn:chiasymp}
\MoveEqLeft
\tilde{\chi}(e_t) \sim 
- \frac{56429761}{884736 \, \pi} \bigg[\log(1-e_t^2)
-\frac{79015440}{56429761} 
+ \frac{2}{3} \gamma_E + \frac{4}{3} \log (2) 
- \frac{2}{3} \log(3) \bigg] \frac{1}{(1-e_t^2)^{5}} 
\approx -20.3023 \frac{\log(1-e_t^2)}{(1-e_t^2)^{5}} + \frac{16.7219}{(1-e_t^2)^{5}} .
\end{autobreak}
\end{align}
Thus, we see that the form of \eqref{eqn:chiTild}, though already verified through direct high-order expansion, is 
validated by the asymptotic analysis. 

The asymptotic analysis confirmed what we already guessed about the closed form for the leading singular term 
(involving $\tilde{F}(e_t)$) in \eqref{eqn:chiTild} for $\tilde{\chi}(e_t)$, since it resembles strongly its energy 
counterpart (Paper I).  In fact, if we consider making a PN expansion in the orbital parameter $1/p$ instead of $y$, 
that specific term in \eqref{eqn:chiTild} with its logarithmic and algebraic divergences is necessary to cancel other 
logarithmically divergent terms in the full 3PN flux.  As a last check, note that the two numerical coefficients in 
\eqref{eqn:chiasymp} compare well with their counterparts in \eqref{eqn:chiTild}, which were found to be approximately 
$-20.3027$ and $+16.7230$, respectively.

\section{Finding new coefficients in the fluxes via mode-by-mode fitting}
\label{sec:lmnfitting}

\subsection{PN expansion from the BHPT viewpoint}

We move now beyond known results to calculate new coefficients at higher order in the PN expansion using 
perturbation theory.  To that end, we use the following general form for the angular momentum flux at infinity
\begin{widetext}
\begin{align}
\label{eqn:angMomFluxInf}
\left\langle \frac{dL}{dt} \right\rangle^\infty &=  
\left\langle \frac{dL}{dt} \right\rangle_\text{N}^\infty 
\biggl[\mathcal{J}_0 + y\mathcal{J}_1 + y^{3/2}\mathcal{J}_{3/2}
+y^2\mathcal{J}_2 + y^{5/2}\mathcal{J}_{5/2}
+ y^3\biggl(\mathcal{J}_3+\mathcal{J}_{3L}\log(y)\biggr)
+ y^{7/2}\mathcal{J}_{7/2}
\notag\\&+ y^4\biggl(\mathcal{J}_4+\mathcal{J}_{4L}\log(y)\biggr)
+ y^{9/2}\biggl(\mathcal{J}_{9/2}+\mathcal{J}_{9/2L}\log(y)\biggr)
+ y^5\biggl(\mathcal{J}_5+\mathcal{J}_{5L}\log(y)\biggr)
\notag\\&+y^{11/2}\biggl(\mathcal{J}_{11/2}+\mathcal{J}_{11/2L}\log(y)\biggr) 
+ y^6\biggl(\mathcal{J}_6
+\mathcal{J}_{6L}\log(y)+\mathcal{J}_{6L^2}\log^2(y)\biggr) +
y^{13/2}\biggl(\mathcal{J}_{13/2}+\mathcal{J}_{13/2L}\log(y)\biggr)
\notag\\& + y^7\biggl(\mathcal{J}_7
+\mathcal{J}_{7L}\log(y)+\mathcal{J}_{7L^2}\log^2(y)\biggr)+
y^{15/2}\biggl(\mathcal{J}_{15/2}+\mathcal{J}_{15/2L}\log(y)+\mathcal{J}_{15/2L^2}\log^2(y)\biggr)
\notag\\& + y^8\biggl(\mathcal{J}_8
+\mathcal{J}_{8L}\log(y)+\mathcal{J}_{8L^2}\log^2(y)\biggr)+
y^{17/2}\biggl(\mathcal{J}_{17/2}+\mathcal{J}_{17/2L}\log(y)+\mathcal{J}_{17/2L^2}\log^2(y)\biggr) +\cdots
\biggr] ,
\end{align}
\end{widetext}
where the Newtonian prefactor, as before, is given by
\begin{equation}
\label{eqn:newtAngMomInf}
\left\langle\frac{dL}{dt}\right\rangle_\text{N}^\infty = \frac{32}{5} \frac{\mu^2}{M} y^{7/2}.
\end{equation}
In the above expansion, each $\mathcal{J}_i = \mathcal{J}_i(e)$ represents an eccentricity flux function for the 
total flux radiated at a PN term scripted by $i$.  Terms in the PN expansion have the form $y^n \log^k(y)$ and the 
script $i$ encodes both the integer or half-integer for power $n$ of $y$ and the integer power $k$ of $\log(y)$.  

There are two changes in this notation over that of Sec.~\ref{sec:LcurrPN}.  First, we transition from using time 
eccentricity $e_t$ to using Darwin eccentricity $e$, which is the natural choice for BHPT calculations (see 
Sec.~\ref{sec:orbits}).  We have recently derived the relationship between $e_t$ and $e$ to all PN orders 
\cite{MunnEvan19b} at lowest order in the mass ratio.  Using that relationship, it is possible to check the results 
of BHPT fitting against the enhancement functions of Section \ref{sec:LcurrPN} through 3PN, thus confirming prior 
work.  Second, in \eqref{eqn:angMomFluxInf} we no longer attempt to separate instantaneous and hereditary contributions 
to the flux functions, since that distinction is not possible with perturbative methods alone.  Therefore, we 
generically use the $\mathcal{J}_i$ notation at all orders to denote the combination. 

\subsection{The original fitting scheme}

When the orbit is wide (i.e., in the PN regime), the representation \eqref{eqn:angMomFluxInf} is a valid expansion 
for the values that would emerge from evaluating the BHPT flux formula \eqref{eqn:fluxNumeric}.  We can use 
\eqref{eqn:fluxNumeric} to derive the analytic form of the functions in \eqref{eqn:angMomFluxInf}.  One way to do this 
is to directly expand the MST solutions analytically and carry the results through to obtain $| C_{lmn}^{\pm} |^2$ as 
a PN expansion (see e.g., \cite{BiniDamo13,BiniDamo14a,BiniDamo14b, BiniDamo14c,Fuji12b,KavaOtteWard15,HoppKavaOtte16}).  
However, Paper I used a different approach, evaluating the fluxes numerically and then determining the analytic 
coefficients.  In this approach, the full flux is calculated to some preset accuracy goal for a large number of orbits 
that vary in $p$ and $e$.  This creates a two-dimensional array of orbital flux values, which can then be fit to the 
form of the PN expansion in $y$ and $e$.  If the fit is performed with enough accuracy, analytic forms for the 
coefficients in the $\mathcal{J}_i$ can be extracted from the highly precise numerical results using an 
integer-relation algorithm \cite{FergBailArno99}.  Using this procedure, Paper I showed some of the analytic dependence 
of the energy flux up to 7PN order, by computing roughly 1700 orbits of varying separation and eccentricity, with 
roughly $\sim7500$ $lmn$ flux components for each orbit and with an overall accuracy of 200 decimal places for the 
flux relative to the quadrupole mode.  We refer the reader to Paper I for details.

It is noteworthy that the two methods (all-analytic and numeric-analytic) are somewhat complementary in nature.  We 
recently began supplementing our fitting results using a new code based on the purely analytic approach.  In 
the process we discovered that while analytic methods are significantly more efficient at reaching high PN order, 
they have more difficulty attaining high orders in the power series in eccentricity.  On the other hand, while the 
fitting approach becomes unwieldy around 8PN to 12PN order, with enough sampling it can successfully extract nearly 
arbitrary orders in $e^2$---at least when the eccentricity power series coefficients are simple (e.g., all rational 
numbers).  However, when the coefficients instead contain complicated combinations of transcendental numbers, like 
$\pi^2$ and $\log(2)$, the integer-relation algorithm struggles to identify the analytic representation of the decimal 
number input, unless much higher numerical precision is utilized.  More complicated combinations of transcendental 
numbers are exactly what occurs in certain higher order (integer) PN terms, like $\mathcal{L}_5(e)$ and 
$\mathcal{L}_6(e)$.  As a result, very few coefficients in the power series in eccentricity were extracted in terms 
like these in Paper I.  

Developments in the purely analytic approach will be described in future work \cite{Munn20}.  In this paper we show
instead a new technique for obtaining a marked improvement in the numerical fitting scheme.  It turns out possible 
to separate key dependences in the eccentricity flux functions and to determine coefficients in a hierarchical 
fashion.  With this modification to the procedure of Paper I we have significantly expanded the understanding of high 
PN order fluxes.  One key to the procedure is to avoid computing and summing all of the $lmn$ modes and then performing 
one single fit.  Instead, we perform the entire fitting process on each $lmn$ mode individually, extracting as many 
analytic coefficients as we can, and then sum all the results into the final PN expansion.  This process has two key 
advantages over the previous one: it turns out possible in a predictable way to reduce the number of modes that 
necessarily must be computed and at the individual $lmn$ mode level we find the existence of regular structure in the 
appearance of transcendental coefficients, which can be exploited in the fitting process.

\subsection{Reducing the number of flux calculations}

As noted above, calculations of the full energy flux for each orbit, characterized by a unique combination of $p$ and 
$e$, required around $7500$ $lmn$ modes, each of which had to be computed to an accuracy sufficient to make the net 
flux accurate to $200$ decimal places.  With over $1700$ combinations of $p$ and $e$ needed for the fit, the total 
number of modes computed numbered over ten million.  However, by fitting each $lmn$ separately, this number can be 
reduced significantly.  This works as follows.  We set first a particular goal in power of eccentricity to be 
reached.  We chose $e^{30}$ for our purposes.  This power of eccentricity becomes a hard limit, sacrificing any 
ability to find analytic coefficients beyond it.  Yet a benefit results, because each $lmn$ mode can be written as a 
power series in eccentricity starting with $e^{2|n|}$, and so no modes with $|n|$ beyond $15$ need be calculated, 
which leaves us with just $31$ $n$ modes for each spherical harmonic $lm$.  

By counting the $lm$ modes that would be needed to reach $7$PN (the goal in Paper I), we find that this would reduce
the necessary mode calculations to about $1450$ for each orbit $(p,e)$, a five-fold reduction.  Additional gain might 
be made by setting the eccentricity expansion goals to vary by PN order, thus potentially lowering this number further.  

\subsection{Transcendental structure and the eulerlog function}

As Paper I showed, in the PN expansion the most significant bottleneck in the extraction of analytic coefficients
is the appearance of transcendental numbers at higher orders.  These begin at 3PN order and increase in complexity 
each 3PN thereafter.  For instance, the circular-orbit limit of the angular momentum flux at 4PN order is given by
\begin{align}
\label{eqn:4circ}
\begin{autobreak}
\MoveEqLeft
\mathcal{J}_{4}^{\rm circ}=-\frac{323105549467}{3178375200}
+\frac{232597 }{4410} \gamma_\text{E} -\frac{1369}{126} \pi^2
+\frac{39931}{294} \log (2)-\frac{47385}{1568} \log (3) . 
\end{autobreak}
\end{align} 

In general, as is clear from the 3PN hereditary function $\tilde{\chi}(e)$ and from recent work \cite{MunnEvan19b} at 
4PN order, the 3PN, 4PN, and 5PN non-log terms will (without isolating any eccentricity singular factors) have the 
form
\begin{align}
\label{eqn:fullFlux345}
\begin{autobreak}
\MoveEqLeft
\mathcal{J}_{q} = 
\sum_{i=1}^{\infty}{e^{2i}\Big(a_i
+b_i \, \pi^2
+c_i \, \gamma_E+d_i \, \log{2}}
+e_i \, \log{3}+f_i \, \log{5}
+g_i \, \log{7}+\cdots\Big), 
\end{autobreak}
\end{align}
where $q \in \{3,4,5\}$ and the coefficients in the sets $(a_i, b_i, c_i, \cdots)$ are all rational numbers that vary 
with $q$.  Note that the natural log of each prime will first appear at some sufficiently high $i$ and then will remain 
present at every PN order thereafter.  Because for a given $i$ each coefficient $(a_i, b_i, c_i, \cdots)$ is different, 
all of these coefficients must be found simultaneously by an integer-relation algorithm using a multi-dimensional 
search vector, e.g., $\{1,\pi^2,\gamma_E,\log(2),\log(3)\}$ (the dimension of the search vector is set by the number 
of unique transcendental numbers plus one for the rationals themselves; i.e., five dimensional in the example given).  
This drastically reduces the ability of the integer-relation algorithm to find analytic coefficients, unless numerical 
precision is raised significantly.  

Fitting by $lmn$ (or even just $lm$) modes aids this effort because of how the transcendental structure of individual 
modes differs from that of the full flux.  To elaborate, it is well known that only the $l=2$, $m=2$ and $l=2$, $m=-2$
modes contribute to the Peters flux $\mathcal{J}_0$.  The PN expansions of all other $lm$ modes begin at higher 
powers of $y$.  Specifically, any given mode will be suppressed by a factor of $y^r$ \cite{JohnMcDa14}, where
\begin{equation} 
r = \begin{cases} 
      l - 2 & l + m \text{ even}, \\
      l - 1 & l + m \text{ odd}. 
   \end{cases}
\end{equation}

This being the case, the PN expansion of an individual $lm$ mode will have a form that differs from 
\eqref{eqn:angMomFluxInf} and is found to be instead
\begin{align}
\label{eqn:angMomFluxlmn} 
&\bigg\langle \frac{dL}{dt} \bigg\rangle_{lm}^{\infty}=
\sum_{n} \frac{m \o}{64\pi}\frac{(l+2)!}{(l-2)!} |C^{+}_{lmn}|^2     
\\ 
\notag 
& \qq =\bigg\langle \frac{dL}{dt} \bigg\rangle_\text{N}^\infty y^r
\bigg[\mathcal{J}_0^{lm} + y \mathcal{J}_1^{lm} + y^{3/2}\mathcal{J}_{3/2}^{lm}
+y^2\mathcal{J}_2^{lm} 
\\ 
\notag 
& \qq \qq + y^{5/2}\mathcal{J}_{5/2}^{lm}
+ y^3\Big(\mathcal{J}_3^{lm}+\mathcal{J}_{3L}^{lm}\log(y)\Big) +\cdots\bigg] .
\end{align}
(Note that there is a slight subtlety in this notation, as $\mathcal{J}_i^{lm}$ does not represent a decomposition
of $\mathcal{J}_i$, but rather the $i$th \emph{relative-order} flux term in the $lm$ mode.)  We can immediately see 
from this expansion that the lowest appearance of transcendental numbers will be in $\mathcal{J}_{3}^{lm}$, which 
contributes to the term $\mathcal{J}_{r+3}$ in the full flux.  As an example, consider the $l=4$, $m=3$ mode, for 
which $r=3$.  This mode will not contain a term with the eccentric transcendental structure \eqref{eqn:fullFlux345} 
until 6PN relative to $\mathcal{J}_0$.  Therefore, in calculating full flux coefficients at, say, 4PN order, this 
mode will only contribute rational numbers, which require less numerical precision to extract.

The PN series for an $lmn$ mode will mirror \eqref{eqn:angMomFluxlmn}, but with $lm \rightarrow lmn$ (the exception 
to this is when $m = -n$; see Sec. \ref{sec:mNegn}).  Therefore, in fitting by $lmn$, rather than place a universal 
precision goal to be met across the board, we vary the number of significant decimal places to which we calculate 
in a planned fashion by mode to account for this changing transcendental structure.  This improvement is a useful but 
fairly modest one, as the precision needs for even a small number of $lmn$ modes quickly become prohibitively expensive 
once the search vector surpasses five terms. 

Fortunately, there is another key difference between the transcendental structure of the full flux and that of its 
$lmn$ modes--the appearance of the eulerlog function.  It is well known and studied \cite{JohnMcDa14} that for the 
$lm$ modes of a circular-orbit flux, at 3/4/5PN (integer) orders the sum of the non-log and log terms has the form
\begin{align}
\label{eqn:lmCircF345}
\begin{autobreak}
\MoveEqLeft
\mathcal{J}_{q}^{lm, \rm circ} + \mathcal{J}_{qL}^{lm, \rm circ}\log y= a_0+b_0 \pi^2
\mkern-36mu+c_0 \Big(\gamma_E+\log{2}+\log{|m|}+\frac{1}{2}\log{y}\Big) \quad (m\ne0) ,
\end{autobreak}
\end{align}
where the rational coefficients $\{a_0,b_0,c_0\}$ vary with both $q$ and $lm$.  Because the same coefficient sits in 
front of all of the transcendentals $\gamma_E,$ $\log2,$ $\log |m|$, and the factor $(1/2)\log y$ 
\cite{DamoIyerNaga09,JohnMcDa14}, these factors are all grouped together into defining the eulerlog function:
\begin{equation}
\text{eulerlog}_m(y)=\gamma_E+\log{2}+\log{|m|}+\frac{1}{2}\log{y} .
\end{equation}
Thus, in these circular-orbit cases, the number of independent rational coefficients reduces to three.  This 
convenient reduction is lost when the modes are summed over $m$, as the log terms will accumulate coefficients
that can no longer be related to one another. 

As it turns out, the $lmn$ modes of the flux for eccentric orbits admit an analog of this eulerlog function.  When 
separated by $lmn$ mode, $\eqref{eqn:fullFlux345}$ and $\eqref{eqn:lmCircF345}$ generalize to
\begin{align}
\label{eqn:3PNform}
\begin{autobreak}
\MoveEqLeft
\mathcal{J}_{q}^{lmn} + \mathcal{J}_{qL}^{lmn}\log y= 
\sum_{i=|n|}^{\infty}{e^{2i}\bigg[a_i+b_i \pi^2}
+c_i \Big(\gamma_E+\log{2}+\log{|m+n|}
+\frac{1}{2}\log{y}\Big)\bigg] ,
\end{autobreak}
\end{align}
for $m\ne -n$.  This allowed us to define a generalized eulerlog function (for $m\ne -n$) that is given by
\begin{equation}
\label{eqn:genEulerlog}
\text{eulerlog}_{m,n}(y)=\gamma_E+\log{2}+\log{|m+n|}+\frac{1}{2}\log{y} .
\end{equation}  
We discovered this function while working with $lmn$ mode fitting, but Nathan Johnson-McDaniel in 2015 was actually 
the first to find the generalized eulerlog function $\text{eulerlog}_{m,n}(y)$ by modifying his $S_{lm}$ 
factorization \cite{JohnMcDa14}.  By using this function while fitting, the search vector required for the 
integer-relation algorithm immediately drops from five or more terms down to three.  

\subsection{Hierarchical fitting: the eulerlog simplification}

Because the generalized eulerlog function includes the $\log(y)$ term, we can improve the fitting process even further.  
Taking any of the 3/4/5PN series again and using the expected general form \eqref{eqn:3PNform} as our model, we note 
that the $\log(y)$ term ($\mathcal{J}_{qL}^{lmn}$) is simply a rational power series in $e^2$ -- one which can be fit 
separately from the more complicated non-log term ($\mathcal{J}_{q}^{lmn}$).  By fitting this log series first we can 
determine the $c_i$ coefficients independently.  Once these coefficients are known, we can return attention to the 
more complicated non-log flux term.  Then, applying knowledge of the eulerlog function, we see that the $c_i$ 
coefficients are also the ones that stand on the combination of transcendentals, $\gamma_E+\log{2}+\log{|m+n|}$.  
With this piece subtracted off, we are left with a remaining search vector with only two terms, $\{1,\pi^2\}$.  

Better still, Johnson-McDaniel's progress on tail factorizations (both circular in \cite{JohnMcDa14} and more recently 
eccentric) suggests that the $\pi^2$ piece is also linked to this eulerlog function.  We have in fact discovered 
empirically that the ratio of $b_i$ to $c_i$ in \eqref{eqn:3PNform} depends exclusively on $l$, allowing a 
generalization of the eulerlog function to what might be called the ``eulerlogpi" function.  This connection was 
used as a further aid in extracting a few more coefficients in some of the high PN terms (shown in 
Secs.~\ref{sec:LResults} and \ref{sec:Eresults}), but we leave discussion of structure beyond the eulerlog function 
to a later paper.

Up to this point we have only discussed simplifications to terms at PN orders 3, 4, and 5.  However, the eulerlog 
function at least partially characterizes the appearance of those particular transcendentals, and their products, at 
even higher orders.  Specifically, for each integer $k \ge 1$, $\mathcal{J}_{3k}^{lmn}$, 
$\mathcal{J}_{3k+1}^{lmn}$, and $\mathcal{J}_{3k+2}^{lmn}$ will all contain a product of $k$ terms, each 
having the form of the square-bracketed portion in \eqref{eqn:3PNform}, though there may also be additional 
transcendentals native to that order (such as $\zeta{(3)}$).  An analog of this fact was also applied in 
\cite{JohnMcDaShahWhit15} to simplify the $lm$ modes of the redshift invariant in the circular-orbit limit.

As an example, consider the sum of the 6PN non-log, log, and log-squared enhancement functions.  This sum can be 
shown to decompose into 
\begin{align}
\begin{autobreak}
\label{eqn:6PNeulog}
\MoveEqLeft
\mathcal{J}_{6}^{lmn}+\mathcal{J}_{6L}^{lmn}\log{y}+\mathcal{J}_{6L^2}^{lmn}(\log{y})^2 
= \sum_{i=|n|}^{\infty}{e^{2i}\bigg[a_i + b_i \, \pi^2+c_i \, \pi^4+d_i \, \zeta(3)}
+ \Big(e_i + f_i \, \pi^2+g_i \, \text{eulerlog}_{m,n}(y)\Big)
\times\Big(h_i + j_i \, \pi^2 + k_i \, \text{eulerlog}_{m,n}(y)\Big)\bigg] .
\end{autobreak}
\end{align}
We can see the aforementioned product involving the eulerlog function, and also the first appearance of the new 
transcendental, $\zeta(3)$.  As it turns out, this expression is more complicated than it has to be and contains 
more than the true number of degrees of freedom.  With a bit of inspection, we can write \eqref{eqn:6PNeulog}
in the simpler form 
\begin{align}
\begin{autobreak}
\MoveEqLeft
\mathcal{J}_{6}^{lmn}+\mathcal{J}_{6L}^{lmn}\log{y}+\mathcal{J}_{6L^2}^{lmn}(\log{y})^2 
= \sum_{i=|n|}^{\infty}{e^{2i}\bigg[a_i + b_i \, \pi^2+c_i \, \pi^4+d_i \, \zeta(3)}
+ \Big(e_i + f_i \, \pi^2+g_i \, \text{eulerlog}_{m,n}(y)\Big)
\times\Big(1+h_i \, \pi^2+ \text{eulerlog}_{m,n}(y)\Big)\bigg] .
\end{autobreak}
\end{align}

After multiplying out the summands, collecting terms, and renaming coefficients, we can separate the terms with 
different powers of $\log(y)$ to obtain
\begin{align} 
\mathcal{J}_{6L^2}^{lmn} &= \sum_{i=|n|}^{\infty}{e^{2i}\Big(\frac{C_i}{4}\Big) }, 
\\
\mathcal{J}_{6L}^{lmn} &= \sum_{i=|n|}^{\infty}{e^{2i}\Big(A_i+B_i \, \pi^2 +C_i \, \beta \Big)}, 
\\
\mathcal{J}_{6}^{lmn} &=\sum_{i=|n|}^{\infty}{e^{2i}\Big(2 A_i \, \beta + 2 B_i ~\pi^2 \, \beta} + C_i \, \beta^2 
\notag 
\\
 &+ D_i+ E_i ~ \pi^2 + F_i ~ \pi^4 + G_i ~ \zeta{(3)} \Big) .   
\end{align}
Here we define $\beta =\gamma_E+\log{2}+\log{|m+n|}$.  Remarkably, by working from the top down and carrying over 
results, we can ascertain the full analytic structure at 6PN, including the hardest non-log term, of each $lmn$ mode 
with a search vector of maximum length of four.  Without this hierarchical procedure the required search vector would 
be $\{1, \pi^2, \gamma_E, \log{2}, \log{|m+n|} 
\pi^2 \gamma_E, \pi^2 \log{2}, \pi \log{|m+n|},\\ \gamma_E \log{2}, \gamma_E \log{|m+n|}, \log{2} \log{|m+n|},
 \pi^4, \gamma_E^2, (\log{2})^2,\\ (\log{|m+n|})^2, \zeta(3) \}$ (i.e., 15 dimensional!), which would likely preclude 
ever finding the analytic fit by direct application of the integer-relation algorithm.  

Though this hierarchical fitting method also works perfectly as just described for 7PN, it (perhaps unexpectedly) 
requires modification at 8PN order.  As can be seen in the circular-orbit results of \cite{JohnMcDa14}, 8PN marks the 
first appearance of the $\log(2y)$ contribution, separate from the eulerlog function.  While the $\log(2y)$ term 
can be separated, it alters the balance between the $\log(y)$ and $\log(2)$ factors and alters the transition from 
\eqref{eqn:3PNform} to \eqref{eqn:6PNeulog}.  The situation at 8PN can be salvaged by introducing a 5th search vector 
component of $2\, \beta -\log(2)$, though doing so significantly decreases the ability to find complete analytic 
fits without increasing numerical precision.

\subsection{Modes with $m=-n$}
\label{sec:mNegn}

Omitted from the above analysis is what to do with all modes for which $m=-n$.  In these cases the eulerlog 
function, as previously defined in \eqref{eqn:genEulerlog}, is divergent and not useful.  The reason these flux 
components represent a special case is that the frequency $\omega=m\Omega_{\vp}+n\Omega_{r}$ of each of these 
$m=-n$ modes almost vanishes and appears at one PN order higher than neighboring modes, because 
$\Omega_{\vp} \rightarrow \Omega_r$ in the Newtonian limit.  As a result, the lowest power of $y$ appearing in each 
of these modes is $2l+1$ higher ($2l+2$ for the energy flux) than it would be otherwise (thus yielding total order
$3l-1$ for $l+m$ even and $3l$ for $l+m$ odd relative to $\mathcal{J}_0$).  Hence, there are no contributions until 
5PN order.  Unlike cases where $m \ne -n$, these modes produce no contribution at $1.5$PN and so no 
combinations of transcendentals appear until $5$PN order.  When transcendentals do appear, they are characterized 
by a different eulerlog function of the form
\begin{equation}
\text{eulerlog}_{m,-m}(y)=\gamma_E+\log{6}+\log{m}+\frac{3}{2}\log{y} .
\end{equation}

Thus, if we restrict attention to the $m=-n$ contributions, the structure of the sum of the non-log and log 
terms that contribute at the 5/6/7PN level relative to the lowest order for that $l$ is
\begin{align}
\label{eqn:5PNmnform}
\begin{autobreak}
\MoveEqLeft
\mathcal{J}_{5/6/7}^{lm-m} + \mathcal{J}_{5L/6L/7L}^{lm-m}\log y= 
\sum_{i=|n|}^{\infty}{e^{2i}\bigg[a_i+b_i \pi^2}
+c_i \Big(\gamma_E+\log{6}+\log{m}+\frac{3}{2}\log{y}\Big)\bigg] .
\end{autobreak}
\end{align}
The net effect is that the first appearance of such a series will be at $10$PN relative to $\mathcal{J}_0$.

To round out this discussion, note that no $m=0$ modes contribute to the angular momentum flux (see 
\eqref{eqn:fluxNumeric}), while only those modes with $m=n=0$ vanish in the energy flux case.  Finally, in both 
fluxes $lmn$ modes are equal to $l$,-$m$,-$n$ modes.

\subsection{Validation}
\label{sec:Val}

Irrespective of finding substantial improvements in fitting, the same issue is still faced here as it was in 
Paper I--how do we know that the analytic results emerging from an ``experimental mathematics'' technique are not 
simply a coincidence?  After all, it is always possible to represent a floating point number of any precision as 
a rational number or as a transcendental times a rational number.  An integer-relation algorithm can potentially 
return multiple solutions for the same numeric input.  We need some means of testing to be confident that a retrieved 
rational number, or sum of rationals times transcendentals, is in fact the correct one which would emerge from a 
first-principles analytic calculation.  Fortunately, several validation tests exist.

The simplest, as noted in Paper I, can be applied to any derived rational coefficient.  For any rational number
in irreducible form with the coprime integers having $N_N$ digits in the numerator and $N_D$ digits in the 
denominator, define the \textit{fractional complexity} $f$ by $f=N_D+N_N$.  Say it is suspected that a rational with 
complexity $f$ represents a given decimal number, and that the decimal number and rational agree to $N$ digits of 
precision.  Then, the probability that the rational number lies within the uncertainty range of the decimal 
number but is a mere coincidence is roughly given by $\mathcal{P}\simeq10^{f-N}$ \cite{ShahFrieWhit14}.  
Using this test alone, 
many of the coefficients that we obtain, for the subset of PN series that only have rational coefficients, have 
\textit{extremely} small chances of coincidence--as low as $10^{-300}$ in some cases.  Moving one step further, 
at certain PN orders involving the eulerlog simplification, like $9/2$PN and $11/2$PN, we can use this formula 
successively to confirm that the procedure is not producing results that are mere coincidence.  To be more specific, 
with the hierarchical procedure we first check that the log series, with its expected rational coefficients, has 
essentially zero chance of coincidence.  Then we apply the step of removing the eulerlog term with its transcendentals 
from the non-log flux and use the integer-relational algorithm to determine remaining rational coefficients.  These 
are then subjected to the same test, with the requirement that they too have near vanishing chance of coincidence.  
If the extraction fails, the fractional complexity of the derived rationals will be large and the probabilities of 
coincidence will jump many orders of magnitude.  This is the usual indication we get when we reach the limit of being 
able to determine a high PN order fit.

For series that are not purely rational (e.g., 6PN) and one of the integer-relation fits is being made in multiple 
dimensions, verification requires a slightly subtler technique, as the above expression cannot account for the 
greater dimension of the search vector space.  One possible extension would be to add up all the digits of all the 
coprimes in the rational numbers to use as a substitute formula for $\mathcal{P}$.  However, such a probability test 
is not as well validated as its single rational counterpart.

A useful second test, available for all emergent rational numbers, is to examine the prime factorization of each 
term's overall denominator, as these factorizations exhibit a certain universal behavior.  In any given power series in 
eccentricity, this behavior involves the largest prime $p_i$ that appears in the common denominator of the coefficient 
of the $e^{2i}$ term.  An analytic inspection of the RWZ formalism reveals that this prime should remain within an 
order of magnitude of the limit $p_i \sim i$, though with somewhat larger $p_i$ occurring at higher PN order.  Every 
single rational number denominator that we have encountered through 9PN has $p_i \leq 29$.  In number theory, 
large integers whose prime factorization only involves small primes, or powers of small primes, up to $p_i$ are called 
$p_i$-smooth numbers.  Furthermore, number theory considerations show \cite{Gran08} that the probability that a randomly 
selected denominator with $d$ digits will have all prime factors $\leq p_i$ is less than
\begin{equation}
\frac{1}{10^{d+1}-10^d}\binom{\lfloor(d+1)\log(10)/\log(2)\rfloor+\pi(p_i)}{\pi(p_i)} ,
\end{equation}
where $\binom{a}{b}$ is the binomial coefficient and $\pi(p_i)$ is the prime counting function (the number of primes 
$\leq p_i$).  This condition has been utilized also in \cite{JohnMcDaShahWhit15}.  As an example, in the $202$ mode, 
the coefficient of $e^6$ at 6PN order has a denominator equal to $1150293142462464000$.  For this number, $d=20$ 
and the maximum prime is $p_3=17$.  Therefore, the probability of an integer of this size being 17-smooth is of order 
$10^{-11}$.  Another quick check, as reflected in this example, is that denominators for high powers of $e^2$ and 
$y$ will be characterized by powers of $10$ (though this does not apply for modes where $5$ divides $|m+n|$).

\begin{figure*}
\hspace{-2.4em}\includegraphics[scale=.71]{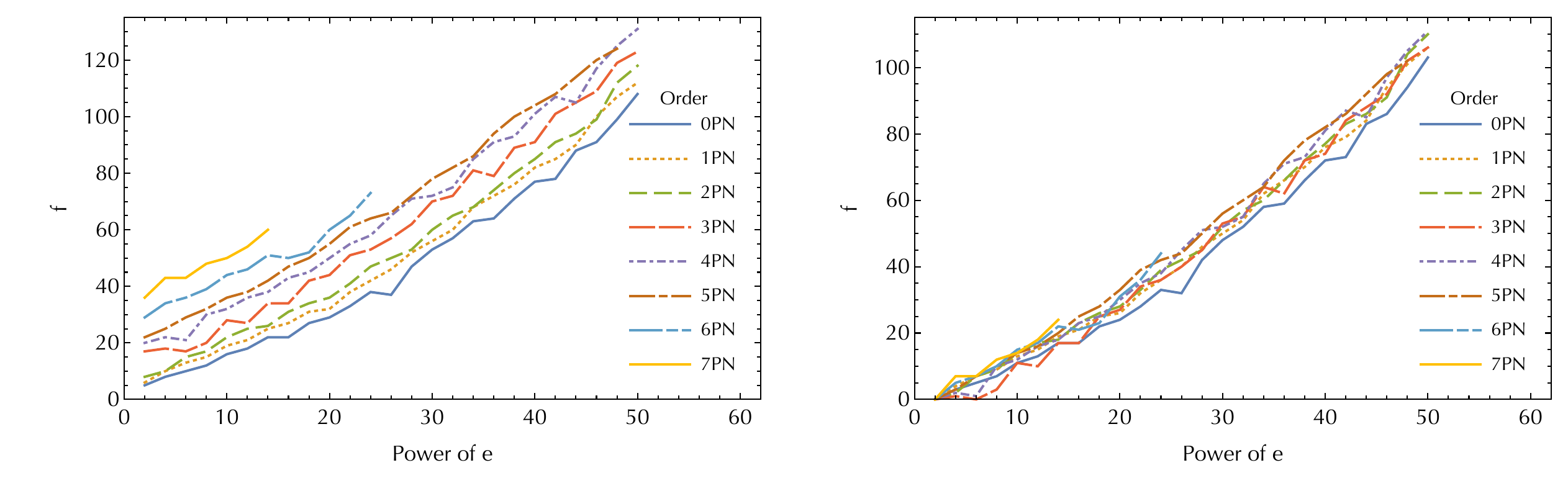}
\caption{
Increase of the fractional complexity $f$ with eccentricity power series exponent, separated by PN order.  
For orders 3PN and above, the purely rational coefficient is used.  On the left, we plot data from the $lmn = (2,2,1)$ 
mode for multiple PN orders.  On the right, each trend has been normalized by subtracting the first value of 
$f$ in the sequence, leading to a more universal growth in complexity.
\label{fig:fracComps}}
\end{figure*}

Finally, we have identified a fairly intuitive universal trend characterizing the PN series of $lmn$ modes, which 
involves the fractional complexity directly.  Consider a given PN order and the power series in eccentricity at that 
order.  It can be easily seen that the fractional complexity of the rational numbers in the eccentricity series 
generally increases with powers of $e$ (i.e., the rational numbers get more ``complicated'' for higher powers of 
$e^2$).  But as we vary the PN order the precise values of $f$ at the same power of $e$ differ greatly, both because 
of a different power of $y$ and because transcendentals may or may not be attached to the particular rational number.  
We can see both aspects of this behavior on the left side of Fig.~\ref{fig:fracComps}.  There we plot $f$ versus 
the exponent of eccentricity $e$ extracted from a set of different series.  The first behavior is clearly evident as 
all of the curves have roughly the same shape.  However, it is also clear that the curves differ significantly as 
each has effectively an offset corresponding to PN order.  We then normalize these trends by taking each sequence and 
subtracting off the first value of $f$ in each sequence.  The results are plotted on the right side, which illustrates 
a more universal behavior.  This universal trend in fractional complexity serves as a third test for the 
integer-relation algorithm results.  We compare the $f$ values of rational numbers extracted in a series to this 
trend, looking for any values that represent clear outliers.  

\subsection{The roadmap for fitting by $lmn$ modes}
\label{sec:FullProc}

Here we present the full procedural roadmap for extracting eccentricity coefficients in the high PN order series 
from BHPT flux calculations.  An analogous roadmap was given in Paper I, with much more detail on the BHPT steps, 
and should be consulted as well.

\begin{itemize}[leftmargin=*]

\item \emph{Solve orbit equations for given $p$ and $e$.}  Given a set of 
orbital parameters, we find $t_p (\chi)$, $\vp_p (\chi)$, and $r_p(\chi)$ to 
high accuracy at locations equally spaced in $\chi$ using SSI.  From these 
functions we also obtain the orbital frequencies $\O_r$ and $\O_\vp$.  

\item \emph{Obtain homogeneous solutions to the FD RWZ master equation for given $lmn$ mode.}  
We find the (normalized) homogeneous solutions to \eqref{eqn:masterInhomogFD} using the MST 
formalism and transformations outlined in Paper I.  All quantities are computed with some pre-determined, 
mode-dependent accuracy goal; in this paper that goal ranged from $450$ decimals for the 
$20$ and $22$ modes to $150$ decimals for $l=8$ and above.

\item \emph{Form $lmn$ flux contribution.}  Form $C_{lmn}^+$ by applying the
exponentially-convergent SSI technique to \eqref{eqn:EHSC}.  The coefficient 
$C_{lmn}^+$ feeds into a single positive-definite term in the sum 
\eqref{eqn:fluxNumeric}.  Unlike in Paper I, here we leave the flux data in this $lmn$ component form.

\end{itemize}

\noindent The next steps involve the PN-side computations:

\begin{itemize}[leftmargin=*]
\item \emph{Select fitting goals for $y$ and $e$}.  We set a hard limit of
$8.5$PN order in $y$ and $30$th order in $e$.  Thus, we compute $lmn$ 
fluxes for $l\le 10$, $0 \le m \le l$ (and only even $m$ for $l=10$), and $-15\le n \le 15$.  Because the
fitting is done separately for each mode, we are not restricted to any universal choices of $p$ and $e$,
and we can wildly vary our accuracy goals with $l, m,$ and $n$.  For optimal results, we increased
the number of $p$ and $e$ values for low $l$ and $m$, computing as many as $2750$ orbits
for the $220$ mode and as few as $1300$ orbits for all $l=10$ modes. In general, 
$e$ values ranged from $10^{-5}$ to $0.1$, while $p$ values ranged anywhere from $10^{8}$ 
through $10^{55}$, depending on mode.  The values of $y$ are derived from $p$ and $e$.

\item \emph{Use expected form of the expansion in $y$}.  Known results for circular fluxes 
on Schwarzschild backgrounds 
allow us to surmise the expected terms in the $y$-expansion, shown in 
Eqn.~\eqref{eqn:angMomFluxInf}.  In mode-by-mode fitting
this form is adjusted by an overall factor of $y^r$, where $r=l-2$ if $l+m$ is even
and $r=l-1$ if $l+m$ is odd.

\item \emph{Fit for terms on powers of $y$ and log(y)}.  
We use \textsc{Mathematica's} \texttt{NonlinearModelFit} function to obtain 
numerical values for the coefficients $\mathcal{J}_{7/2}^{lmn}$, $\mathcal{J}_4^{lmn}$, etc.
  We perform this fit 
separately for each of the values of $e$ in a mode's dataset.  

\item \emph{Fit each model for chosen $\mathcal{J}_i^{lmn}(e)$ using eccentricity-dependent data,
starting with the highest power of log}.  
The function \texttt{NonlinearModelFit} is again used to 
find the unknown coefficients in each eccentricity function expansion.
The eccentricity coefficient models allow us to perform hierarchical fitting.  As 
lower order e coefficients are firmly determined in analytic form (see next 
step), they can be eliminated in the fitting model to allow new, higher-order 
ones to be included.  In keeping with the
eulerlog procedure, we first fit the highest $\log(y)$ power appearing at any given order.
 
\item \emph{Attempt to determine analytic form of $e^2$ coefficients}.
Because we have chosen the highest power of log at this order, the fitted series will necessarily
be rational.  We use \textsc{Mathematica's} function \texttt{RootApproximant} (hereafter RA),
which finds simple fractional representations for rational coefficients only.  As we progress to 
more complicated terms, transcendentals will begin to appear,
and we will require \textsc{Mathematica's} \texttt{FindIntegerNullVector} (hereafter FINV), 
which is an implementation of the PSLQ integer-relation algorithm. 

\item \emph{Assess the validity of the analytic coefficients}.
A rational or irrational number, or combination thereof, predicted by RA or FINV 
to represent a given decimal number has a certain probability of being a 
coincidence (Note: the output will be a very accurate 
\emph{representation} of the input decimal number regardless).  The specifics of this determination,
as well as various additional consistency checks, are 
given in Section \ref{sec:Val}.  With the analytic coefficients we obtain, 
in no case is the probability of coincidence larger than $10^{-6}$, 
and in many cases the probability is as low as $10^{-300}$.  It is also important that the analytic 
output of PSLQ not change when the number of significant digits in the input 
is varied (within some range).

\item \emph{If necessary, move sequentially down in powers of log at the same order, fitting each new term
via the corresponding eulerlog simplification}.
Once the rational series is taken to the limits of precision for a given $\log(y)^k$, it can be multiplied by the 
appropriate eulerlog factors and subtracted off the fit data for the $\log(y)^{k-1}$ term.  As shown in previous 
sections, this new series will be more complicated than that for $\log(y)^k$, but it will typically still be 
tractable.  In this way, we fit all log powers (including the non-log terms with power $0$) for a single PN 
order (power of $y$) together.  We perform this fit for all unknown PN orders through $8.5$PN.  We are also able 
to retrieve the term $\mathcal{J}_{9L3}$, which admits a closed form expression.

\item \emph{Sum over $lmn$ modes.}  We repeat the steps above for each $lmn$ component of the flux.  Then, we 
reconstruct the total flux at each PN order by summing: first over $-15 \le n \le 15$, then over $0 \le m \le l$, 
and then over $2 \le l \le 10$.  Finally, with the full flux summed at a given PN order, a predicted eccentricity 
singular factor is divided out of the expression, often leaving a convergent power series.  Note that this procedure 
allows us to save our data by both $lm$ and $lmn$ modes.  Segregating results by $lm$ modes can allow the use of 
known or expected PN forms to turn some truncated infinite series into exact, closed-form or simplified expressions 
with appropriate eccentricity singular factors. 
\end{itemize}

\mbox{}

\section{Angular momentum radiation: New coefficients through 8.5PN order}
\label{sec:LResults}

\subsection{Results}
We now provide the understanding we have gained of the high-order terms in the PN expansions of the angular 
momentum flux (at lowest order in the mass ratio $\nu$).  The expansion is given sequentially to $8.5$PN order 
relative to the leading (Newtonian) term.  We attempted to take all eccentricity power series to $e^{30}$ 
analytically, but we substituted numeric forms when necessary.  As a result, we have accurate mixed analytic/numeric 
results to $e^{30}$ for all PN orders except the $8$PN non-log term, which was only completed to $e^{10}$.  In some 
cases, we have additional numeric results beyond $e^{30}$ from the full-flux fitting method of Paper I.  We also 
give the $9$PN $(\log y)^3$ term, which has a simple closed form \cite{MunnEvan19a}.

Because coefficients in many of the non-rational enhancement functions grow excessively large at high powers of 
$e^2$, we generally only provide here in print the first few terms for those series in analytic form.  However, 
expansions relevant to the discussion section \ref{sec:LDiscussion} are given to as many orders as necessary.  
Additionally, the 8PN non-log term, for which no new coefficients could be extracted, is given in approximate 
numeric form to 20 decimals of precision.  We indicate the highest \emph{analytic} power of $e^2$ found for each 
PN order via a Greek constant shown at the end of each series.  Note that this is in contrast to Paper I, where 
listed constants represented known numeric coefficients.  Even though we can report here only a subset of the full 
analytic structure that we have discovered, our full results are published in a \textsc{Mathematica} notebook on 
the Black Hole Perturbation Toolkit website \cite{BHPTK18} for easy retrieval.  

\begin{widetext}

We start with a second presentation of all of the flux terms up to and including 3PN, this time in our 
$\mathcal{J}(e)$ notation.  These enhancement functions can be found by fitting; however, they are more easily 
derived by simply recasting the functions of Sec.~\ref{sec:LcurrPN} in terms of the Darwin eccentricity $e$ using 
Eq.~(4.38) of Paper I.  Computing them in both ways allows for an independent check of our methodology.  In either 
case, we find that $\tilde{\vp}$, $\tilde{F}$, and $\tilde{\chi}$ remain the same functions, just with the replacement 
$e_t \rightarrow e$.  On the other hand, coefficients in $\tilde{\psi}$ are changed by the transformation, which 
is reflected in the subscript $D$ (for Darwin eccentricity).  Note that all terms in \eqref{eqn:J0} through 
\eqref{eqn:J3L} can be expanded to arbitrary order in $e^2$, using the techniques described earlier for summing 
the various enhancement functions.  Hence, we only show a few terms in these cases (see \cite{BHPTK18}).  We find
\begin{align}
\label{eqn:J0}
\mathcal{J}_0 &= \frac{1}{(1-e^2)^{2}}\left(1+\frac{7}{8}e^2\right),
\\
\mathcal{J}_1 &=  -\frac{1}{(1-e^2)^{3}}\biggl(\frac{1247}{336}
+\frac{2777}{336}e^2+\frac{5713}{2688}e^4\biggr),\\
\mathcal{J}_{3/2} &=  4\pi\tilde{\phi}(e),\\
\mathcal{J}_2 &=  \frac{1}{(1-e^2)^{4}}\left(
-\frac{135431}{9072}-\frac{190087 }{6048}e^2+\frac{192133}{24192} e^4
+\frac{3499}{2304} e^6
\right)+\frac{10}{(1-e^2)^{5/2}}\bigg(1+\frac{7}{8}e^2\bigg),\\
\mathcal{J}_{5/2} &= 
-\frac{8191}{672}\pi\tilde{\psi}_D(e)
\\
\notag
&= -\frac{8191}{672}\frac{\pi}{(1-e^2)^{9/2}}
\biggl(
1
+\frac{102121}{16382} e^2
+\frac{3557227}{524224} e^4
+\frac{6395111}{9436032} e^6
+\frac{5968651}{603906048} e^8
-\frac{59360743}{30195302400} e^{10}
+\cdots\biggr),\\
\mathcal{J}_3 &  = 
\frac{1}{(1-e^2)^5}\biggl[
\frac{2017023341}{9979200}
+\frac{3081414883}{2494800}e^2
+\frac{1949678087}{1900800} e^4
+\frac{621661339}{4435200} e^6
+\frac{29539919}{5677056} e^8
\notag\\&\qquad -
\sqrt{1-e^2}\left(\frac{379223}{5040}+\frac{138673}{630}e^2
-\frac{196073}{5760}e^4-\frac{36405}{896}e^6\right)
\biggr]
+
\left(
\frac{16}{3}\pi^2-\frac{1712}{105}\gamma_\text{E}-\frac{116761}{3675}
\right)\tilde{F}(e)
\notag\\&\qquad -
\frac{1712}{105}\log \left(\frac{8 \left(1-e^2\right)}{1
+\sqrt{1-e^2}} \right)\tilde{F}(e)
-\frac{1712}{105}\tilde{\chi}(e) ,
\\
\label{eqn:J3L}
\mathcal{J}_{3L} 
&=  -\frac{1}{\left(1-e^2\right)^{5}}
\left(\frac{856}{105}+\frac{24503 e^2}{420}+\frac{11663 e^4}{280}+\frac{2461 e^6}{1120}\right). 
\end{align}

From this point, we present new coefficients found by fitting.  We start with the 3.5PN enhancement function, which 
was computed to $e^{30}$.
\begin{align}
\begin{autobreak}
\MoveEqLeft
\mathcal{J}_{7/2} = \frac{\pi}{(1-e^2)^{11/2}}\bigg(-\frac{16285}{504}-\frac{370255}{1008}e^{2}-\frac{11888119}{48384}e^{4}+\frac{6476904953}{20901888}e^{6}+\frac{8877357035}{167215104}e^{8}+\frac{186159455101}{26754416640}e^{10}
+\frac{34468729921289}{9631589990400}e^{12}+\frac{7790078031395741}{3775583276236800}e^{14}+\frac{319549718350556899}{241637329679155200}e^{16}+\frac{23734478429688515533}{26096831605348761600}e^{18}
+\frac{857752832161057782787}{1304841580267438080000}e^{20}+\frac{1662293559107552959945669}{3368231065863680163840000}e^{22}+\frac{208313298078191084760346703}{545653432669916186542080000}e^{24}
+\frac{7142452708941483674392066761043}{23607150111031253894556549120000}e^{26}+\frac{2263710549385314213293062077437587}{9254002843524251526666167255040000}e^{28}
+\frac{1845877280291874535352833774291427}{9177523481181075894214380748800000}e^{30} 
+\cdots \bigg)
\end{autobreak}.
\end{align}

Next are the 4PN and 4PN log terms.  The 4PN non-log series was found to $e^{30}$ through eulerlog simplifications. 
We show the series up to $e^{10}$ to illustrate some of its structure, which will be discussed in the next 
subsection.  The remainder of the series can be found at \cite{BHPTK18}.
\begin{align}
\begin{autobreak}
\MoveEqLeft
\mathcal{J}_4  =  \frac{1}{(1-e^2)^{6}}
\bigg[-\frac{323105549467}{3178375200}+\frac{232597\gamma_E}{4410}-\frac{1369\pi^2}{126}+\frac{39931\log (2)}{294}-\frac{47385\log (3)}{1568}
+ \bigg(-\frac{25591550692117}{12713500800}
+\frac{3482879\gamma_E}{4410}-\frac{22495\pi^2}{126}
-\frac{744809\log (2)}{490}+\frac{8684577\log (3)}{3920}\bigg)e^{2}
+ \bigg(-\frac{230437128487837}{25427001600}+\frac{34971299\gamma_E}{17640}
-\frac{259969\pi^2}{504}+\frac{1133219251\log (2)}{17640}-\frac{9701228007\log (3)}{501760}
-\frac{3173828125\log (5)}{301056}\bigg)e^{4}+ \bigg(-\frac{511692926097851}{50854003200}
+\frac{6578731\gamma_E}{7056}-\frac{89393\pi^2}{336}-\frac{28743092759\log (2)}{45360}
+\frac{4730808321\log (3)}{501760}+\frac{314655859375\log (5)}{1161216}\bigg)e^{6}
+ \bigg(-\frac{228428222760809}{67805337600}+\frac{2503623\gamma_E}{62720}
-\frac{21495\pi^2}{1792}+\frac{3273460573169\log (2)}{725760}+\frac{3963317295231\log (3)}{2621440}
-\frac{5065761265234375\log (5)}{2080899072}-\frac{6297785676455\log (7)}{14155776}\bigg)e^{8}
+ \bigg(-\frac{15338989354349}{11070259200}-\frac{246231168185717\log (2)}{7938000}
-\frac{12237402512884383\log (3)}{802816000}+\frac{6381944296484375\log (5)}{520224768}
+\frac{12620489342050037\log (7)}{1327104000}\bigg)e^{10}
+ \cdots + \alpha_{30}e^{30} +\cdots \biggr]
\end{autobreak}.
\end{align}
The 4PN log term, found earlier \cite{Fors16}, revealed an exact, closed-form expression:
\begin{align}
\mathcal{J}_{4L} =  \frac{1}{(1-e^2)^{6}}\biggl(
\frac{232597}{8820}+\frac{3482879 e^2}{8820}+\frac{34971299 e^4}{35280}
+\frac{6578731 e^6}{14112}+\frac{2503623 e^8}{125440}
\biggr).
\end{align}

The 4.5PN functions were both found to $e^{30}$.  We give the non-log function to $e^{10}$
in order to contrast it with $\mathcal{J}_4$ above, with the remainder posted at \cite{BHPTK18}.
\begin{align}
\begin{autobreak}
\MoveEqLeft
\mathcal{J}_{9/2} =  \frac{\pi}{(1-e^2)^{13/2}}\bigg[\frac{265978667519}{745113600}-\frac{6848\gamma_E}{105}
-\frac{13696\log (2)}{105}+ \bigg(\frac{119401899839}{19353600}
-\frac{27713\gamma_E}{30}-\frac{36487\log (2)}{210}
-\frac{234009\log (3)}{140}\bigg)e^{2}
+ \bigg(\frac{38291634777373}{2167603200}-\frac{1474139\gamma_E}{840}
-\frac{6556639\log (2)}{168}+\frac{2106081\log (3)}{112}\bigg)e^{4}
+ \bigg(\frac{15381024734595431}{1287556300800}
-\frac{71788333\gamma_E}{120960}
+\frac{47822794963\log (2)}{120960}
-\frac{172932651\log (3)}{1792}-\frac{5224609375\log (5)}{48384}\bigg)e^{6}
+ \bigg(\frac{3913633755471997}{3139184885760}
-\frac{513707\gamma_E}{27648}-\frac{2513356038497\log (2)}{967680}
-\frac{17373530187\log (3)}{71680}
+\frac{496337890625\log (5)}{387072}\bigg)e^{8}
+ \bigg(\frac{1983010121518334771}{2354388664320000}
-\frac{690257\gamma_E}{55296000}
+\frac{5129311678694857\log (2)}{387072000}
+\frac{346105863017991\log (3)}{57344000}
-\frac{87339794921875\log (5)}{12386304}-\frac{507989081563901\log (7)}{221184000}\bigg)e^{10}
+ \cdots + \beta_{30}e^{30} 
+\cdots \bigg]
\end{autobreak},\\

\begin{autobreak}
\MoveEqLeft
\mathcal{J}_{9/2L} =  \frac{\pi}{(1-e^2)^{13/2}}
\bigg(-\frac{3424}{105}-\frac{27713}{60}e^{2}-\frac{1474139}{1680}e^{4}-\frac{71788333}{241920}e^{6}-\frac{513707}{55296}e^{8}-\frac{690257}{110592000}e^{10}
-\frac{32944979}{9289728000}e^{12}
+\frac{1896198253}{14566293504000}e^{14}+\frac{217366002683}{699182088192000}e^{16}+\frac{20237480138479}{1812279972593664000}e^{18}
-\frac{1186535077588513}{181227997259366400000}e^{20}
-\frac{570980043986842753}{350857402694133350400000}e^{22}-\frac{6584105389810998751}{25261732993977601228800000}e^{24}
-\frac{3047031016950659382647}{68307726015715433722675200000}e^{26}
-\frac{175631935156003401313237}{13388314299080225009644339200000}e^{28}-\frac{80704977292116623135976991}{12852781727117016009258565632000000}e^{30}
+\cdots \bigg)
\end{autobreak}.
\end{align}

Analytic coefficients were also found to $e^{30}$ in the 5PN non-log series.  There is no novel behavior in the 
appearance of $\gamma_E$ and $\pi^2$ in the series, so we truncate at $e^8$ and leave the rest to \cite{BHPTK18}. 
\begin{align}
\begin{autobreak}
\MoveEqLeft
\mathcal{J}_{5} =  \frac{1}{(1-e^2)^{7}}\bigg[
-\frac{2500861660823683}{2831932303200}+\frac{916628467\gamma_E}{7858620}
-\frac{424223\pi^2}{6804}-\frac{83217611\log (2)}{1122660}
+\frac{47385\log (3)}{196}
+ \bigg(-\frac{7848030223872703}{471988717200}
+\frac{769099141\gamma_E}{523908}-\frac{2387269\pi^2}{2268}
+\frac{2778275573\log (2)}{124740}
-\frac{33084207\log (3)}{5390}
-\frac{76708984375\log (5)}{25147584}\bigg)e^{2}
+ \bigg(-\frac{60115129871947373}{1373058086400
}-\frac{453258311\gamma_E}{476280}-\frac{4569965\pi^2}{2268}
-\frac{8765389513\log (2)}{21384}
-\frac{3015086409\log (3)}{2759680}
+\frac{72979638671875\log (5)}{402361344}\bigg)e^{4}
+ \bigg(\frac{51952994948318117}{3624873348096}
-\frac{373576122307\gamma_E}{31434480}
+\frac{43547141\pi^2}{27216}
+\frac{457831837310951\log (2)}{94303440}
+\frac{10321124212899\log (3)}{5519360}
-\frac{5005044628796875\log (5)}{1810626048}
-\frac{152212635349397\log (7)}{295612416}\bigg)e^{6}
+ \bigg(\frac{8413909247102002313}{161105482137600}
-\frac{232447680943\gamma_E}{37255680}
+\frac{49944247\pi^2}{32256}-\frac{44427069728132087\log (2)}{823011840}-
\frac{1262436103060623\log (3)}{44154880}
+\frac{277340233759765625\log (5)}{12875563008}
+\frac{82923917976541537\log (7)}{4729798656}\bigg)e^{8}
+ \cdots + \gamma_{30}e^{30}+\cdots \bigg]
\end{autobreak}.
\end{align}

The 5PN log flux revealed a closed-form representation with a two polynomial structure reminiscent of the 2PN 
flux.  Note the recurrence of the 3PN log term (see \cite{MunnEvan19a}).
\begin{align}
\begin{autobreak}
\MoveEqLeft
\mathcal{J}_{5L} =  \frac{1}{(1-e^2)^{7}}
\bigg(\frac{4119951667}{15717240}+\frac{1977018931}{1047816}e^{2}-\frac{1472497511}{952560}e^{4}-\frac{433137913057}{62868960}e^{6}
-\frac{202342374943}{74511360}e^{8}
-\frac{7349401019}{74511360}e^{10}\bigg)
-\frac{25}{\left(1-e^2\right)^{11/2}}
 \left(\frac{856}{105}+\frac{24503 e^2}{420}+\frac{11663 e^4}{280}+\frac{2461 e^6}{1120}\right)
\end{autobreak}.
\end{align}

Like those at 4.5PN non-log and log terms, the two 5.5PN enhancement functions were obtained through $e^{30}$.  
The 5.5PN non-log has a structure that parallels that of the 4.5PN non-log flux, so it is only given to $e^{8}$ 
(see \cite{BHPTK18}).
\begin{align}
\begin{autobreak}
\MoveEqLeft
\mathcal{J}_{11/2} =  \frac{1}{(1-e^2)^{15/2}}\bigg[\frac{8399309750401}{101708006400}
+\frac{177293\gamma_E}{1176}+\frac{8521283\log (2)}{17640}
-\frac{142155\log (3)}{784}+ \bigg(-\frac{317038110775093}{31294771200}
+\frac{64081361\gamma_E}{11760}
-\frac{406889101\log (2)}{35280}
+\frac{119451753\log (3)}{7840}\bigg)e^{2}
+ \bigg(-\frac{348610408725721199}{3254656204800}
+\frac{10614822977\gamma_E}{376320}
+\frac{224456603713\log (2)}{376320}
-\frac{38840680413\log (3)}{250880}
-\frac{15869140625\log (5)}{150528}\bigg)e^{4}
+ \bigg(-\frac{27027476102569882391}{117167623372800}
+\frac{74351130037\gamma_E}{2257920}
-\frac{9038099531309\log (2)}{1354752}
-\frac{13841114031\log (3)}{100352}
+\frac{24891232421875\log (5)}{8128512}\bigg)e^{6}
+ \bigg(-\frac{2224416660039631007669}{14997455791718400}
+\frac{401363859989\gamma_E}{48168960}
+\frac{163924038199663\log (2)}{2949120}
+\frac{193784040880587\log (3)}{9175040}
-\frac{32000395138671875\log (5)}{1040449536}
-\frac{44084499735185\log (7)}{7077888}\bigg)e^{8}
+\cdots+\delta_{30}e^{30}+\cdots \bigg]
\end{autobreak},\\

\begin{autobreak}
\MoveEqLeft
\mathcal{J}_{11/2L} =  \frac{\pi}{(1-e^2)^{15/2}}\bigg(\frac{177293}{2352}+\frac{64081361}{23520}e^{2}
+\frac{10614822977}{752640}e^{4}+\frac{74351130037}{4515840}e^{6}
+\frac{401363859989}{96337920}e^{8}
+\frac{415995121057}{3715891200}e^{10}
-\frac{7209622999493}{8323596288000}e^{12}
+\frac{72341367473941}{349591044096000}e^{14}
-\frac{42609390600435763}{1879401453060096000}e^{16}
-\frac{100722749765702921}{101487678465245184000}e^{18}
+\frac{12431692376218945133}{81190142772196147200000}e^{20}
-\frac{87071794412050090903}{6549338183623822540800000}e^{22}
-\frac{24657404147085482364377}{3772418793767321783500800000}e^{24}
+\frac{1024340334562524085072459}{1275077552293354762823270400000}e^{26}
+\frac{6248139326602860647729211263}{5536582897835022280911382118400000}e^{28}
+\frac{10800846845298064086150500419}{18455276326116740936371273728000000}e^{30}+\cdots\bigg)\end{autobreak}.
\end{align}

The 6PN order reveals the first significant increase in complexity, limiting output in the non-log series to $e^{20}$ 
even with our new methods.  Coefficients are given to $e^{12}$ to aid the discussion in Sec.~\ref{sec:LDiscussion}, 
but a few are skipped in the middle of the 6PN non-log series for the sake of brevity.  The 6PN log term, meanwhile, 
was found to $e^{30}$, but is also abbreviated here.  All of the now-known coefficients for both of these series 
are available at \cite{BHPTK18}.  
\begin{align}
\begin{autobreak}
\MoveEqLeft
\mathcal{J}_{6} =  \frac{1}{(1-e^2)^{8}}\bigg[\frac{2067586193789233570693}{602387400044430000}
-\frac{246137536815857\gamma_E}{157329572400}
+\frac{1465472\gamma_E^2}{11025}
+\frac{3803225263\pi^2}{10478160}-\frac{27392\gamma_E\pi^2}{315}
-\frac{256\pi^4}{45}-\frac{271272899815409\log (2)}{157329572400}
+\frac{5861888\gamma_E\log (2)}{11025}-\frac{54784}{315}\pi^2\log (2)
+\frac{5861888\log^2(2)}{11025}
-\frac{437114506833\log (3)}{789268480}
-\frac{37744140625\log (5)}{260941824}-\frac{27392\zeta (3)}{105}
+ \bigg(\frac{333496126867093189441241}{2409549600177720000}
-\frac{7595864167160341\gamma_E}{157329572400}
+\frac{37312291\gamma_E^2}{11025}
+\frac{131859072299\pi^2}{10478160}
-\frac{697426\gamma_E\pi^2}{315}
-\frac{6518\pi^4}{45}
-\frac{753377050209181\log (2)}{6293182896}
+\frac{7304462\gamma_E\log (2)}{2205}
-\frac{68266}{63}\pi^2\log (2)
-\frac{19841117\log^2(2)}{11025}-\frac{1908385569124767\log (3)}{27624396800}
+\frac{25038963\gamma_E\log (3)}{2450}
-\frac{234009}{70}\pi^2\log (3)
+\frac{25038963\log (2)\log (3)}{2450}
+\frac{25038963\log^2(3)}{4900}
+\frac{4771622294921875\log (5)}{115075344384}
-\frac{697426\zeta (3)}{105}\bigg)e^{2}
+ \bigg(\frac{11732182856513046341196869}{12850931200947840000}
-\frac{101582265565497851\gamma_E}{419545526400}
+\frac{48268984\gamma_E^2}{3675}
+\frac{2007227812021\pi^2}{27941760}
-\frac{902224\gamma_E\pi^2}{105}
-\frac{8432\pi^4}{15}-\frac{62869385779677563\log (2)}{419545526400}
+\frac{3516949616\gamma_E\log (2)}{11025}
-\frac{32868688}{315}\pi^2\log (2)
+\frac{926636264\log^2(2)}{1575}
+\frac{288069901518860361\log (3)}{160723763200}
-\frac{1276987113\gamma_E\log (3)}{9800}
+\frac{11934459}{280}\pi^2\log (3)
-\frac{1276987113\log (2)\log (3)}{9800}
-\frac{1276987113\log^2(3)}{19600}
-\frac{9599818019775390625\log (5)}{7364822040576}
-\frac{3669865047185939\log (7)}{16700276736}
-\frac{902224\zeta (3)}{35}\bigg)e^{4}
+ \cdots
+ \bigg(\frac{1102884951466368489130807}{10544353805905920000}
+\frac{342273999229759\gamma_E}{57373747200}
+\frac{1614309\gamma_E^2}{31360}
+\frac{32395236497\pi^2}{16558080}
-\frac{15087\gamma_E\pi^2}{448}-\frac{141\pi^4}{64}
-\frac{3767593808715133006785437\log (2)}{411917425920000}
-\frac{27170243004666463\gamma_E\log (2)}{158760000}
+\frac{253927504716509\pi^2\log (2)}{4536000}
-\frac{94122888541066079\log^2(2)}{317520000}
+\frac{6158909116569246725891349\log (3)}{5657476464640000}
-\frac{84905458917700797\gamma_E\log (3)}{1003520000}
+\frac{793508961847671\pi^2\log (3)}{28672000}
-\frac{176618348868400317\log (2)\log (3)}{1003520000}
-\frac{84905458917700797\log^2(3)}{2007040000}
-\frac{2860433896932411829109375\log (5)}{9898320822534144}
+\frac{11948216650390625\gamma_E\log (5)}{130056192}
-\frac{558327880859375\pi^2\log (5)}{18579456}
+\frac{11948216650390625\log (2)\log (5)}{130056192}
+\frac{11948216650390625\log^2(5)}{260112384}
+\frac{24952742409203475686255233\log (7)}{8657423459942400}
+\frac{54354831727337407\gamma_E\log (7)}{1658880000}
-\frac{3555923570947307\pi^2\log (7)}{331776000}
+\frac{54354831727337407\log (2)\log (7)}{1658880000}
+\frac{54354831727337407\log^2(7)}{3317760000}
-\frac{45261\zeta (3)}{448}\bigg)e^{10}
+ \bigg(\frac{965874068977331961500021}{8392444865925120000}
-\frac{11778703354456943\gamma_E}{852409958400}
+\frac{815219098163\pi^2}{170311680}
+\frac{4370148427543824812401338563\log (2)}{46605514475520000}
+\frac{15544521547964903\gamma_E\log (2)}{12757500}
-\frac{145275902317429\pi^2\log (2)}{364500}
+\frac{420172500588128243\log^2(2)}{178605000}
-\frac{3268367789575842149871375567\log (3)}{90519623434240000}
+\frac{2324949929559348129\gamma_E\log (3)}{4014080000}
-\frac{21728504014573347\pi^2\log (3)}{114688000}
+\frac{939690587607417837\log (2)\log (3)}{802816000}
+\frac{2324949929559348129\log^2(3)}{8028160000}
+
\frac{191286506617120840992788490625\log (5)}{22805731175118667776}
-\frac{1664837216064453125\gamma_E\log (5)}{4682022912}
+\frac{77796131591796875\pi^2\log (5)}{668860416}
-
\frac{1664837216064453125\log (2)\log (5)}{4682022912}
-\frac{1664837216064453125\log^2(5)}{9364045824}
-\frac{459130554615555174602630946209\log (7)}{24933379564634112000}
-\frac{1032741802819410733\gamma_E\log (7)}{2211840000}
+\frac{67562547847998833\pi^2\log (7)}{442368000}-
\frac{1032741802819410733\log (2)\log (7)}{2211840000}
-\frac{1032741802819410733\log^2(7)}{4423680000}-
\frac{5720393206911557758236103\log (11)}{4579133069721600}\bigg)e^{12}
+\cdots+
\epsilon_{20}e^{20}+\cdots \bigg]
\end{autobreak},\\

\begin{autobreak}
\MoveEqLeft
\mathcal{J}_{6L} =  \frac{1}{(1-e^2)^{8}}\bigg[-\frac{246137536815857}{314659144800}
+\frac{1465472\gamma_E}{11025}
-\frac{13696\pi^2}{315}+\frac{2930944\log (2)}{11025}
+ \bigg(-\frac{7595864167160341}{314659144800}
+\frac{37312291\gamma_E}{11025}-\frac{348713\pi^2}{315}
+\frac{3652231\log (2)}{2205}
+\frac{25038963\log (3)}{4900}\bigg)e^{2}
+ \bigg(-\frac{101582265565497851}{839091052800}
+\frac{48268984\gamma_E}{3675}
-\frac{451112\pi^2}{105}+\frac{1758474808\log (2)}{11025}
-\frac{1276987113\log (3)}{19600}\bigg)e^{4}
+ \bigg(-\frac{188707966764313411}{1258636579200}
+\frac{105365147\gamma_E}{8820}
-\frac{984721\pi^2}{252}-\frac{712154281537\log (2)}{396900}+\frac{120161983437\log (3)}{313600}+\frac{559033203125\log (5)}{1016064}\bigg)e^{6}
+ \bigg(-\frac{1036935631457042173}{40276370534400}
+\frac{42578831\gamma_E}{17640}-\frac{397933\pi^2}{504}
+\frac{1596888808397\log (2)}{113400}
+\frac{4849120691469\log (3)}{2508800}
-\frac{59816552734375\log (5)}{8128512}\bigg)e^{8}
+ \bigg(\frac{342273999229759}{114747494400}
+\frac{1614309\gamma_E}{31360}-\frac{15087\pi^2}{896}
-\frac{27170243004666463\log (2)}{317520000}
-\frac{84905458917700797\log (3)}{2007040000}
+\frac{11948216650390625\log (5)}{260112384}
+\frac{54354831727337407\log (7)}{3317760000}\bigg)e^{10}
+ \bigg(-\frac{11778703354456943}{1704819916800}
+\frac{15544521547964903\log (2)}{25515000}
+\frac{2324949929559348129\log (3)}{8028160000}
-\frac{1664837216064453125\log (5)}{9364045824}
-\frac{1032741802819410733\log (7)}{4423680000}\bigg)e^{12}
+\cdots+ \zeta_{30}e^{30}+\cdots
\biggr]
\end{autobreak},
\end{align}

The 6PN $\log^2$ flux yielded another closed form expression (for its origin see \cite{MunnEvan19a})
\begin{align}
\mathcal{J}_{6L^2} =  \frac{1}{(1-e^2)^{8}}\biggl(
\frac{366368}{11025}+\frac{37312291 e^2}{44100}
+\frac{12067246 e^4}{3675}+\frac{105365147 e^6}{35280}
+\frac{42578831 e^8}{70560}+\frac{1614309 e^{10}}{125440}
\biggr).
\end{align}

Mirroring 4.5PN and 5.5PN in appearance, analysis of the 6.5PN non-log flux yielded analytic coefficients to 
$e^{28}$ (abbreviated here with the rest found at \cite{BHPTK18}).  The 6.5PN log series was extracted to $e^{30}$.
\begin{align}
\begin{autobreak}
\MoveEqLeft
\mathcal{J}_{13/2} =  \frac{1}{(1-e^2)^{17/2}}
\bigg[-\frac{81605095538444363}{20138185267200}
+\frac{300277177\gamma_E}{436590}
-\frac{42817273\log (2)}{71442}
+\frac{142155\log (3)}{98}
+ \bigg(-\frac{234251633966628833}{1917922406400}
+\frac{3018730571\gamma_E}{249480}
+\frac{2945961630581\log (2)}{15717240}
-\frac{200620071\log (3)}{4312}
-\frac{383544921875\log (5)}{12573792}\bigg)e^{2}
+ \bigg(-\frac{774717636162954505499}{1288843857100800}
-\frac{1025646313\gamma_E}{3104640}
-\frac{1054969026721721\log (2)}{251475840}
-\frac{125742420117\log (3)}{689920}
+\frac{199777099609375\log (5)}{100590336}\bigg)e^{4}
+ \bigg(-\frac{3686697161827337895751}{29827529264332800}
-\frac{547729534754263\gamma_E}{2586608640}
+\frac{118655627259703271\log (2)}{2011806720}
+\frac{69674180291811\log (3)}{2759680}
-\frac{61926330507109375\log (5)}{1810626048}
-\frac{1065488447445779\log (7)}{147806208}\bigg)e^{6}
+\cdots+ \eta_{28}e^{28} +\cdots \bigg]
\end{autobreak},\\

\begin{autobreak}
\MoveEqLeft
\mathcal{J}_{13/2L} =  \frac{\pi}{(1-e^2)^{17/2}}\bigg(\frac{300277177}{873180}
+\frac{3018730571}{498960}e^{2}
-\frac{1025646313}{6209280}e^{4}-\frac{547729534754263}{5173217280}e^{6}
-\frac{1847568691294327}{13168189440}e^{8}
-\frac{150836673548029393}{4213820620800}e^{10}-\frac{341056100428493993}{151697542348800}e^{12}-\frac{900692084654430303761}{1308239605216051200}e^{14} 
-\frac{183817754019571452481}{478441912764727296}e^{16}
-\frac{1369950195790953661052617}{5651595094533341184000}e^{18}
-\frac{60937088128013175521329153}{369083761275646771200000}e^{20}
-\frac{2076214844718290480358259887751}{17506380964826477651558400000}e^{22}
-\frac{67058838175907561436011530774049}{756275657680503834547322880000}e^{24}
-\frac{399586784196479229606707351966520311}{5842769652480235338902745907200000}e^{26}
-\frac{19706198618409033215085911333142483997}{364376361963767403862480335667200000}e^{28}
-\frac{1333103852525373905617630701770154129101}{30538209383630030037998351941632000000}e^{30} 
+\cdots \bigg)
\end{autobreak}.
\end{align}

The 7PN non-log series, of similar complexity to its 6PN counterpart, was extracted to $e^{12}$.  Only the first 
3 coefficients are listed here.  The 7PN log term was obtained to $e^{26}$, but its presentation here is 
truncated at $e^{14}$.  See \cite{BHPTK18} for complete expressions. 
\begin{align}
\begin{autobreak}
\MoveEqLeft
\mathcal{J}_{7} =  \frac{1}{(1-e^2)^{9}}\bigg[\frac{58327313257446476199371189}{8332222517414555760000}
+\frac{9640384387033067\gamma_E}{17896238860500}
-\frac{52525903\gamma_E^2}{154350}
+\frac{2621359845833\pi^2}{2383781400}
+\frac{531077\gamma_E\pi^2}{6615}
-\frac{9523\pi^4}{945}
+\frac{19402232550751339\log (2)}{17896238860500}
-\frac{471188717\gamma_E\log (2)}{231525}
+\frac{128223}{245}\pi^2\log (2)
-\frac{5811697\log^2(2)}{2450}
-\frac{6136997968378863\log (3)}{1256910054400}
+\frac{1848015\gamma_E\log (3)}{2744}
-\frac{142155}{392}\pi^2\log (3)
+\frac{1848015\log (2)\log (3)}{2744}
+\frac{1848015\log^2(3)}{5488}
+\frac{9926708984375\log (5)}{5088365568}
+\frac{531077\zeta (3)}{2205}
+ \bigg(-\frac{217658436746027815895102341}{1666444503482911152000}
+\frac{24231077015148314777\gamma_E}{143169910884000}
-\frac{2214256717\gamma_E^2}{92610}
-\frac{1329680222711\pi^2}{866829600}
+\frac{14572037\gamma_E\pi^2}{1323}
+\frac{210017\pi^4}{945}
-\frac{11466432970124391527\log (2)}{143169910884000}
+\frac{8341589759\gamma_E\log (2)}{231525}
-\frac{69704909\pi^2\log (2)}{2205}
+\frac{63780712447\log^2(2)}{463050}
+\frac{2351107897519100859\log (3)}{3591171584000}
-\frac{939847941\gamma_E\log (3)}{9800}
+\frac{9620613}{280}\pi^2\log (3)
-\frac{939847941\log (2)\log (3)}{9800}
-\frac{939847941\log^2(3)}{19600}
-\frac{1612267989365234375\log (5)}{6981237559296}
-\frac{611078988636949\log (7)}{21201523200}
+\frac{14572037\zeta (3)}{441}\bigg)e^{2}
+ \bigg(-\frac{130192781785212682155024739573}{33328890069658223040000}
+\frac{591633756214144946231\gamma_E}{286339821768000}
-\frac{763796786629\gamma_E^2}{3704400}
-\frac{86202827232472\pi^2}{297972675}
+\frac{5789871557\gamma_E\pi^2}{52920}
+\frac{6673217\pi^4}{1512}
+\frac{4393870365431735509303\log (2)}{286339821768000}
-\frac{117429592367\gamma_E\log (2)}{24696}
+\frac{19166514113\pi^2\log (2)}{10584}
-\frac{10950400503703\log^2(2)}{1234800}
-\frac{42089890286054028933\log (3)}{3093932441600}
+\frac{4964131679733\gamma_E\log (3)}{4390400}
-\frac{49938060789\pi^2\log (3)}{125440}
+\frac{4964131679733\log (2)\log (3)}{4390400}
+\frac{4964131679733\log^2(3)}{8780800}
-\frac{50857646229982421875\log (5)}{1172847909961728}
+\frac{1031494140625\gamma_E\log (5)}{1580544}
-\frac{79345703125\pi^2\log (5)}{225792}
+\frac{1031494140625\log (2)\log (5)}{1580544}
+\frac{1031494140625\log^2(5)}{3161088}
+\frac{18836032355690667229\log (7)}{4070692454400}
+\frac{5789871557\zeta (3)}{17640}\bigg)e^{4}
 +\cdots+\theta_{12}e^{12}+\cdots \bigg]
\end{autobreak},\\

\begin{autobreak}
\MoveEqLeft
\mathcal{J}_{7L} =  \frac{1}{(1-e^2)^{9}}\bigg[\frac{9640384387033067}{35792477721000}
-\frac{52525903\gamma_E}{154350}+\frac{531077\pi^2}{13230}-\frac{471188717\log (2)}{463050}
+\frac{1848015\log (3)}{5488}
+ \bigg(\frac{24231077015148314777}{286339821768000}
-\frac{2214256717\gamma_E}{92610}
+\frac{14572037\pi^2}{2646}+\frac{8341589759\log (2)}{463050}
-\frac{939847941\log (3)}{19600}\bigg)e^{2}
+ \bigg(\frac{591633756214144946231}{572679643536000}
-\frac{763796786629\gamma_E}{3704400}
+\frac{5789871557\pi^2}{105840}
-\frac{117429592367\log (2)}{49392}
+\frac{4964131679733\log (3)}{8780800}
+\frac{1031494140625\log (5)}{3161088}\bigg)e^{4}
+ \bigg(\frac{7068212226017284341287}{2290718574144000}
-\frac{824747159647\gamma_E}{1852200}
+\frac{6687480911\pi^2}{52920}
+\frac{482678846963689\log (2)}{16669800}
+\frac{109318137597\log (3)}{627200}
-\frac{580527388671875\log (5)}{42674688}\bigg)e^{6}
+ \bigg(\frac{1118293348517845456727}{366514971863040}
-\frac{3296625810013\gamma_E}{11854080}
+\frac{27848763749\pi^2}{338688}
-\frac{150747095221023577\log (2)}{533433600}
-\frac{242530132831033581\log (3)}{2247884800}
+\frac{3529852121076171875\log (5)}{21849440256}
+\frac{573098496557405\log (7)}{21233664}\bigg)e^{8}
+ \bigg(\frac{7417800287709479830427}{7330299437260800}
-\frac{515546659387\gamma_E}{11854080}
+\frac{4476597491\pi^2}{338688}
+\frac{34281927681591020089\log (2)}{13335840000}
+\frac{11043010597723065117\log (3)}{8028160000}
-\frac{2515592654376953125\log (5)}{2427715584}
-\frac{33276227221291377337\log (7)}{39813120000}\bigg)e^{10}
+ \bigg(\frac{782647393237829745047}{2443433145753600}
-\frac{8161170019\gamma_E}{10536960}
+\frac{72242227\pi^2}{301056}
-\frac{24632208831281134768831\log (2)}{960180480000}
-
\frac{13240749939950955181941\log (3)}{1798307840000}
+\frac{13741105065468587890625\log (5)}{3146319396864}
+\frac{222041600754634101785281\log (7)}{22932357120000}\bigg)e^{12}
+ \bigg(\frac{1679007616758134752753}{6515821722009600}
+\frac{79294797830738769387229\log (2)}{367569090000}
-\frac{202498419750717826393947\log (3)}{22029271040000}
-\frac{176421957727012978515625\log (5)}{19271206305792}
-\frac{122463544084536896035417\log (7)}{1911029760000}\bigg)e^{14}
+\cdots + \kappa_{26}e^{26}+\cdots\bigg]
\end{autobreak}.
\end{align}

Like its 6PN counterpart, the $\log^2$ piece was also found to have a closed-form expression (see \cite{MunnEvan19b})
\begin{align}
\begin{autobreak}
\MoveEqLeft
\mathcal{J}_{7L^2} =  \frac{1}{(1-e^2)^{9}}\bigg(\frac{52525903}{617400}+\frac{2214256717}{370440}e^{2}+\frac{763796786629}{14817600}e^{4}+\frac{824747159647}{7408800}e^{6}
+\frac{3296625810013}{47416320}e^{8}
+\frac{515546659387}{47416320}e^{10}+\frac{8161170019}{42147840}e^{12}\bigg)
\end{autobreak}.
\end{align}

The 7.5PN half-integer series presented difficulties, allowing us to find analytic coefficients only through 
$e^{12}$.  The corresponding $\log$ and $\log^2$ terms were found to $e^{26}$ and $e^{28}$, respectively.
\begin{align}
\begin{autobreak}
\MoveEqLeft
\mathcal{J}_{15/2} =  \frac{\pi}{(1-e^2)^{19/2}}\bigg[\frac{51603801120086143145449}{8567287467298560000}
-\frac{3025414963439009\gamma_E}{559394035200}
+\frac{5861888\gamma_E^2}{11025}-\frac{1465472\pi^2}{11025}
-\frac{1999998476702377\log (2)}{5034546316800}
+\frac{23447552\gamma_E\log (2)}{11025}
+\frac{23447552\log^2(2)}{11025}-\frac{1311343520499\log (3)}{394634240}
-\frac{188720703125\log (5)}{130470912}-\frac{109568\zeta (3)}{105}
+ \bigg(\frac{2662696956467596386499309}{3426914986919424000}
-\frac{349953536858546087\gamma_E}{1118788070400}
+\frac{50616029\gamma_E^2}{2205}-\frac{50616029\pi^2}{8820}
-\frac{11140310485001952479\log (2)}{10069092633600}
+\frac{336165538\gamma_E\log (2)}{11025}
-\frac{1911983\log^2(2)}{11025}-\frac{12592554481394127\log (3)}{27624396800}
+\frac{75116889\gamma_E\log (3)}{1225}
+\frac{75116889\log (2)\log (3)}{1225}
+\frac{75116889\log^2(3)}{2450}
+\frac{7153146044921875\log (5)}{16439334912}
-\frac{946094\zeta (3)}{21}\bigg)e^{2}
+ \bigg(\frac{1215046582634472109846437257}{126532245670871040000}
-\frac{11416899892367577827\gamma_E}{4130909798400}
+\frac{7015042729\gamma_E^2}{44100}
-\frac{7015042729\pi^2}{176400}
+\frac{974649950407627431787\log (2)}{161105482137600}
+\frac{57310041769\gamma_E\log (2)}{22050}
+\frac{5883446467\log^2(2)}{1260}
+\frac{15692774935495349169\log (3)}{883980697600}
-\frac{4281662673\gamma_E\log (3)}{4900}
-\frac{4281662673\log (2)\log (3)}{4900}
-\frac{4281662673\log^2(3)}{9800}
-\frac{56175438510986328125\log (5)}{3682411020288}
-\frac{25689055330301573\log (7)}{8350138368}
-\frac{65561147\zeta (3)}{210}\bigg)e^{4}
+ \bigg(\frac{565749089288149621030397977099}{17765127292190294016000}
-\frac{3720660313410463236349\gamma_E}{579979735695360}
+\frac{372612960091\gamma_E^2}{1270080}
-\frac{372612960091\pi^2}{5080320}
-\frac{1787089659108846763543\log (2)}{5114459750400}
-\frac{2035608947081\gamma_E\log (2)}{64800}
-\frac{80645525885093\log^2(2)}{1270080}
-\frac{642174379247781899307\log (3)}{1767961395200}
+\frac{454231827783\gamma_E\log (3)}{78400}
+\frac{454231827783\log (2)\log (3)}{78400}
+\frac{454231827783\log^2(3)}{156800}
+\frac{60188792123471152140625\log (5)}{463983788556288}
+\frac{2795166015625\gamma_E\log (5)}{254016}
+\frac{2795166015625\log (2)\log (5)}{254016}
+\frac{2795166015625\log^2(5)}{508032}
+\frac{1421510488851228906067\log (7)}{6763612078080}
-\frac{3482364113\zeta (3)}{6048}\bigg)e^{6}
 + \cdots+\lambda_{12}e^{12} \bigg]
\end{autobreak},\\

\begin{autobreak}
\MoveEqLeft
\mathcal{J}_{15/2L} =  \frac{\pi}{(1-e^2)^{19/2}}\bigg[-\frac{3025414963439009}{1118788070400}
+\frac{5861888\gamma_E}{11025}
+\frac{11723776\log (2)}{11025}
+ \bigg(-\frac{349953536858546087}{2237576140800}
+\frac{50616029\gamma_E}{2205}+\frac{168082769\log (2)}{11025}
+\frac{75116889\log (3)}{2450}\bigg)e^{2}
+ \bigg(-\frac{11416899892367577827}{8261819596800}
+\frac{7015042729\gamma_E}{44100}
+\frac{57310041769\log (2)}{44100}
-\frac{4281662673\log (3)}{9800}\bigg)e^{4}
+ \bigg(-\frac{3720660313410463236349}{1159959471390720}
+\frac{372612960091\gamma_E}{1270080}
-\frac{2035608947081\log (2)}{129600}
+\frac{454231827783\log (3)}{156800}
+\frac{2795166015625\log (5)}{508032}\bigg)e^{6}
+ \bigg(-\frac{855337222347819322823827}{494916041126707200}
+\frac{8066792325467\gamma_E}{50803200}
+\frac{7496582286172507\log (2)}{50803200}
+\frac{6992405846343\log (3)}{250880}
-\frac{47517822265625\log (5)}{580608}\bigg)e^{8}
+\cdots+ \xi_{26}e^{26} +\cdots \bigg]
\end{autobreak},\\

\begin{autobreak}
\MoveEqLeft
\mathcal{J}_{15/2L^2} =  \frac{\pi}{(1-e^2)^{19/2}}\bigg(\frac{1465472}{11025}+\frac{50616029}{8820}e^{2}+\frac{7015042729}{176400}e^{4}+\frac{372612960091}{5080320}e^{6}
+\frac{8066792325467}{203212800}e^{8}
+\frac{438339815188777}{81285120000}e^{10}
+\frac{27165778367659}{325140480000}e^{12}+\frac{139559840953}{169940090880000}e^{14}-\frac{138994608139273}{73414119260160000}e^{16}
-\frac{510325343998375097}{190289397122334720000}e^{18}
+\frac{29294767139126946059}{19028939712233472000000}e^{20}
-\frac{160261051102927034773}{1473601091315360071680000}e^{22}
-\frac{144909050598554594415739}{2652481964367648129024000000}e^{24}
-\frac{3780135848816146128183067}{1434462246330024108176179200000}e^{26}
+\frac{347794649410341565383017707}{281154600280684725202531123200000}e^{28}+\cdots \bigg)
\end{autobreak}.
\end{align}

The 8PN non-log flux was the least successful term to analyze and allowed only confirmation of the (known) circular
orbit limit \cite{Fuji12a}.  Numeric coefficients were obtained to $e^{10}$, which we present here.  The 8PN log 
function yielded coefficients to $e^{18}$.  We list these here to $e^6$ with the remainder given at \cite{BHPTK18}.
\begin{align}
\begin{autobreak}
\MoveEqLeft
\mathcal{J}_{8} \approx  \frac{1}{(1-e^2)^{10}}\bigg(-
\frac{2206020140875740874945597498877}{63104087235639138048360000}
+\frac{17328950668070007334987\gamma_E}{1084297320079974000}
-\frac{3428849385499\gamma_E^2}{2723011830}
-\frac{18584197930153871\pi^2}{4247898454800}
+\frac{1397063663\gamma_E\pi^2}{1178793}
+\frac{2192471\pi^4}{25515}
-\frac{4773986555637567504053\log (2)}{1084297320079974000}
+\frac{15332591650681\gamma_E\log (2)}{6807529575}
-\frac{11366135381}{5893965}\pi^2\log (2)
+\frac{106165554403193\log^2(2)}{13615059150}
+\frac{8479423463263174971\log (3)}{213674709248000}
-\frac{1848015\gamma_E\log (3)}{343}
+\frac{142155}{49}\pi^2\log (3)
-\frac{1848015\log (2)\log (3)}{343}
-\frac{1848015\log^2(3)}{686}
-\frac{83415474560546875\log (5)}{8477217036288}
-\frac{2025852318599963\log (7)}{2948939136000}
+\frac{1397063663\zeta (3)}{392931} - 
 1954977.501298132062640986690 e^2 - 
 26349959.944946765641790484057 e^4 - 
 87126977.786788602494976694986 e^6 - 
 83445624.027185442658036338727 e^8 - 
 3243044.237540247549987011144 e^{10} +\cdots \bigg)
\end{autobreak},\\

\begin{autobreak}
\MoveEqLeft
\mathcal{J}_{8L} =  \frac{1}{(1-e^2)^{10}}\bigg[\frac{17254929304352547776587}{2168594640159948000}-\frac{3428849385499 \gamma_E }{2723011830}
+\frac{1397063663 \pi ^2}{2357586}+\frac{15332591650681 \log (2)}{13615059150}
-\frac{1848015 \log (3)}{686}
+ \bigg(\frac{6902005678706412730657}{26772773335308000}
-\frac{3056423284787 \gamma_E }{605113740}
+\frac{8584514299 \pi ^2}{523908}
-\frac{4058693142384893 \log (2)}{9076706100}
+\frac{2504263499487 \log (3)}{16601200}
+\frac{602549072265625 \log (5)}{8713637856}\bigg)e^{2}
+ \bigg(\frac{20464697512527803859881}{35697031113744000}
+\frac{5591213449564669 \gamma_E }{12102274800}
+\frac{222333848491 \pi ^2}{10478160}
+\frac{60797852720869631 \log (2)}{4034091600}
-\frac{9604994143653 \log (3)}{531238400}
-\frac{120096731201171875 \log (5)}{19916886528}\bigg)e^{4}
+ \bigg(-\frac{111072047800937176690152571}{11565838080853056000}
+\frac{236506814452713983 \gamma_E }{72613648800}
-\frac{32262959758303 \pi ^2}{62868960}
-\frac{16818201812995485953 \log (2)}{72613648800}
-\frac{1428503830753023 \log (3)}{16601200}
+\frac{6972609105435390625 \log (5)}{51214851072}
+\frac{1673882350937318809 \log (7)}{73164072960}\bigg)e^{6}
+\cdots + \sigma_{18}e^{18}+\cdots \bigg]
\end{autobreak}.
\end{align}
 
The 8PN $\log^2$ term, meanwhile, was found to have an exact closed-form expression.  This form is similar to that 
of the $\mathcal{J}_2$ and the $\mathcal{J}_{5L}$ fluxes
\begin{align}
\begin{autobreak}
\MoveEqLeft
\mathcal{J}_{8L^2} =  \frac{1}{(1-e^2)^{10}}\bigg(-\frac{3581369037215}{2178409464}
-\frac{80146723840979}{2420454960}e^{2}+\frac{1666319275502269}{48409099200}e^{4}
+\frac{255322553526353183}{290454595200}e^{6}
+\frac{49182918759586933}{30981823488}e^{8}
+\frac{49243901204481373}{61963646976}e^{10}+\frac{634196505488069863}{6196364697600}e^{12}
+\frac{47480389267723}{30599331840}e^{14}\bigg)
+\frac{40}{(1-e^{2})^{17/2}}\bigg(\frac{366368}{11025}+\frac{37312291}{44100}e^{2}+\frac{12067246}{3675}e^{4}+\frac{105365147}{35280}e^{6}
+\frac{42578831}{70560}e^{8}+\frac{1614309}{125440}e^{10}\bigg)
\end{autobreak}.
\end{align}

The 8.5PN eccentricity functions were similarly troublesome, yielding coefficients to $e^2,$ $e^{16},$ and $e^{20}$, 
respectively.  We truncated the presentation of the 8.5PN log series at $e^6$ and leave the remainder of it to 
\cite{BHPTK18}.
\begin{align}
\begin{autobreak}
\MoveEqLeft
\mathcal{J}_{17/2} =  \frac{\pi}{(1-e^2)^{21/2}}\bigg[\frac{60050471374198816098730954501}{1083453442264445091840000}
-\frac{16654515688953719 \gamma_E }{2020034016000}
-\frac{91049249 \gamma_E ^2}{132300}+\frac{91049249 \pi ^2}{529200}
-\frac{11256322928659829467 \log (2)}{381786429024000}
-\frac{116527141 \gamma_E  \log (2)}{17150}
-\frac{1632801787 \log ^2(2)}{185220}
-\frac{19606939404628941 \log (3)}{628455027200}
+\frac{5544045 \gamma_E  \log (3)}{1372}
+\frac{5544045 \log (2) \log (3)}{1372}
+\frac{5544045 \log ^2(3)}{2744}+\frac{49633544921875 \log (5)}{2544182784}
-\frac{84807 \zeta (3)}{70}
+ \bigg(-\frac{225941369691757950007727558579}{3500388044238976450560000}
+\frac{106048473884633692003 \gamma_E }{117472747392000}
-\frac{276926667149 \gamma_E ^2}{1852200}
+\frac{276926667149 \pi ^2}{7408800}
-\frac{596136690001068277 \log (2)}{13052527488000}
+\frac{307019986547 \gamma_E  \log (2)}{926100}
+\frac{2006541325939 \log ^2(2)}{1852200}
+\frac{67576336886522478429 \log (3)}{12569100544000}
-\frac{44095962681 \gamma_E  \log (3)}{68600}
-\frac{44095962681 \log (2) \log (3)}{68600}
-\frac{44095962681 \log ^2(3)}{137200}
-\frac{2935275808466796875 \log (5)}{1163539593216}
-\frac{4277552920458643 \log (7)}{10600761600}
+\frac{495943519 \zeta (3)}{2940}\bigg)e^{2}+ \cdots \bigg]
\end{autobreak},\\

\begin{autobreak}
\MoveEqLeft
\mathcal{J}_{17/2L} =  \frac{\pi}{(1-e^2)^{21/2}}\bigg[\bigg(-\frac{16654515688953719}{4040068032000}
-\frac{91049249 \gamma_E }{132300}
-\frac{116527141 \log (2)}{34300}
+\frac{5544045 \log (3)}{2744}
+ \bigg(\frac{106048473884633692003}{234945494784000}
-\frac{276926667149 \gamma_E }{1852200}
+\frac{307019986547 \log (2)}{1852200}
-\frac{44095962681 \log (3)}{137200}\bigg)e^{2}
+ \bigg(\frac{769384178879100689605873}{73302994372608000}
-\frac{132263217227483 \gamma_E }{59270400}
-\frac{426945632248393 \log (2)}{19756800}
+\frac{2672662827537 \log (3)}{627200}
+\frac{5157470703125 \log (5)}{1580544}\bigg)e^{4}
+ \bigg(\frac{138182998655659305884296627}{2638907797413888000}
-\frac{8915432624231827 \gamma_E }{1066867200}
+\frac{309724239257528941 \log (2)}{1066867200}
+\frac{4882416222447 \log (3)}{351232}
-\frac{12873561103515625 \log (5)}{85349376}\bigg)e^{6}
+\cdots + \rho_{16}e^{16} + \cdots \bigg]
\end{autobreak},\\

\begin{autobreak}
\MoveEqLeft
\mathcal{J}_{17/2L^2} =  \frac{1}{(1-e^2)^{21/2}}\bigg(-\frac{91049249}{529200}-\frac{276926667149}{7408800}e^{2}-\frac{132263217227483}{237081600}e^{4}-\frac{8915432624231827}{4267468800}e^{6}
-\frac{138995825711906599}{54623600640}e^{8}-\frac{4664651987016654047}{4551966720000}e^{10}
-\frac{294210339077129459393}{2621932830720000}e^{12}
-\frac{377078959571059380349}{256949417410560000}e^{14}
-\frac{145031943920041772287}{28191021795901440000}e^{16}+\frac{45645187675975688916133}{31968618716552232960000}e^{18}
-\frac{5672267216086086404759471}{25574894973241786368000000}e^{20}+\cdots\bigg)
\end{autobreak}.
\end{align}

Finally, we find that the 9PN $\log^3$ term can be expressed in a nice closed form.  For further understanding of 
its origin, see \cite{MunnEvan19a}.
\begin{align}
\mathcal{J}_{9L^3} &=  \frac{1}{(1-e^2)^{11}}\bigg(\frac{313611008}{3472875}
+\frac{44220377171}{6945750}e^{2}+\frac{112166162123}{1543500}e^{4}
+\frac{623241851293}{2646000}e^{6}+\frac{1354930634161}{5292000}e^{8} \qquad \notag \\
&+\frac{52244408821}{564480}e^{10}+\frac{17088124807}{1881600}e^{12}
+\frac{96778397}{903168}e^{14}\bigg) .
\end{align}
\end{widetext}

\subsection{Discussion}
\label{sec:LDiscussion}

A careful review of the above results reveals several patterns at the
various PN orders, some expected, some rather surprising.  Starting at the top,
we immediately notice that the 2PN function $\mathcal{J}_2$ has a curious form.
In Sec. \ref{sec:LcurrPN}, we presented a result for this term using the time eccentricity.  That 
function ($\mathcal{N}_2$) was similar in structure, also being a sum of two finite series
with different eccentricity singular factors.  However, in the time eccentricity expression, the factors
were only separated by a half-power of $(1-e_t^2)$.  In the conversion from time eccentricity
to Darwin eccentricity, the subdominant series becomes $(10-5/4e^2-35/4e^4)$, from 
which another factor of $(1-e^2)$ can be removed, leaving a multiple of the Peters term
$(1+7/8e^2)$.  

The more fundamental reason for this behavior remains unknown.  
However, we have since found that the 2PN energy flux term $\mathcal{I}_2$ is characterized by a similar
simplification. Paper I represented $\mathcal{I}_2$ with two finite series prefaced by the 
singular factors $1/(1-e^2)^{11/2}$ and $1/(1-e^2)^5$.
As we will see in the next section, the subdominant series can be reduced via 
\begin{align}
\begin{autobreak}
\MoveEqLeft
\frac{1}{(1-e^2)^{5}}\bigg(\frac{35}{2} 
+ \frac{1715}{48} e^2 
  - \frac{2975}{64} e^4 - \frac{1295}{192} e^6\bigg)
   = \frac{35}{2(1-e^2)^{4}} \bigg(1 
+ \frac{73}{24} e^2 
  + \frac{37}{96} e^4 \bigg)
\end{autobreak}
\end{align}
to obtain a comparable result.  

Furthermore, and perhaps most remarkably, analogous forms occur in all found enhancement functions
with this dominant-subdominant structure.  In both the angular momentum and energy
regimes, we note that the $5L$ and $8L^2$ terms contain series from the $3L$ and $6L^2$ terms, respectively.

In total this improved fitting method yielded five new closed-form expressions---$\mathcal{J}_{5L}, 
\mathcal{J}_{6L2}, \mathcal{J}_{7L2}, \mathcal{J}_{8L2},$ and $\mathcal{J}_{9L3}$.  On first glance, it is interesting 
that such closed representations all involve the first few appearances of a new power of logarithm in the expansion.  
However, given recent work on logarithmic series \cite{JohnMcDa14,JohnMcDaShahWhit15,GoldRoss10}, this is 
not surprising.   It was this empirical observation that led us to study the origins of
these logarithms in the PN expansion more closely, culminating in the characterization of several infinite sets of
logarithms in the PN expansion.  For instance, the first appearance of each new power of logarithm is part of a 
set termed the \emph{leading logarithms}.  We have since shown that all leading logarithms can be described by
simple Fourier mass quadrupole summations, just like those given in Sec.~\ref{sec:heredExps}.  We refer the reader
to \cite{MunnEvan19a} for more details.

The 4PN enhancement function is also a case of interest.  As with the full flux at
3PN, we see that the transcendentals $\gamma_E$ and $\pi^2$ vanish identically after
a certain order in $e$ (here $e^8$).  The specific polynomial prefacing $\gamma_E$ is
proportional to the 4PN log term.  Using this fact, one might think it possible to fit our series at 4PN to the form of 
the full 3PN flux, giving most of the exact series.  
All that would remain is the 4PN equivalent of $\tilde{\chi}(e)$, which would likely result from the 1PN correction
to the tail-of-tails and (tail$)^2$ terms that generate $\tilde{\chi}$ at 3PN.  This suspicion turns out to be
correct, and the relevant details, along with a compact form for $\mathcal{J}_4$, will be presented in an 
upcoming paper \cite{MunnEvan19b}.  A similar effect occurs in the 6PN Log term, and its compact form was 
found in \cite{MunnEvan19a}.  Note that $\mathcal{J}_{7L}$ also shows finite series in $\gamma_E$ and 
$\pi^2$.  The 5PN function does not, resulting from the fact that $\mathcal{J}_{5L}$ has the aforementioned 
dominant-subdominant singular factor structure.  An exact form for that function will be saved for future 
investigations. 

Moving one step further, we can see a similar simplification in the 6PN integer (non-log)
term.  The 6PN Log enhancement factor has $\gamma_E$ and $\pi^2$ series that 
terminate at $e^{10}$.  If we compare the 6PN integer series coefficient of $e^{10}$ to that
of $e^{12}$, we spot a difference: The latter does not contain $\gamma_E^2$, $\pi^4$, or $\gamma_E*\pi^2$. 
Indeed, these expressions quadratic in the relevant transcendentals vanish, just as their linear counterparts are 
eliminated in $\mathcal{J}_{6L}$.  The $\zeta(3)$ piece also vanishes at that order, for reasons  
related to higher order tail integrals \cite{MunnEvan19a}.  Unfortunately, the dataset was not accurate enough
to extract the 7PN integer series beyond $e^{12}$; however, we can infer that this series
will likely follow a similar pattern, losing all $\gamma_E^2$, $\pi^4$, $\gamma_E*\pi^2$, and $\zeta(3)$
dependence at $e^{14}$ and beyond.

\begin{widetext}
\section{Update: energy flux radiated to infinity}
\label{sec:Eresults}

We now briefly review past work on the energy flux before presenting new results obtained
through the procedures developed in this paper.  Arun, Blanchet, Iyer, and Qusailah derived 3PN 
relative-order expansions for the energy flux to infinity for eccentric orbits on Schwarzschild backgrounds
in \cite{ArunETC08a} and \cite{ArunETC08b}.  Then, in Paper I, we used flux comparisons 
to find new analytic and numeric $e$ coefficients from $3.5$ to $7$PN order.  
Now we work to 8.5PN order (along with the 9PN $\log^3$ term).

Through 3PN we can preserve the split between instantaneous and hereditary terms:
\begin{align}
\label{eqn:energyfluxInf}
\left\langle \frac{dE}{dt} \right\rangle_\infty =&  
\left\langle \frac{dE}{dt} \right\rangle_\text{N}^\infty 
\biggl[\mathcal{I}_0 + y\mathcal{I}_1
+y^{3/2}\mathcal{K}_{3/2}+y^2\mathcal{I}_2+y^{5/2}\mathcal{K}_{5/2}
+ y^3\left(\mathcal{I}_3 + \mathcal{K}_3\right)
+\mathcal{L}_{7/2}y^{7/2}
+y^4\Bigl(\mathcal{L}_4+\log(y)\mathcal{L}_{4L}\Bigr)
\notag\\&
+y^{9/2}\Bigl(\mathcal{L}_{9/2}+\log(y)\mathcal{L}_{9/2L}\Bigr)
+y^5\Bigl(\mathcal{L}_5
+\log(y)\mathcal{L}_{5L}\Bigr)
+y^{11/2}\Bigl(\mathcal{L}_{11/2}+\log(y)\mathcal{L}_{11/2L}\Bigr)
\notag\\& 
+ y^6\Bigl(\mathcal{L}_6 + \log(y)\mathcal{L}_{6L}
+ \log^2(y)\mathcal{L}_{6L^2} \Bigr)
+y^{13/2}\Bigl(\mathcal{L}_{13/2}+\log(y)\mathcal{L}_{13/2L}\Bigr)
\notag\\& 
+ y^7\Bigl(\mathcal{L}_7 + \log(y)\mathcal{L}_{7L}
+ \log^2(y)\mathcal{L}_{7L^2} \Bigr)
+y^{15/2}\biggl(\mathcal{L}_{15/2}+\log(y)\mathcal{L}_{15/2L}+\log^2(y)\mathcal{L}_{15/2L^2}\biggr)
 \notag\\&+ y^8\biggl(\mathcal{L}_8
+\log(y)\mathcal{L}_{8L}+\log^2(y)\mathcal{L}_{8L^2}\biggr)+y^{17/2}\biggl(\mathcal{L}_{17/2}+\log(y)\mathcal{L}_{17/2L}+\log^2(y)\mathcal{L}_{17/2L^2}\biggr)+\cdots
\biggr],
\end{align}
where

\begin{equation}
\label{eqn:newtFluxInf}
\left\langle\frac{dE}{dt}\right\rangle_\text{N}^\infty
= \frac{32}{5} \left(\frac{\mu}{M}\right)^2 y^5,
\end{equation}
and

\begin{align}
\label{eqn:infOldEnh}
\mathcal{I}_0 &= \frac{1}{(1-e^2)^{7/2}}
{\left(1+\frac{73}{24}~e^2 + \frac{37}{96}~e^4\right)},\\
\mathcal{I}_1 &= \frac{1}{(1-e^2)^{9/2}}
\left(-\frac{1247}{336}-\frac{15901}{672} e^2-\frac{9253}{384} e^4
-\frac{4037}{1792} e^6
\right),\\
\begin{split}
\mathcal{I}_2 &= 
\frac{1}{(1-e^2)^{11/2}}
{\left(-\frac{203471}{9072}-\frac{1430873 }{18144}e^2+\frac{2161337}{24192} e^4
+\frac{231899}{2304} e^6+\frac{499451}{64512} e^8\right)}
\\
&\hspace{65ex} + \frac{35}{2(1-e^2)^{4}} \left(1 
+ \frac{73}{24} e^2 
  + \frac{37}{96} e^4 \right) ,
\end{split}\\

\begin{autobreak}
\MoveEqLeft
\mkern15mu \mathcal{I}_3 = \frac{1}{(1-e^2)^{13/2}}
\bigg[
\frac{2193295679}{9979200}+\frac{55022404229 }{19958400}e^2
+\frac{68454474929 }{13305600}e^4 
+\frac{40029894853}{26611200} e^6
-\frac{32487334699 }{141926400}e^8
-\frac{233745653 }{11354112}e^{10}
+ \sqrt{1-e^2} \bigg(
-\frac{14047483}{151200}-\frac{75546769}{100800}e^2-\frac{210234049}{403200}e^4
+\frac{1128608203}{2419200} e^6
+\frac{617515}{10752} e^8\bigg) \bigg]   
\mkern555mu
+\frac{1712}{105} \log\bigg[\frac{y}{y_0} \frac{1+\sqrt{1-e^2}}{2(1-e^2)}\bigg]F(e),
\end{autobreak}\\

\mathcal{K}_{3/2} &= \frac{4\pi}{(1-e^2)^5}
\biggl(1+\frac{1375}{192} e^2+\frac{3935}{768} e^4
+\frac{10007}{36864} e^6
+\frac{2321 }{884736}e^8
-\frac{237857 }{353894400}e^{10}
\notag\\&\hspace{32ex}
+\frac{182863 }{4246732800}e^{12}+
\frac{4987211 }{6658877030400}e^{14}
-\frac{47839147 }{35514010828800}e^{16}
+\cdots
\biggr),\\
\mathcal{K}_{5/2}&=\frac{-\pi}{(1-e^2)^6}
\biggl(
\frac{8191}{672}+\frac{62003}{336} e^2
+\frac{20327389}{43008} e^4
+\frac{87458089}{387072} e^6
+\frac{67638841}{7077888} e^8
+\frac{332887 }{25804800}e^{10}
\notag\\&\hspace{25ex}
-\frac{482542621}{475634073600} e^{12}
+\frac{43302428147}{69918208819200} e^{14}
-\frac{2970543742759 }{35798122915430400}e^{16}+\cdots
\biggr),\\
\mathcal{K}_3 &= -\frac{1712}{105}\,\chi(e) +
\left[
-\frac{116761}{3675} + \frac{16}{3} \,\pi^2 -\frac{1712}{105}\,\gamma_\text{E} -
  \frac{1712}{105}\log\left(\frac{4y^{3/2}}{y_0}\right)\right]\, F(e),
\end{align}
are functions of the Darwin eccentricity $e$ which were known at
the time of our previous work (the original functions were given in terms of the time
eccentricity, but we quote the converted forms given in Paper I).
 The $\mathcal{I}$'s encode instantaneous 
corrections to the radiated energy, whereas the $\mathcal{K}$'s are
hereditary terms which include effects from the entire orbital history
of the particle.  Note that like $\tilde{\psi}$, $\mathcal{K}_{5/2}$ has now been extracted to $e^{120}$.

$F(e)$ and $\chi(e)$ are given by

\be
F(e) = \frac{1}{(1-e^2)^{13/2}} 
\bigg( 1+ \frac{85}{6} e^2 + \frac{5171}{192} e^4 + 
\frac{1751}{192} e^6 + \frac{297}{1024} e^8\bigg),
\label{eqn:capFe}
\ee
and
\begin{align}
\label{eqn:chiExp}
\chi (e) = - \frac{3}{2} F(e) \log(1 - e^2) &+
\frac{1}{(1-e^2)^{13/2}} \, 
\Bigg\{ 
\left[-\frac{3}{2} - \frac{77}{3} \log(2) + \frac{6561}{256} \log(3) \right] 
e^2 \\ \notag
&+
\left[-22 + \frac{34855}{64} \log(2) - \frac{295245}{1024} \log(3) \right] 
e^4 
\\ \notag
&+
\left[-\frac{6595}{128} - \frac{1167467}{192} \log(2) + 
\frac{24247269}{16384} \log(3) + \frac{244140625}{147456} \log(5) \right] 
e^6 
\\ \notag
&+
\left[-\frac{31747}{768} + \frac{122348557}{3072} \log(2) + 
\frac{486841509}{131072} \log(3) - \frac{23193359375}{1179648} \log(5) \right] 
e^8 
+ \cdots 
\Bigg\} . 
\end{align}

Now we move beyond 3PN order. Like the $\mathcal{J}_i$ of the previous section, 
these $\mathcal{L}_i(e)$ functions are
calculated using $lmn$ fitting.  Coefficients are given to $8.5$PN order in $y$ 
and varying orders in eccentricity as needed (full results will be posted to the BHP Toolkit).  
Here, the use of a Roman letter (e.g. $b_{30}$)
denotes the highest power for which we retrieved an analytic form.  
As with the angular momentum, we are not able
to distinguish between instantaneous and hereditary terms at this level, so
the $\mathcal{L}$'s generically include both contributions.  A subset of these coefficients
were first produced in Paper I.

\begin{align}
\begin{autobreak}
\MoveEqLeft
\mathcal{L}_{7/2}= \frac{\pi}{(1-e^2)^7} \bigg(-\frac{16285}{504}
-\frac{21500207}{48384}e^{2}-\frac{3345329}{48384}e^{4}
+\frac{111594754909}{41803776}e^{6}+\frac{82936785623}{55738368}e^{8}
+\frac{11764982139179}{107017666560}e^{10}
+\frac{216868426237103}{9631589990400}e^{12}
+\frac{30182578123501193}{2517055517491200}e^{14}
+\frac{351410391437739607}{48327465935831040}e^{16}
+\frac{1006563319333377521717}{208774652842790092800}e^{18}
+\frac{138433556497603036591}{40776299383357440000}e^{20}
+\frac{16836217054749609972406421}{6736462131727360327680000}e^{22}
+\frac{2077866815397007172515220959}{1091306865339832373084160000}e^{24}
+\frac{6702459208696786891810972264771}{4496600021148810265629818880000}e^{26}
+\frac{1003693903183075635039911792668567}{841272985774931956969651568640000}e^{28}
+\frac{2160389373606905789762084388554056897}{2220960682445820366399880141209600000}e^{30}
+\cdots \biggr)
\end{autobreak},\label{eqn:Ienh72}\\

\begin{autobreak}
\MoveEqLeft
\mathcal{L}_4 =\frac{1}{(1-e^2)^{15/2}}\bigg[\frac{323105549467}{3178375200}
+\frac{232597 \gamma_E}{4410}-\frac{1369 \pi ^2}{126}+\frac{39931 \log (2)}{294}-\frac{47385 \log (3)}{1568}
+ \bigg(-\frac{128412398137}{23543520} 
+\frac{4923511\gamma_E }{2940}-\frac{104549 \pi ^2}{252} -\frac{343177 \log (2)}{252}+\frac{55105839 \log (3)}{15680}\bigg)e^2
 +\bigg(-\frac{981480754818517}{25427001600} +\frac{142278179 \gamma_E }{17640}
   -\frac{1113487 \pi ^2}{504}+\frac{762077713 \log (2)}{5880}
   -\frac{2595297591 \log (3)}{71680}
   -\frac{15869140625 \log (5)}{903168}\bigg)e^4 
   +\bigg(-\frac{874590390287699}{12713500800}
   +\frac{318425291 \gamma_E}{35280}-\frac{881501 \pi ^2}{336}
   -\frac{90762985321 \log (2)}{63504}+\frac{31649037093 \log
   (3)}{1003520}+\frac{10089048828125 \log (5)}{16257024}\bigg)e^6  
   + \bigg(-\frac{588262620227803}{15647385600}+\frac{1256401651 \gamma_E}{564480}
   -\frac{3609941 \pi   ^2}{5376}+\frac{60196618062379 \log (2)}{5080320}
   +\frac{103481536492359 \log (3)}{25690112} 
   -\frac{13689354185546875 \log (5)}{2080899072}
   -\frac{44084499735185 \log
   (7)}{42467328}\bigg)e^8 
   + \bigg(-\frac{2506068425640457}{271221350400}
   +\frac{7220691 \gamma_E }{125440}
   -\frac{63771 \pi ^2}{3584}
   -\frac{7802806392729223 \log (2)}{84672000} 
   -\frac{304325022941627589 \log (3)}{6422528000}
   +\frac{52671902802734375 \log (5)}{1387266048}
   +\frac{99735805288105217 \log (7)}{3538944000}\bigg)e^{10} 
   +\bigg(-\frac{205790655085493}{33210777600}
   +\frac{9036104785041317 \log(2)}{11430720}
   +\frac{4775869078725402009 \log (3)}{20552089600}
   -\frac{4795866464849609375 \log (5)}{33294385152} 
   -\frac{179568613346696178017 \log (7)}{611529523200}\bigg)e^{12}
  +\cdots
   + a_{30}e^{30} +\cdots\biggr]
 \end{autobreak},\label{eqn:Ienh4} \\

 \mathcal{L}_{4L} &= \frac{1}{(1-e^2)^{15/2}}
\biggl(
\frac{232597}{8820} + \frac{4923511}{5880}e^2
+ \frac{142278179}{35280}e^4+
\frac{318425291}{70560}e^6
+ \frac{1256401651}{1128960}e^8
+\frac{7220691}{250880}e^{10}
\biggr),\label{eqn:Ienh4L}\\

\begin{autobreak}
\MoveEqLeft
\mathcal{L}_{9/2} = \frac{\pi}{(1-e^2)^{8}}\bigg(\frac{265978667519}{745113600}
-\frac{6848\gamma_E}{105}-\frac{13696\log (2)}{105}
+\bigg(\frac{5031659060513}{447068160}-\frac{418477\gamma_E}{252}
-\frac{1024097\log (2)}{1260}
-\frac{702027\log (3)}{280}\bigg)e^2
+\bigg(\frac{4137488075571679}{71530905600}-\frac{32490229\gamma_E}{5040}
-\frac{56349731\log (2)}{720}+\frac{35803377\log (3)}{1120}\bigg)e^4
+\bigg(\frac{119161057323769}{1609445376}-\frac{283848209\gamma_E}{48384}
+\frac{212985174443\log (2)}{241920}-\frac{481356513\log (3)}{2560}
-\frac{26123046875\log (5)}{96768}\bigg)e^6
+\bigg(\frac{916628147773341301}{65922882600960}
-\frac{1378010735\gamma_E}{1161216}
-\frac{8023715124847\log (2)}{1161216}
-\frac{135922489587\log (3)}{143360}
+\frac{2795166015625\log (5)}{774144}\bigg)e^8
+\cdots+b_{30}b^{30}+\cdots\bigg]
\end{autobreak},\label{eqn:Ienh92}\\

\begin{autobreak}
\MoveEqLeft
\mathcal{L}_{9/2L} =\frac{\pi}{(1-e^2)^8}\bigg(-\frac{3424}{105}-\frac{418477}{504}e^{2}
-\frac{32490229}{10080}e^{4}-\frac{283848209}{96768}e^{6}
-\frac{1378010735}{2322432}e^{8}-\frac{59600244089}{4644864000}e^{10}
+\frac{482765917}{7962624000}e^{12}-\frac{532101153539}{29132587008000}e^{14}
+\frac{576726373021}{199766310912000}e^{16}-\frac{98932878601597}{3624559945187328000}e^{18}
-\frac{56946683948951263}{1087367983556198400000}e^{20}
-\frac{90233805781037113}{60146983318994288640000}e^{22}
+\frac{73049155670984045033}{50523465987955202457600000}e^{24}
+\frac{30834120217438664094539}{81969271218858520467210240000}e^{26}
+\frac{4892777190662608136893709}{80329885794481350057866035200000}e^{28}
+\frac{625894086470885360433206659}{77116690362702096055551393792000000}e^{30}+\cdots\bigg)
\end{autobreak},\label{eqn:Ienh92L}\\

\begin{autobreak}
\MoveEqLeft
\mathcal{L}_5 =\frac{1}{(1-e^2)^{17/2}}\bigg[-\frac{2500861660823683}{2831932303200}+\frac{916628467 \gamma_E }{7858620}-\frac{424223 \pi ^2}{6804}-\frac{83217611 \log (2)}{1122660}+\frac{47385 \log (3)}{196}
+ \bigg(-\frac{121566202635820681}{5663864606400}+\frac{11627266729 \gamma_E }{15717240}
-\frac{16845407 \pi ^2}{13608}+\frac{41528347547 \log (2)}{1428840}
-\frac{1380946887 \log (3)}{137984}-\frac{383544921875 \log (5)}{100590336}\bigg)e^{2}
+ \bigg(-\frac{886493383307889029}{15103638950400}-\frac{84010607399 \gamma_E }{5239080}
-\frac{14848651 \pi ^2}{9072}-\frac{3992455076567 \log (2)}{5239080}
+\frac{61777429029 \log (3)}{2759680}+\frac{120783447265625 \log (5)}{402361344}\bigg)e^{4}
+ \bigg(\frac{11463059954793067}{53495769600}
-\frac{67781855563 \gamma_E }{816480}
+\frac{111910879 \pi ^2}{7776}
+\frac{1925006801181043 \log (2)}{188606880}
+\frac{153356656665033 \log (3)}{44154880}
-\frac{30664709673671875 \log (5)}{5267275776}
-\frac{1065488447445779 \log (7)}{1182449664}\bigg)e^{6}
+\cdots+c_{30}e^{30}+\cdots\bigg]
\end{autobreak},\label{eqn:Ienh5}\\

\mathcal{L}_{5L}&=\frac{1}{\left(1-e^2\right)^{17/2}}\Big(\frac{5080948627}{15717240}
+\frac{117123377449 }{31434480}e^2-\frac{4199642054 }{654885}e^4
-\frac{78989239933 }{1632960}e^6-\frac{88593702010771 }{2011806720}e^8\notag \\ 
&-\frac{261925436695 }{29804544}e^{10} -\frac{245975507 }{1290240}e^{12}\Big) -\frac{65}{2 \left(1-e^2\right)^7} \Big(\frac{856}{105}+\frac{7276 }{63}e^2+\frac{553297 }{2520}e^4+\frac{187357 }{2520}e^6+\frac{10593 }{4480}e^8\Big),\label{eqn:Ienh5L}\\

\begin{autobreak}
\MoveEqLeft
\mathcal{L}_{11/2} = \frac{\pi}{(1-e^2)^9}\bigg[\frac{8399309750401}{101708006400}
+\frac{177293\gamma_E}{1176}+\frac{8521283\log (2)}{17640}
-\frac{142155\log (3)}{784}+ \bigg(-\frac{6454125584294467}{203416012800}
+\frac{197515529\gamma_E}{17640}-\frac{195924727\log (2)}{17640}
+\frac{1909251\log (3)}{80}\bigg)e^{2}+ \bigg(-\frac{354252739653461867}{813664051200}
+\frac{22177125281\gamma_E}{225792}
+\frac{1349842104869\log (2)}{1128960}
-\frac{14094701055\log (3)}{50176}
-\frac{79345703125\log (5)}{451584}\bigg)e^{4}
+ \bigg(-\frac{3220604701659665695}{2343352467456}
+\frac{362637121649\gamma_E}{1693440}
-\frac{551674667051\log (2)}{37632}
-\frac{28712823381\log (3)}{125440}+\frac{14134345703125\log (5)}{2032128}\bigg)e^{6}
+\cdots+d_{30}e^{30}+\cdots\bigg]
\end{autobreak},\label{eqn:Ienh112}\\

\begin{autobreak}
\MoveEqLeft
\mathcal{L}_{11/2L} = \frac{\pi}{(1-e^2)^9}
\bigg(\frac{177293}{2352}+\frac{197515529}{35280}e^{2}
+\frac{22177125281}{451584}e^{4}
+\frac{362637121649}{3386880}e^{6}
+\frac{175129893794507}{2601123840}e^{8}
+\frac{137611940506079}{13005619200}e^{10}
+\frac{75058973874797}{396361728000}e^{12}
+\frac{1045783525483}{131096641536000}e^{14}
+\frac{44925442631482501}{1879401453060096000}e^{16}
-\frac{801339891963050743}{50743839232622592000}e^{18}
+\frac{719061383331468255529}{243570428316588441600000}e^{20}+
\frac{14491034377225531751}{421028883232960020480000}e^{22}
-\frac{48380310946786430680357}{1028841489209269577318400000}e^{24}
+\frac{328896042939144986202607}{117098958884083600667443200000}e^{26}
+\frac{12426147326832099974747530661}{5536582897835022280911382118400000}e^{28}
+\frac{1156253390804057519850290651}{5623092005613694504050622464000000}e^{30}+\cdots\bigg)
\end{autobreak},\label{eqn:Ienh112L}\\

\begin{autobreak}
\MoveEqLeft
\mathcal{L}_6 = \frac{1}{(1-e^2)^{19/2}}\bigg[\frac{2067586193789233570693}{602387400044430000}
-\frac{246137536815857\gamma_E}{157329572400}
+\frac{1465472\gamma_E^2}{11025}
+\frac{3803225263\pi^2}{10478160}
-\frac{27392\gamma_E\pi^2}{315}
-\frac{256\pi^4}{45}-\frac{271272899815409\log (2)}{157329572400}
+\frac{5861888\gamma_E\log (2)}{11025}
-\frac{54784}{315}\pi^2\log (2)
+\frac{5861888\log^2(2)}{11025}
-\frac{437114506833\log (3)}{789268480}
-\frac{37744140625\log (5)}{260941824}-\frac{27392\zeta (3)}{105}
+ \bigg(\frac{620642724143587842589757}{2409549600177720000}
-\frac{25915820507512391\gamma_E}{314659144800}
+\frac{189812971\gamma_E^2}{33075}
+\frac{8630456095\pi^2}{381024}
-\frac{3547906\gamma_E\pi^2}{945}
-\frac{33158\pi^4}{135}
-\frac{1204827593616887\log (2)}{6421615200}
+\frac{36018554\gamma_E\log (2)}{4725}
-\frac{336622}{135}\pi^2\log (2)
-\frac{57245\log^2(2)}{1323}
-\frac{425707669538577\log (3)}{4249907200}
+\frac{75116889\gamma_E\log (3)}{4900}
-\frac{702027}{140}\pi^2\log (3)
+\frac{75116889\log (2)\log (3)}{4900}
+\frac{75116889\log^2(3)}{9800}
+\frac{1735378662109375\log (5)}{32878669824}
-\frac{3547906\zeta (3)}{315}\bigg)e^{2}
+ \bigg(\frac{866764151375288467902617}{321273280023696000}
-\frac{56861331626354501\gamma_E}{83909105280}
+\frac{1052380631\gamma_E^2}{26460}
+\frac{106659145841\pi^2}{508032}
-\frac{9835333\gamma_E\pi^2}{378}
-\frac{91919\pi^4}{54}
-\frac{469561807262423641\log (2)}{419545526400}
+\frac{42983885171\gamma_E\log (2)}{66150}
-\frac{401718553\pi^2\log (2)}{1890}
+\frac{6177731563\log^2(2)}{5292}
+\frac{4967869967044739217\log (3)}{1767961395200}
-\frac{4281662673\gamma_E\log (3)}{19600}
+\frac{40015539}{560}\pi^2\log (3)
-\frac{4281662673\log (2)\log (3)}{19600}
-\frac{4281662673\log^2(3)}{39200}
-\frac{1749708882763671875\log (5)}{818313560064}
-\frac{25689055330301573\log (7)}{83501383680}
-\frac{9835333\zeta (3)}{126}\bigg)e^{4}
+\cdots
+ \bigg(\frac{24612086555137636537042301}{131593535497705881600}-\frac{477961162088755717\gamma_E}{7160243650560}
+\frac{32323997924497\pi^2}{1430618112}
-\frac{61278163606788680414049737704313\log (2)}{31971382930206720000}
-\frac{339392544622900323521\gamma_E\log (2)}{8751645000}
+\frac{3171892940400937603\pi^2\log (2)}{250047000}
-\frac{4532425889525064665801\log^2(2)}{52509870000}
-\frac{15568492847979888930357\gamma_E\log (3)}{3147038720000}
+\frac{9519685411620604508156942363799\log (3)}{8870923096555520000}
-\frac{46776237102385425837621\log (2)\log (3)}{3147038720000}
+\frac{145499933158690550751\pi^2\log (3)}{89915392000}
-\frac{25647883085450625849\log^2(3)}{1573519360000}
+\frac{20971917520162841796875\gamma_E\log (5)}{5506058944512}
-\frac{2425918968367925016852218629596875\log (5)}{6704884965484888326144}
+\frac{20971917520162841796875\log (2)\log (5)}{5506058944512}
-\frac{979996145802001953125\pi^2\log (5)}{786579849216}
+\frac{20971917520162841796875\log^2(5)}{11012117889024}
+\frac{77148041218710802588787\gamma_E\log (7)}{5733089280000}
+\frac{11539161795601951836966750333833\log (7)}{49866759129268224000}
-\frac{5047068117111921664687\pi^2\log (7)}{1146617856000}
+\frac{77148041218710802588787\log (2)\log (7)}{5733089280000}
+\frac{77148041218710802588787\log^2(7)}{11466178560000}
+\frac{8938746466657465062086011285151\log (11)}{76115758848933888000}\bigg)e^{14}
+\cdots+f_{20}e^{20}+\cdots\bigg]
\end{autobreak},\label{eqn:Ienh6}\\

\begin{autobreak}
\MoveEqLeft
\mathcal{L}_{6L} = \frac{1}{(1-e^2)^{19/2}}\bigg[-\frac{246137536815857}{314659144800}
+\frac{1465472\gamma_E}{11025}-\frac{13696\pi^2}{315}
+\frac{2930944\log (2)}{11025}
+ \bigg(-\frac{25915820507512391}{629318289600}
+\frac{189812971\gamma_E}{33075}
-\frac{1773953\pi^2}{945}
+\frac{18009277\log (2)}{4725}
+\frac{75116889\log (3)}{9800}\bigg)e^{2}
+ \bigg(-\frac{56861331626354501}{167818210560}
+\frac{1052380631\gamma_E}{26460}-\frac{9835333\pi^2}{756}
+\frac{42983885171\log (2)}{132300}-\frac{4281662673\log (3)}{39200}\bigg)e^{4}
+ \bigg(-\frac{710806279550045831}{1006909263360}
+\frac{9707068997\gamma_E}{132300}
-\frac{90720271\pi^2}{3780}-\frac{519508209691\log (2)}{132300}
+\frac{454281905709\log (3)}{627200}+\frac{2795166015625\log (5)}{2032128}\bigg)e^{6}
+ \bigg(-\frac{10213351238593603069}{40276370534400}
+\frac{8409851501\gamma_E}{211680}-\frac{78596743\pi^2}{6048}
+\frac{117139032193219\log (2)}{3175200}
+\frac{6991554521601\log (3)}{1003520}-\frac{47517822265625\log (5)}{2322432}\bigg)e^{8}
+ \bigg(\frac{3985515397336843519}{26850913689600}
+\frac{4574665481\gamma_E}{846720}
-\frac{42753883\pi^2}{24192}-\frac{252510878807655859\log (2)}{952560000}
-\frac{576360297584196039\log (3)}{4014080000}
+\frac{223101765869140625\log (5)}{1560674304}
+\frac{380483822091361849\log (7)}{6635520000}\bigg)e^{10}
+ \bigg(\frac{50719954422267749}{3254656204800}
+\frac{6308399\gamma_E}{75264}-\frac{294785\pi^2}{10752}
+\frac{2887481794238961637\log (2)}{1270080000}
+\frac{17322463230547056201\log (3)}{16056320000}
-\frac{1297619485595703125\log (5)}{2080899072}
-\frac{2663386754639532943\log (7)}{2949120000}\bigg)e^{12}
+ \bigg(-\frac{477961162088755717}{14320487301120}
-\frac{339392544622900323521\log (2)}{17503290000}
-\frac{15568492847979888930357\log (3)}{6294077440000}
+\frac{20971917520162841796875\log (5)}{11012117889024}
+\frac{77148041218710802588787\log (7)}{11466178560000}\bigg)e^{14}
+\cdots+g_{30}e^{30}+\cdots\bigg]
\end{autobreak},\label{eqn:Ienh6L}\\

\mathcal{L}_{6L^2} &= \frac{1}{(1-e^2)^{19/2}}\biggl(
\frac{366368}{11025} + \frac{189812971}{132300}e^2
+\frac{1052380631}{105840}e^4
+\frac{9707068997}{529200}e^6
\notag\\&\
\qquad \qquad \qquad \qquad \qquad \qquad \qquad \qquad \qquad \qquad \qquad
+\frac{8409851501}{846720}e^8
+\frac{4574665481}{3386880}e^{10}
 +\frac{6308399}{301056}e^{12}
\biggr),\label{eqn:Ienh6L2}\\

\begin{autobreak}
\MoveEqLeft
\mathcal{L}_{13/2}=\frac{\pi}{(1-e^2)^{10}}\bigg[-\frac{81605095538444363}{20138185267200}
+\frac{300277177\gamma_E}{436590}-\frac{42817273\log (2)}{71442}
+\frac{142155\log (3)}{98}
+ \bigg(-\frac{486006274042153993}{3098182348800}
+\frac{99375022631\gamma_E}{13970880}
+\frac{30885453339487\log (2)}{125737920}-\frac{26221716657\log (3)}{344960}
-\frac{1917724609375\log (5)}{50295168}\bigg)e^{2}
+ \bigg(-\frac{978074410273210177483}{1288843857100800}
-\frac{206420323339\gamma_E}{1164240}
-\frac{35044764797711\log (2)}{4490640}
-\frac{896501601\log (3)}{12320}
+\frac{82230224609375\log (5)}{25147584}\bigg)e^{4}
+ \bigg(\frac{1759614571265146017649}{652477202657280}
-\frac{52528099035138203\gamma_E}{36212520960}
+\frac{490814480869706621\log (2)}{4023613440}
+\frac{1040915745740691\log (3)}{22077440}
-\frac{2078689555036328125\log (5)}{28970016768}
-\frac{7458419132120453\log (7)}{591224832}\bigg)e^{6}
+\cdots+h_{28}e^{28}+\cdots\bigg]
\end{autobreak},\label{eqn:Ienh132}\\

\begin{autobreak}
\MoveEqLeft
\mathcal{L}_{13/2L} = \frac{\pi}{(1-e^2)^{10}}\bigg(\frac{300277177}{873180}+\frac{99375022631}{27941760}e^{2}-\frac{206420323339}{2328480}e^{4}-\frac{52528099035138203}{72425041920}e^{6}
-\frac{133623698374169077}{96566722560}e^{8}-\frac{13064004066588147059}{16855282483200}e^{10}
-\frac{1963639930072973146717}{16686729658368000}e^{12}
-\frac{33400949279751680423063}{4360798684053504000}e^{14}
-\frac{179371445578657546009993}{59805239095590912000}e^{16}
-\frac{637047737965052868548277511}{361702086050133835776000}e^{18}
-\frac{3460275187517318400615660587}{3014184050417781964800000}e^{20}
-\frac{27996584084597317460228073711577}{35012761929652955303116800000}e^{22}
-\frac{19726180767340366267420639753777}{33762306146451064042291200000}e^{24}
-\frac{24140999291524879880417880052762466303}{54532516756482196496425628467200000}e^{26}-
\frac{5533246861404900450857176595606015862331}{16032559926405765769949134769356800000}e^{28}
-\frac{39240045588213441329120436124666666397117}{142511643790273473510658975727616000000}e^{30}+\cdots\bigg)
\end{autobreak},\label{eqn:Ienh132L}\\

\begin{autobreak}
\MoveEqLeft
\mathcal{L}_7 = \frac{1}{(1-e^2)^{21/2}}\bigg[\frac{58327313257446476199371189}{8332222517414555760000}
+\frac{9640384387033067 \gamma_E }{17896238860500}
-\frac{52525903 \gamma_E ^2}{154350}
+\frac{2621359845833 \pi ^2}{2383781400}
+\frac{531077 \gamma_E  \pi ^2}{6615}
-\frac{9523 \pi ^4}{945}
+\frac{19402232550751339 \log (2)}{17896238860500}
-\frac{471188717 \gamma_E  \log (2)}{231525}
+\frac{128223}{245} \pi ^2 \log (2)
-\frac{5811697 \log ^2(2)}{2450}
-\frac{6136997968378863 \log (3)}{1256910054400}
+\frac{1848015 \gamma_E  \log (3)}{2744}
-\frac{142155}{392} \pi ^2 \log (3)
+\frac{1848015 \log (2) \log (3)}{2744}
+\frac{1848015 \log ^2(3)}{5488}
+\frac{9926708984375 \log (5)}{5088365568}
+\frac{531077 \zeta (3)}{2205}
+ \bigg(-\frac{1833694744307038499536301503}{3332889006965822304000}
+\frac{5361621824744487121 \gamma_E }{14316991088400}
-\frac{8436767071 \gamma_E ^2}{185220}
-\frac{131503074649 \pi ^2}{3531528}
+\frac{20170061 \gamma_E  \pi ^2}{882}+\frac{283391 \pi ^4}{378}
+\frac{2977365445451226901 \log (2)}{71584955442000}
+\frac{8661528101 \gamma_E  \log (2)}{463050}
-\frac{440469373 \pi ^2 \log (2)}{13230}
+\frac{154654591013 \log ^2(2)}{926100}
+\frac{12486523458893227371 \log (3)}{12569100544000}
-\frac{21008472903 \gamma_E  \log (3)}{137200}
+\frac{208895679 \pi ^2 \log (3)}{3920}
-\frac{21008472903 \log (2) \log (3)}{137200}
-\frac{21008472903 \log ^2(3)}{274400}
-\frac{80233643837890625 \log (5)}{268509136896}
-\frac{4277552920458643 \log (7)}{127209139200}
+\frac{20170061 \zeta (3)}{294}\bigg)e^{2}
+ \bigg(-\frac{425327739088776761686492357}{27207257199720998400}
+\frac{69100209694441952051 \gamma_E }{10226422206000}
-\frac{66537493061 \gamma_E ^2}{105840}
-\frac{3452732996641507 \pi ^2}{2724321600}
+\frac{530883301 \gamma_E  \pi ^2}{1512}
+\frac{17810521 \pi ^4}{1080}
+\frac{565134631654755855073 \log (2)}{14316991088400}
-\frac{18504183154799 \gamma_E  \log (2)}{1852200}
+\frac{63724032709 \pi ^2 \log (2)}{17640}
-\frac{13381694922467 \log ^2(2)}{740880}
-\frac{9013628727200023913673 \log (3)}{402211217408000}
+\frac{1804462952967 \gamma_E  \log (3)}{878080}
-\frac{17687032263 \pi ^2 \log (3)}{25088}
+\frac{1804462952967 \log (2) \log (3)}{878080}
+\frac{1804462952967 \log ^2(3)}{1756160}
-\frac{722647442175224609375 \log (5)}{2345695819923456}
+\frac{5157470703125 \gamma_E  \log (5)}{4741632}
-\frac{396728515625 \pi ^2 \log (5)}{677376}
+\frac{5157470703125 \log (2) \log (5)}{4741632}
+\frac{5157470703125 \log ^2(5)}{9483264}
+\frac{130895018390638453 \log (7)}{20102184960}
+\frac{530883301 \zeta (3)}{504}\bigg)e^{4}
+\cdots+j_{12}e^{12}+\cdots\bigg]
\end{autobreak},\label{eqn:Ienh7}\\

\begin{autobreak}
\MoveEqLeft
\mathcal{L}_{7L} = 
\frac{1}{(1-e^2)^{21/2}}\bigg[
\frac{9640384387033067}{35792477721000}
-\frac{52525903 \gamma_E }{154350}
+\frac{531077 \pi ^2}{13230}
-\frac{471188717 \log (2)}{463050}
+\frac{1848015 \log (3)}{5488}
+ \bigg(\frac{5361621824744487121}{28633982176800}
-\frac{8436767071 \gamma_E }{185220}
+\frac{20170061 \pi ^2}{1764}
+\frac{8661528101 \log (2)}{926100}
-\frac{21008472903 \log (3)}{274400}\bigg)e^{2}
+ \bigg(\frac{69100209694441952051}{20452844412000}
-\frac{66537493061 \gamma_E }{105840}
+\frac{530883301 \pi ^2}{3024}
-\frac{18504183154799 \log (2)}{3704400}
+\frac{1804462952967 \log (3)}{1756160}
+\frac{5157470703125 \log (5)}{9483264}\bigg)e^{4}
+ \cdots + \bigg(\frac{392956261308991697579}{222130285977600}
-\frac{6441767405 \gamma_E }{4214784}
+\frac{290022625 \pi ^2}{602112}
+\frac{79758263769894173174170363 \log (2)}{94097687040000}
-\frac{4735538949816648321845247 \log (3)}{176234168320000}
-\frac{11353999155433772705078125 \log (5)}{308339300892672}
-\frac{87075197521422359501707 \log (7)}{339738624000}\bigg)e^{14}
+ \bigg(\frac{14816695856807173325147}{9477558868377600}
-\frac{5064686588825332952885407 \log (2)}{840157920000}
+\frac{42721403084890740280304298693 \log (3)}{22557973544960000}
-\frac{205760911201132587653505859375 \log (5)}{631478888228192256}
+\frac{29799003038979039956177798137 \log (7)}{23482733690880000}
+\frac{201417183487589839275762436609 \log (11)}{3157394441140961280}\bigg)e^{16}
+ \cdots+k_{26}e^{26}+\cdots\bigg]
\end{autobreak},\label{eqn:Ienh7L}\\

\begin{autobreak}
\MoveEqLeft
\mathcal{L}_{7L^2} = -\frac{1}{(1-e^2)^{21/2}}
\bigg(\frac{52525903}{617400}+\frac{8436767071}{740880}e^{2}+\frac{66537493061}{423360}e^{4}+\frac{16839575984743}{29635200}e^{6}+\frac{22951910431067}{33868800}e^{8}
+\frac{18225509849041}{67737600}e^{10}+\frac{2633534008997}{90316800}e^{12}+\frac{6441767405}{16859136}e^{14}\bigg)
\end{autobreak},\label{eqn:Ienh7L2}\\

\begin{autobreak}
\MoveEqLeft
\mathcal{L}_{15/2} = \frac{1}{(1-e^2)^{11}}\bigg[\frac{51603801120086143145449}{8567287467298560000}
-\frac{3025414963439009\gamma_E}{559394035200}
+\frac{5861888\gamma_E^2}{11025}-\frac{1465472\pi^2}{11025}
-\frac{1999998476702377\log (2)}{5034546316800}
+\frac{23447552\gamma_E\log (2)}{11025}
+\frac{23447552\log^2(2)}{11025}
-\frac{1311343520499\log (3)}{394634240}
-\frac{188720703125\log (5)}{130470912}
-\frac{109568\zeta (3)}{105}+ \bigg(\frac{617542475472651187592698603}{411229798430330880000}
-\frac{439734196881760549\gamma_E}{839091052800}
+\frac{2479658767\gamma_E^2}{66150}
-\frac{2479658767\pi^2}{264600}-\frac{106026671002494841\log (2)}{64545465600}
+\frac{1916917519\gamma_E\log (2)}{33075}
+\frac{791435023\log^2(2)}{66150}
-\frac{4420920979736127\log (3)}{6906099200}
+\frac{225350667\gamma_E\log (3)}{2450}
+\frac{225350667\log (2)\log (3)}{2450}
+\frac{225350667\log^2(3)}{4900}
+\frac{15905145751953125\log (5)}{28768836096}
-\frac{23174381\zeta (3)}{315}\bigg)e^{2}
+ \bigg(\frac{1484918820873890610249964661}{54830639790710784000}
-\frac{126350957261075251487\gamma_E}{17900609126400}
+\frac{22643958139\gamma_E^2}{52920}
-\frac{22643958139\pi^2}{211680}
+\frac{133444424175863332003\log (2)}{53701827379200}
+\frac{706231828327\gamma_E\log (2)}{132300}
+\frac{2468502941543\log^2(2)}{264600}
+\frac{62442239264166123\log (3)}{2296053760}
-\frac{2028156003\gamma_E\log (3)}{1400}
-\frac{2028156003\log (2)\log (3)}{1400}
-\frac{2028156003\log^2(3)}{2800}
-\frac{2354821775634765625\log (5)}{94420795392}
-\frac{179823387312111011\log (7)}{41750691840}
-\frac{211625777\zeta (3)}{252}\bigg)e^{4}
+\cdots+l_{12}e^{12}+\cdots\bigg]
\end{autobreak},\label{eqn:Ienh152}\\

\begin{autobreak}
\MoveEqLeft
\mathcal{L}_{15/2L} = 
\frac{1}{(1-e^2)^{11}}\bigg[-\frac{3025414963439009}{1118788070400}+\frac{5861888\gamma_E}{11025}
+\frac{11723776\log (2)}{11025}+ \bigg(-\frac{439734196881760549}{1678182105600}
+\frac{2479658767\gamma_E}{66150}
+\frac{1916917519\log (2)}{66150}
+\frac{225350667\log (3)}{4900}\bigg)e^{2}
+ \bigg(-\frac{126350957261075251487}{35801218252800}
+\frac{22643958139\gamma_E}{52920}+\frac{706231828327\log (2)}{264600}
-\frac{2028156003\log (3)}{2800}\bigg)e^{4}
+ \bigg(-\frac{36144975344017995555691}{2899898678476800}
+\frac{17616792537263\gamma_E}{12700800}
-\frac{428644895504209\log (2)}{12700800}
+\frac{1676533845591\log (3)}{313600}
+\frac{13975830078125\log (5)}{1016064}\bigg)e^{6}
+ \bigg(-\frac{1465091734136920643784967}{134977102125465600}
+\frac{459691434479657\gamma_E}{304819200}
+\frac{23409352359075029\log (2)}{60963840}
+\frac{237441706804107\log (3)}{2508800}
-\frac{1830833740234375\log (5)}{8128512}\bigg)e^{8}+\cdots+m_{26}e^{26}+\cdots\bigg]
\end{autobreak},\label{eqn:Ienh152L}\\

\begin{autobreak}
\MoveEqLeft
\mathcal{L}_{15/2L^2} =\frac{1}{(1-e^2)^{11}}\bigg(\frac{1465472}{11025}
+\frac{2479658767}{264600}e^{2}
+\frac{22643958139}{211680}e^{4}
+\frac{17616792537263}{50803200}e^{6}
+\frac{459691434479657}{1219276800}e^{8}
+\frac{66494784224478367}{487710720000}e^{10}
+\frac{78360178393945783}{5852528640000}e^{12}
+\frac{1444655514143830483}{9176764907520000}e^{14}
+\frac{51015640024026887}{146828238520320000}e^{16}
-\frac{43873302622896741181}{380578794244669440000}e^{18}
+\frac{373288343491048076867}{16310519753342976000000}e^{20}
-\frac{286026594234455117352479}{221040163697304010752000000}e^{22}
-\frac{2464347696391370853689}{6062815918554624294912000}e^{24}
+\frac{1224305061272403640352089951}{43033867389900723245285376000000}e^{26}
+\frac{120162136825359369885614962913}{8434638008420541756075933696000000}e^{28}+\cdots\bigg)\end{autobreak},\label{eqn:Ienh152L2}\\

\begin{autobreak}
\MoveEqLeft
\mathcal{L}_{8} \approx \frac{1}{(1-e^2)^{21/2}}
\bigg(-
\frac{2206020140875740874945597498877}{63104087235639138048360000}
+\frac{17328950668070007334987\gamma_E}{1084297320079974000}
-\frac{3428849385499\gamma_E^2}{2723011830}
-\frac{18584197930153871\pi^2}{4247898454800}
+\frac{1397063663\gamma_E\pi^2}{1178793}
+\frac{2192471\pi^4}{25515}
-\frac{4773986555637567504053\log (2)}{1084297320079974000}
+\frac{15332591650681\gamma_E\log (2)}{6807529575}
-\frac{11366135381}{5893965}\pi^2\log (2)
+\frac{106165554403193\log^2(2)}{13615059150}
+\frac{8479423463263174971\log (3)}{213674709248000}
-\frac{1848015\gamma_E\log (3)}{343}
+\frac{142155}{49}\pi^2\log (3)
-\frac{1848015\log (2)\log (3)}{343}
-\frac{1848015\log^2(3)}{686}
-\frac{83415474560546875\log (5)}{8477217036288}
-\frac{2025852318599963\log (7)}{2948939136000}
+\frac{1397063663\zeta (3)}{392931}- 3445110.45223167809957813155 e^2 - 
 63011640.2589502111479578408 e^4 - 
 273933223.6521104390237430479 e^6 - 
 325300545.71499163564006669284 e^8 + 
 77909913.97444552477119497207 e^{10}\bigg)
\end{autobreak},\label{eqn:Ienh8}\\

\begin{autobreak}
\MoveEqLeft
\mathcal{L}_{8L} = \frac{1}{(1-e^{2})^{23/2}}\bigg[\frac{17254929304352547776587}{2168594640159948000}
-\frac{3428849385499 \gamma_E }{2723011830}
+\frac{1397063663 \pi ^2}{2357586}
+\frac{15332591650681 \log (2)}{13615059150}
-\frac{1848015 \log (3)}{686}
+ \bigg(\frac{131085309923714183816419}{619598468617128000}
+\frac{21497081974969 \gamma_E }{555716700}
+\frac{39542529067 \pi ^2}{3367980}
-\frac{15451532104941719 \log (2)}{27230118300}
+\frac{34563425601321 \log (3)}{132809600}
+\frac{3012745361328125 \log (5)}{34854551424}\bigg)e^{2}
+ \bigg(-\frac{671462220891497074433309}{160636640011848000}
+\frac{26436888128127791 \gamma_E }{12102274800}
-\frac{3066363010741 \pi ^2}{10478160}
+\frac{1122122451633337997 \log (2)}{36306824400}
-\frac{67683747020751 \log (3)}{75891200}
-\frac{200964925537109375 \log (5)}{19916886528}\bigg)e^{4}
+ \bigg(-\frac{47316764670092138351403131}{680343416520768000}
+\frac{1290798565697019809 \gamma_E }{72613648800}
-\frac{236118406034659 \pi ^2}{62868960}
-\frac{109330311453653376797 \log (2)}{217840946400}
-\frac{126238870317372129 \log (3)}{772710400}
+\frac{6190036871141024609375 \log (5)}{20076221620224}
+\frac{11717176456561231663 \log (7)}{292656291840}\bigg)e^{6}
+\cdots+n_{18}e^{18}+\cdots\bigg]
\end{autobreak},\label{eqn:Ienh8L}\\

\begin{autobreak}
\MoveEqLeft
\mathcal{L}_{8L^2} = \frac{1}{(1-e^2)^{23/2}}\bigg(-\frac{20621469398683}{10892047320}-\frac{873082546975007}{15560067600}e^{2}+\frac{8493235174147961}{48409099200}e^{4}+\frac{1236049323927605309}{290454595200}e^{6}
+\frac{28536838567917568709}{2323636761600}e^{8}+\frac{2528975648094153077}{221298739200}e^{10}+\frac{3228304767170880073}{885194956800}e^{12}
+\frac{4075229663605721917}{12392729395200}e^{14}+\frac{60344732583283}{16319643648}e^{16}\bigg)+\frac{95}{2(1-e^{2})^{10}} \bigg(\frac{366368}{11025}+\frac{189812971}{132300}e^{2}+\frac{1052380631}{105840}e^{4}
+\frac{9707068997}{529200}e^{6}+\frac{8409851501}{846720}e^{8}+\frac{4574665481}{3386880}e^{10}
+\frac{6308399}{301056}e^{12}\bigg)
\end{autobreak},\label{eqn:Ienh8L2}\\

\begin{autobreak}
\MoveEqLeft
\mathcal{L}_{17/2} = \frac{\pi }{(1-e^2)^{12}}\bigg[\frac{60050471374198816098730954501}{1083453442264445091840000}
-\frac{16654515688953719 \gamma_E }{2020034016000}
-\frac{91049249 \gamma_E ^2}{132300}+\frac{91049249 \pi ^2}{529200}
-\frac{11256322928659829467 \log (2)}{381786429024000}
-\frac{116527141 \gamma_E  \log (2)}{17150}
-\frac{1632801787 \log ^2(2)}{185220}
-\frac{19606939404628941 \log (3)}{628455027200}
+\frac{5544045 \gamma_E  \log (3)}{1372}
+\frac{5544045 \log (2) \log (3)}{1372}
+\frac{5544045 \log ^2(3)}{2744}
+\frac{49633544921875 \log (5)}{2544182784}
-\frac{84807 \zeta (3)}{70}
+
 \bigg(-\frac{15036308338855532675849798486713}{6205233351150912798720000}
+\frac{223778092210802539141 \gamma_E }{104123571552000}
-\frac{65904560053 \gamma_E ^2}{231525}+\frac{65904560053 \pi ^2}{926100}
+
\frac{140869805086990295761 \log (2)}{163622755296000}
+\frac{8623992122 \gamma_E  \log (2)}{33075}
+\frac{310965709387 \log ^2(2)}{231525}
+\frac{12278674760615248437 \log (3)}{1571137568000}
-\frac{1743776019 \gamma_E  \log (3)}{1715}
-\frac{1743776019 \log (2) \log (3)}{1715}
-\frac{1743776019 \log ^2(3)}{3430}
-\frac{13175186181640625 \log (5)}{4040068032}
-\frac{29942870443210501 \log (7)}{63604569600}
+\frac{831792958 \zeta (3)}{2205}\bigg)e^{2}+\cdots\bigg]
\end{autobreak},\label{eqn:Ienh172}\\

\begin{autobreak}
\MoveEqLeft
\mathcal{L}_{17/2L} = 
\frac{\pi}{(1-e^2)^{12}}\bigg[-\frac{16654515688953719}{4040068032000}
-\frac{91049249 \gamma_E }{132300}-\frac{116527141 \log (2)}{34300}
+\frac{5544045 \log (3)}{2744}
+ \bigg(\frac{223778092210802539141}{208247143104000}
-\frac{65904560053 \gamma_E }{231525}+\frac{4311996061 \log (2)}{33075}
-\frac{1743776019 \log (3)}{3430}\bigg)e^{2}
+ \bigg(\frac{1224651117880706056076827}{36651497186304000}
-\frac{25340338934531 \gamma_E }{3951360}
-\frac{2672975873695021 \log (2)}{59270400}
+\frac{33275639432241 \log (3)}{4390400}
+\frac{25787353515625 \log (5)}{4741632}\bigg)e^{4}
+ \bigg(\frac{2447980141428133025519227}{10308233583648000}
-\frac{19825641677397587 \gamma_E }{533433600}
+\frac{335478276938768813 \log (2)}{533433600}
+\frac{124568886652983 \log (3)}{4390400}
-\frac{15233813134765625 \log (5)}{42674688}\bigg)e^{6}
+\cdots+n_{16}e^{16}+\cdots\bigg]
\end{autobreak},\label{eqn:Ienh172L}\\

\begin{autobreak}
\MoveEqLeft
\mathcal{L}_{17/2L^2} =\frac{\pi}{(1-e^2)^{12}}\bigg(-\frac{91049249}{529200}-\frac{65904560053}{926100}e^{2}-\frac{25340338934531}{15805440}e^{4}-\frac{19825641677397587}{2133734400}e^{6}
-\frac{5146045059705234151}{273118003200}e^{8}-\frac{49591538734543178399}{3413975040000}e^{10}-\frac{10634841767381874605891}{2621932830720000}e^{12}
-\frac{126041444979520074540149}{385424126115840000}e^{14}
-\frac{131123991469047848941021}{39467430514262016000}e^{16}
+\frac{44427013734959710303}{285434095683502080000}e^{18}
+\frac{8359183360274467950273439}{25574894973241786368000000}e^{20}+\cdots \bigg)
\end{autobreak},\label{eqn:Ienh172L2}\\

\begin{autobreak}
\MoveEqLeft
\mathcal{L}_{9L^3} = \frac{1}{(1-e^{2})^{25/2}}\bigg(-\frac{313611008}{3472875}-\frac{105607281901}{10418625}e^{2}-\frac{1882969493752}{10418625}e^{4}-\frac{5247811027411}{5556600}e^{6}
-\frac{1237392658483}{694575}e^{8}
-\frac{81460556106397}{63504000}e^{10}
-\frac{10573124400217}{31752000}e^{12}-\frac{599796978359}{24192000}e^{14}-\frac{89428139}{387072}e^{16}\bigg)
\end{autobreak}.\label{eqn:Ienh9L3}
\end{align}

\end{widetext}

\section{Conclusions}

As expected, the results in the energy flux at infinity closely mirror those of the angular momentum, with the 
terms $\mathcal{L}_2, \mathcal{L}_4, \mathcal{L}_{5L}, \mathcal{L}_6, \mathcal{L}_{6L}, 
\mathcal{L}_{7L},$ and $\mathcal{L}_{8L2}$ all repeating the trends noted in their $\mathcal{J}$ counterparts.
Overall, we have found new exact forms for five enhancement functions in each regime -- $5L,$ $6L^2,$ 
$7L^2,$ $8L^2,$ and $9L^3$ -- with numerous more 
coefficients added to those terms with no closed forms.  We immediately conclude that $lmn$ fitting, 
along with the eulerlog simplification, is a viable method of extracting PN coefficients in the fluxes of 
eccentric-orbit EMRIs.  Similar techniques should be possible in the conservative sector and could feasibly
lead to improved expansions for certain quantities like the generalized redshift invariant 
\cite{Detw08, JohnMcDaShahWhit15, HoppKavaOtte16}.

Unfortunately, these methods do appear to reach some limitations around the 8PN integer
series.  After the eulerlog simplifications are performed, the 8PN integer
term still requires that a search vector of length 5 be fit, $\{1,\pi^2,\pi^4,\zeta(3),2 \beta-\log(2)\}$.
We find that such a search vector requires around $140-170+$ decimals
of accuracy to yield a correct result, depending on the fractional complexities involved.  
Maintaining that accuracy to such high order would necessitate flux calculations of $500$ decimals
or more.  Worse, the 9PN non-log series would have a search vector length greater than 6, 
compounding these difficulties by an order of magnitude.  Thus, even if we could increase
the precision and obtain $e$ coefficients in $\mathcal{J}_8$, we see that the non-log
series will become prohibitively expensive at 9PN and beyond.

However, many enhancement functions beyond the scope of this paper are still within reach.  
This possibility stems from the fact that each full PN order has some power or powers of log with a 
short search vector (possibly after simplification).  The most fruitful have the 
vector $\{1\}$, representing a rational series in $e^2$.  A coefficient in such a series only requires about 
$10+f$ decimals of accuracy to extract for fractional complexity $f$.  Length-2 search vectors are a little more
cumbersome, but we find they still offer consistent results with moderate accuracy, say $70$ or so
decimals.  Further still, even high-order terms with search vectors of length 3 will permit some measure of success.
For example, despite almost no yield in $\mathcal{J}_8$, two coefficients
were found in $\mathcal{J}_{17/2}$, with vector $\{1,\pi^2,\zeta(3)\}$. Thus, it is these terms, with search vectors 
under length-4, which could be susceptible to the unmodified methods of this paper, feasibly to 12PN or beyond.  

However, it would be more useful not to repeat these methods, but to improve them.  Indeed, as briefly 
mentioned in Sec.~\ref{sec:lmnfitting} the simplifications we have performed can be extended
by utilizing eccentric-orbit analogs of the tail factorizations employed in \cite{JohnMcDa14}.  
Johnson-McDaniel relayed to us a \textsc{Mathematica} notebook containing examples of an $S_{lmn}$ factorization, 
with which all terms could be reduced to simple rational series until 8PN.  This would likely allow for the computation
of eccentricity coefficients across several more PN orders.  

Furthermore, by also applying purely analytic expansions of the MST solutions, we will greatly enhance our
ability to determine high PN contributions at lower orders in $e^{2}$.  Because the orbit-averaged fluxes
compose the greatest contribution to the gravitational-wave phase \cite{HindFlan08,FlanHind12}, and because 
accurate waveforms are sought in nearly all regions of parameter space (large and small $y$ and $e$), expansions 
are required to very high PN order to simulate the range of possible EMRIs down to merger 
\cite{Fuji12b,JohnMcDaShahWhit15}.  These ideas will be explored fully in future work.

\acknowledgments

We thank Nathan Johnson-McDaniel, Leor Barack, Luc Blanchet for helpful 
discussions.  This work was supported in part by NSF grants PHY-1506182 and 
PHY-1806447.  C.R.E.~acknowledges support from the Bahnson Fund at the 
University of North Carolina-Chapel Hill.  This work was also supported by the
North Carolina Space Grant.

\appendix

\section{Flux expansions using the semi-latus rectum}

Although we have chosen to follow the PN convention of using 
$y=(M \O_{\vp})^{2/3}$ as our expansion parameter, in some ways 
this is not the most natural variable to use. To see this,
consider the flux expression including only the lowest-order 
enhancement factor,
\begin{align}
\left\langle \frac{dE}{dt} \right\rangle_{\rm 0PN} &= 
\frac{32}{5} \left(\frac{\mu}{M}\right)^2 \, \notag \\
& \times \frac{y^5}{(1-e^2)^{7/2}}
{\left(1+\frac{73}{24}~e^2 + \frac{37}{96}~e^4\right)}.
\end{align}
Note that at this order the eccentricities $e$ and $e_t$ are equivalent.
The primary drawback of this expression is the singular factor 
$(1-e^2)^{-7/2}$, which causes the flux to diverge
as $e \to 1$. 
This factor can be traced back to the choice of $x$ as a PN variable.
However, we may choose to write $x$ as an expansion in
$p^{-1}$, which to leading order goes as
\be
\label{eqn:xOfp}
y = (1-e^2)/p + \mathcal{O}(p^{-2}).
\ee
Re-expressed as an expansion in $1/p$, the 0PN flux is then
\begin{align}
\label{eqn:flux0PNp}
\left\langle \frac{dE}{dt} \right\rangle_{\rm 0PN} &= 
\frac{32}{5} \left(\frac{\mu}{M}\right)^2 \, \notag \\
& \times \frac{(1-e^2)^{3/2}}{p^5}
{\left(1+\frac{73}{24}~e^2 + \frac{37}{96}~e^4\right)}.
\end{align}
In this expression, the flux no longer diverges as $e \to 1$,
but instead it goes to zero, which is also not ideal.

Far more useful is to compute not the flux, but rather 
the total energy radiated during one radial libration.
We find this by multiplying the flux by the radial period.
Expanded in $p^{-1}$, the radial period carries a factor
of $(1-e^2)^{-3/2}$ which 
exactly cancels the offending term in Eqn.~\eqref{eqn:flux0PNp},
leaving
\begin{align}
T_r \left\langle \frac{dE}{dt} \right\rangle_{\rm 0PN}
& = \frac{64 \pi}{5} \frac{\mu^2}{M} \frac{1}{p^{7/2}}
\left(1+\frac{73}{24}~e^2 + \frac{37}{96}~e^4\right).
\end{align}
Thus, for a given PN order, the energy radiated is finite
and nonzero in the limit $e \to 1$.

The total energy radiated in one radial period can be written in
a form reminiscent of Eqn.~\eqref{eqn:energyfluxInf}
\begin{align}
 T_r \left\langle \frac{dE}{dt} \right\rangle
&= 
\frac{64 \pi}{5} \frac{\mu^2}{M} \frac{1}{p^{7/2}} 
\l \bar{\mathcal{L}}_0 
+ \frac{\bar{\mathcal{L}}_1}{p^1} 
+ \frac{\bar{\mathcal{L}}_{3/2}}{p^{3/2}} + \cdots \r,
\end{align}
with further terms coming every half-order in $p$.
Through 2.5PN the $\mathcal{L}$ terms are
\begin{align}
\bar{\mathcal{L}}_0 &=
1+\frac{73}{24}~e^2 + \frac{37}{96}~e^4 \\
\bar{\mathcal{L}}_1 &=
-\frac{239}{336} - \frac{5065}{672} e^2 - \frac{211}{128} e^4
+ \frac{2393}{5376} e^6\\
\bar{\mathcal{L}}_{3/2} &=
4 \pi
+\frac{1375 \pi }{48} e^2 +\frac{3935 \pi }{192} e^4 +\frac{10007 \pi }{9216} 
e^6 \nonumber \\
& \hspace{10ex} + \frac{2321 \pi}{221184} e^8 -\frac{237857 \pi}{88473600}e^{10}
+ \cdots
\\
\bar{\mathcal{L}}_{2} &=
-\frac{11623}{4536} -\frac{328673}{4536} e^2 -\frac{18668}{189} e^4\nonumber\\
& \hspace{18ex} -\frac{20477}{2016} e^6 + \frac{61703}{64512}  e^8\\
\bar{\mathcal{L}}_{5/2} &=
-\frac{127 \pi }{672}
-\frac{2461 \pi}{42} e^2
-\frac{5363069 \pi }{43008} e^4 
-\frac{6867607 \pi }{387072} e^6
\nonumber \\
& \hspace{6ex}
+\frac{9437735 \pi }{7077888} e^8
   +\frac{10400743 \pi }{123863040} e^{10}+ \cdots
\end{align}
Note in particular 
that at 2PN the two separate singular pieces [$(1-e^2)^{11/2}$ and $(1-e^2)^{4}$] 
combine to form one nonsingular term.
The 1.5PN and 2.5PN terms are still not closed form, although they too are 
nonsingular

\clearpage

\bibliography{angmom}

\end{document}